\documentclass[manuscript]{acmart}

\usepackage{enumerate}
\usepackage{amsmath}
\usepackage{multirow, makecell}
\usepackage{pifont}
\usepackage{subfig}
\usepackage{graphicx}
\usepackage{colortbl}
\usepackage{balance}
\usepackage{xcolor}
\usepackage[space]{grffile}
\usepackage{lscape}
\usepackage{caption}
\usepackage{commath}
\usepackage{tablefootnote}
\usepackage{tikz}
\usepackage{wrapfig}
\usepackage{parallel}

\AtBeginDocument{%
  }

\newcommand{\changed}[1]{\textcolor{black}{#1}}

\newcommand{\bqa}{Bankura}
\newcommand{\dgp}{Durgapur}
\newcommand{\kgp}{Kharagpur}
\newcommand{\ccu}{Kolkata}
\newcommand{\numsites}{30}
\newcommand{\numoccupants}{46}
\newcommand{\males}{27}
\newcommand{\females}{19}
\newcommand{\datamonths}{six months}
\newcommand{\numcities}{four}

\newcommand{\ourmethod}{\textit{DALTON}}
\newcommand{\cmark}{\ding{51}}
\newcommand{\xmark}{\ding{55}}

\acmConference[Conf acronym 'XX]{Make sure to enter the correct
  conference title from your rights confirmation emai}{June 03--05,
  2018}{Woodstock, NY}

\setcitestyle{sort&compress}
\setcopyright{none}
\renewcommand\footnotetextcopyrightpermission[1]{}

\usepackage{tcolorbox}

\tcbuselibrary{theorems}
\newtcbtheorem
  []
  {lesson}
  {Key Lesson}
  {%
    theorem name,%
    colback=white,
    colframe=black!90,
    fonttitle=\bfseries,
  }
  {tay}

\begin{document}
\title{Exploring Indoor Air Quality Dynamics in Developing Nations: A Perspective from India}

\author{Prasenjit Karmakar}
\email{prasenjitkarmakar52282@gmail.com}
\affiliation{%
	\institution{IIT Kharagpur}
	\country{India}
}

\author{Swadhin Pradhan}
\email{swadhinjeet88@gmail.com}
\affiliation{%
	\institution{Cisco Systems}
	\country{USA}
}

\author{Sandip Chakraborty}
\email{sandipc@cse.iitkgp.ac.in}
\affiliation{%
	\institution{IIT Kharagpur}
	\country{India}
}

\renewcommand{\shortauthors}{Karmakar, et al.}

\begin{abstract}
Indoor air pollution is a major issue in developing countries such as India and Bangladesh, exacerbated by factors like traditional cooking methods, insufficient ventilation, and cramped living conditions, all of which elevate the risk of health issues like lung infections and cardiovascular diseases. With the World Health Organization associating around 3.2 million annual deaths globally to household air pollution, the gravity of the problem is clear. Yet, extensive empirical studies exploring these unique patterns and indoor pollution's extent are missing. To fill this gap, we carried out a \datamonths{} long field study involving over \numsites{} households, uncovering the complexity of indoor air pollution in developing countries, such as the longer lingering time of VOCs in the air or the significant influence of air circulation on the spatiotemporal distribution of pollutants. We introduced an innovative IoT air quality sensing platform, the Distributed Air QuaLiTy MONitor (\ourmethod{}), explicitly designed to meet the needs of these nations, considering factors like cost, sensor type, accuracy, network connectivity, power, and usability. As a result of a multi-device deployment, the platform identifies pollution hot-spots in low and middle-income households in developing nations. It identifies best practices to minimize daily indoor pollution exposure. Our extensive qualitative survey estimates an overall system usability score of $2.04$, indicating an efficient system for air quality monitoring.
\end{abstract}



\maketitle

 \section{Introduction}
\label{sec:intro}

\textbf{Key Motivation:} Indoor air pollution is a significant factor contributing to respiratory and cardiovascular diseases, taking a toll of approximately $3.2$ million lives annually~\cite{whoindoor}. This alarming statistic highlights an urgent need for comprehensive global efforts to tackle this often-neglected crisis~\cite{whodeaths}. It is particularly dire in developing nations such as India, Chad, Bangladesh, etc., where a variety of factors contribute substantially to the problem~\cite{mostpolluted}. In many of these countries, traditional cooking practices often involve burning biomass fuels such as wood, agricultural waste, etc., releasing harmful pollutants like carbon monoxide, nitrogen dioxide, and various organic compounds into homes~\cite{chartier2017comparative}. This issue is exacerbated by inadequate ventilation, particularly in densely populated and cramped areas like slums, where poor airflow allows these pollutants to build up to dangerous levels. The design of indoor spaces, including room structures and floor plans, further complicates the challenge of managing air quality. Far from being a mere discomfort, these indoor air pollutants pose serious health risks, penetrating deep into the lungs and causing diseases like chronic obstructive pulmonary disease (COPD), pneumonia, and lung cancer, and also contributing to cardiovascular diseases such as heart attacks and strokes~\cite{whohealthrisks}. Women and children are particularly vulnerable, spending more time indoors and thus being more exposed to these hazards~\cite{whoindoor}. Therefore, we require a comprehensive system that can consider all aspects of the indoor environment to understand the pollution dynamics and its root causes to actuate preventive counter-measures in time, promoting a healthier life.

\textbf{Uniqueness of Developing Nations:} Unlike first-world nations, developing nations present a plethora of challenges to address indoor air pollution. Various factors are detailed in the following.
(i) \textit{Urbanization and Housing Design}: The challenge of indoor air pollution intensifies in the rapidly urbanizing nations of India and Bangladesh. Here, the proliferation of urban slums is marked by poor housing designs that contribute to stagnant air and elevated pollutant levels. These dense settlements lack proper ventilation infrastructure, leading to a significant accumulation of indoor pollutants~\cite{kulshreshtha2008indoor,anand2022assessment}. Furthermore, there are multiple times people in any indoor space in developing countries rapidly building up carbon dioxide.
(ii) \textit{Economic Constraints and Energy Sources}: In the face of economic limitations, a large segment of the population in developing countries resorts to using affordable but polluting energy sources. The reliance on solid fuels like wood, coal, cow dung cakes, and agricultural waste is prevalent for cooking and heating purposes. These fuels release harmful pollutants when combusted in open fires or traditional stoves. This issue is highlighted in a report by WHO~\cite{whoindoor}, which estimates that over 2.3 billion people globally depend on these fuels, contributing significantly to indoor air pollution.
(iii) \textit{Cultural Practices and Behaviors}: Cultural and traditional practices in many developing countries add another layer to the indoor air pollution problem. Activities such as burning incense and oil lamps, especially prevalent in religious and cultural rituals, can significantly increase indoor air pollution levels. These traditional practices, deeply embedded in the cultural fabric, present unique challenges in mitigating indoor air pollution.
(iv) \textit{Neighbor-Generated Pollution and Urban Design}: The close proximity of buildings in densely populated urban areas of developing countries leads to pollution from one building, easily affecting neighboring homes. This aspect of indoor air pollution is often overlooked but is crucial, especially in areas with poor urban planning. For example, in a congested neighborhood, an open window of \textit{Household1} can allow pollutants to enter from the kitchen exhaust of the adjacent \textit{Household2}. Subsequently, the pollutants gradually span several rooms of the \textit{Household1} according to the room structure and airflow~\cite{cao2020sensor,putri2022spatial}.
(v) \textit{Health Implications and Vulnerable Populations}: The health impacts of indoor air pollution are particularly severe for specific demographics, including children, the elderly, and women, who typically spend more time indoors. The link between indoor air pollution and respiratory diseases in these vulnerable groups is well-established~\cite{whohealthrisks}.

\textbf{Need of Unique System Design:} Thus, the indoor pollution monitoring system must be distributed across several rooms and \textit{collaborative} to explain the pollution sources. As a result, the identified sources must be reported to the user in accordance with the severity of the exposure. Depending on the layout of a room and the activities of its occupants, some pollution sources are more prevalent and result in prolonged accumulation and lingering of pollutants. It is typical for unwashed utensils in the kitchen sink to emit \textit{volatile organic compounds} (VOCs) and ethanol overnight, which spread to the living room and bedrooms, affecting the air quality. The list of \textit{actionable} pollution sources should be tailored according to their long-term impact to improve air quality in a sustainable manner.
Furthermore, the system must be easy to deploy and robust enough to recover from power and network failures with no user intervention, providing a plug-and-play \textit{user experience}. To enrich the event and activity context of the surroundings, a \textit{human-in-the-Loop} design must be employed to engage users in a closed loop low touch dialogue with the underlined air monitoring system, where the system alerts or proposes counter-measures to prevent oblivious pollution exposure. Lastly, such a platform must support remote \textit{service management} to be rolled out in scale so that any fault in the end devices can be handled and firmware fixes can be applied easily across all devices.

\textbf{Need of Custom Indoor Air Pollution Monitor and Detailed Study:} Realizing a global business opportunity~\cite{indoorairmarket}, several enterprises have developed consumer-grade indoor air quality monitors in the last decade. The consumer-grade low-cost air quality monitors available on e-commerce websites (e.g., Amazon, Alibaba, etc.) such as Pallipartners Monitor~\cite{pallipartners}, YVELINES Monitor~\cite{yvelines}, and SmileDrive Portable Monitor~\cite{smiledrive} etc., do not offer interactivity with the end-user and only display the real-time pollutant measurements in the in-build screen. The primary design objective of such devices is ease of use, ignoring the whole human-in-the-loop aspects of such systems. Whereas, commercially matured products like Prana Air Monitor~\cite{pranaair} and Airthings Monitor~\cite{airthings} provide a much better user experience and interactiveness (i.e., pollution analytics, push notifications in case of excessive exposure, and moderate hardware management and maintenance with regular firmware upgrades). These products, however, are isolated single-point monitors that lack user feedback to provide a deeper understanding of their surroundings. As a result, existing air monitoring solutions cannot capture pollution dynamics in large-scale deployments. Moreover, several recent studies have conducted field measurements to understand the distribution of indoor pollutants~\cite{mujan2021development, gao2022understanding, temprano2020indoor, kumar2016real,chartier2017comparative,putri2022spatial, cao2020sensor}. However, these works are either very small-scale (<10 measurement sites) or done in a very controlled manner within the lab setup. Recently developed IoT frameworks~\cite{verma2021sensert,hsu2017community} have employed real-time data streaming over the wireless sensor networks to compute pollution overlays~\cite{ji2019indoor,brazauskas2021real,brazauskas2021data} for buildings; but, these works are only tuned against commercial or educational environments. Therefore, comprehensive in-the-wild empirical studies exploring the unique indoor air pollution patterns, such as spread, lingering, accumulation, ventilation, etc., in the scope of developing nations are lacking in the literature.

\begin{figure}
	\captionsetup[subfigure]{}
	\begin{center}
		\subfloat[Cooking\label{fig:cooking}]{
			\includegraphics[width=0.24\columnwidth,keepaspectratio]{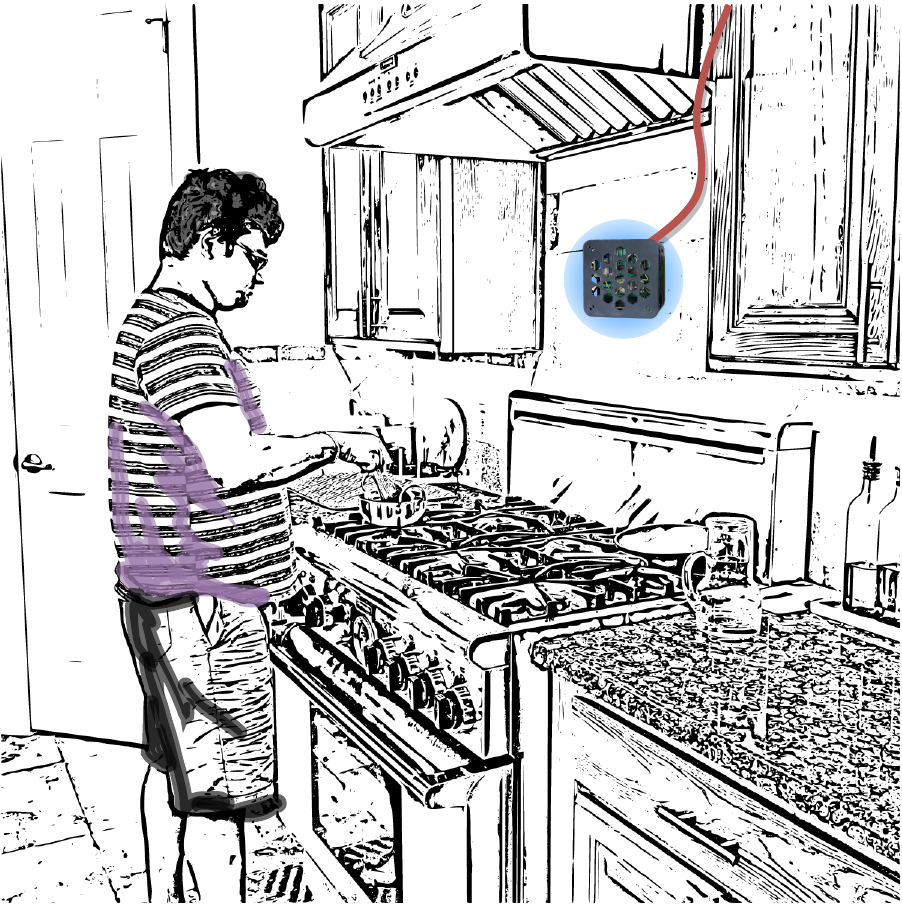}
		}
		\subfloat[Living \label{fig:living}]{
			\includegraphics[width=0.24\columnwidth,keepaspectratio]{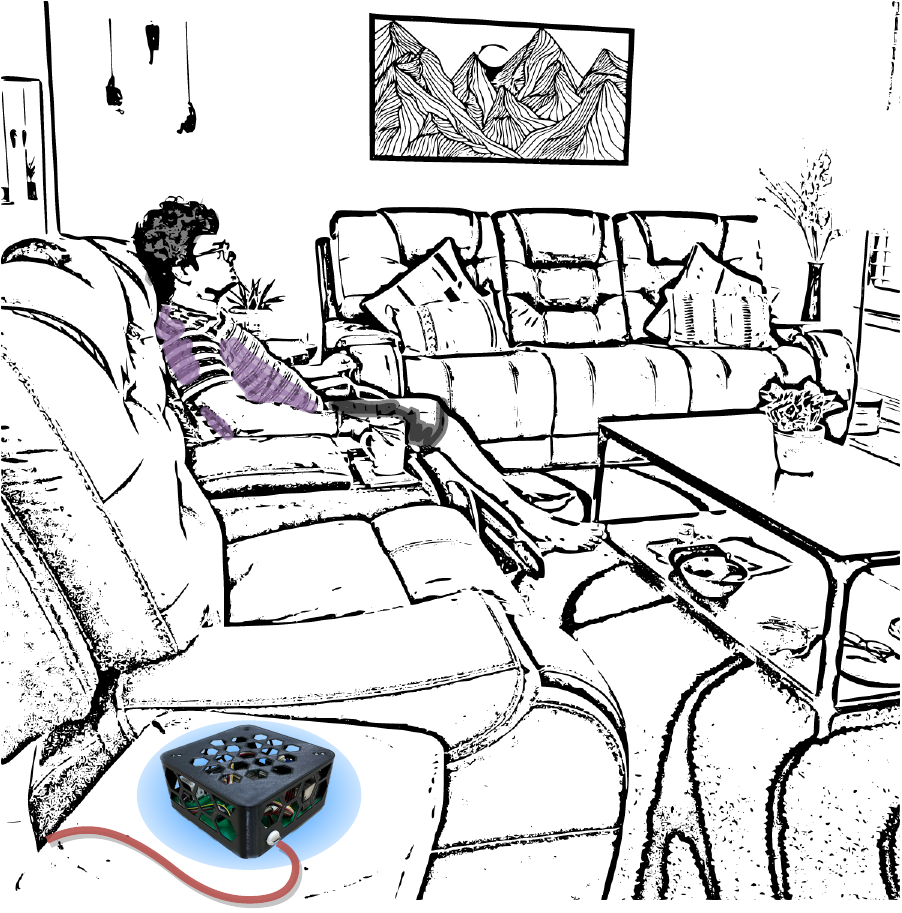}
		}
		\subfloat[Working \label{fig:working}]{
			\includegraphics[width=0.24\columnwidth,keepaspectratio]{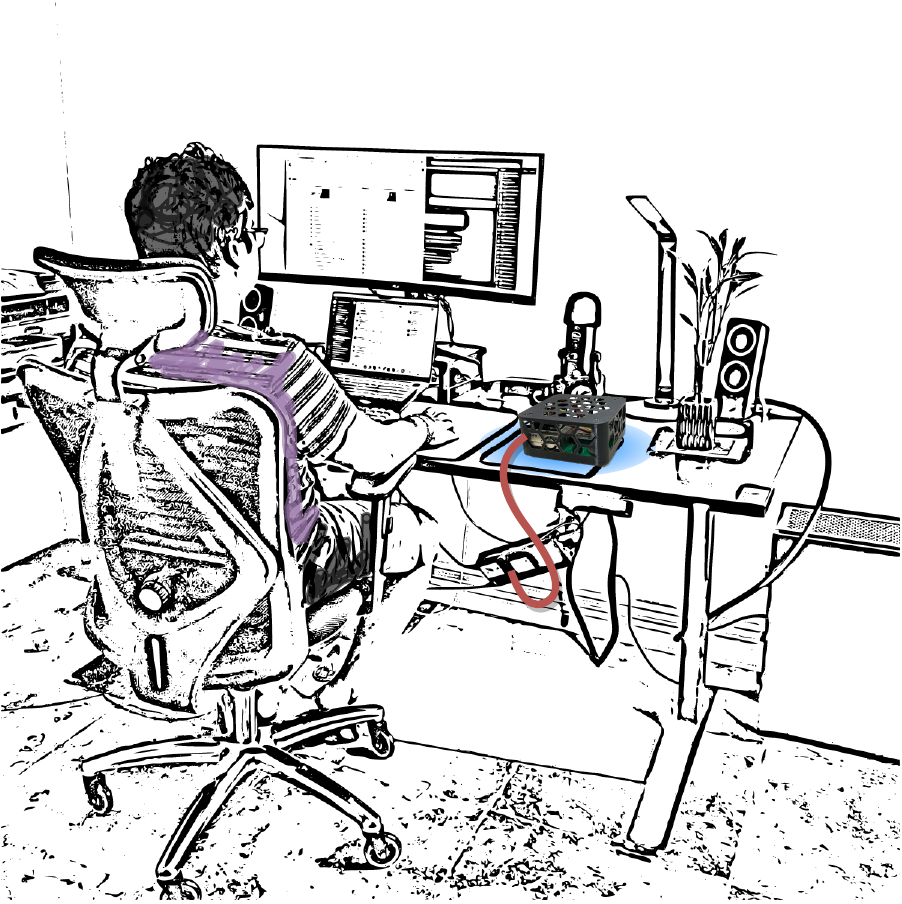}
		}
		\subfloat[\ourmethod{} module \label{fig:daim}]{
			\includegraphics[width=0.24\columnwidth,keepaspectratio]{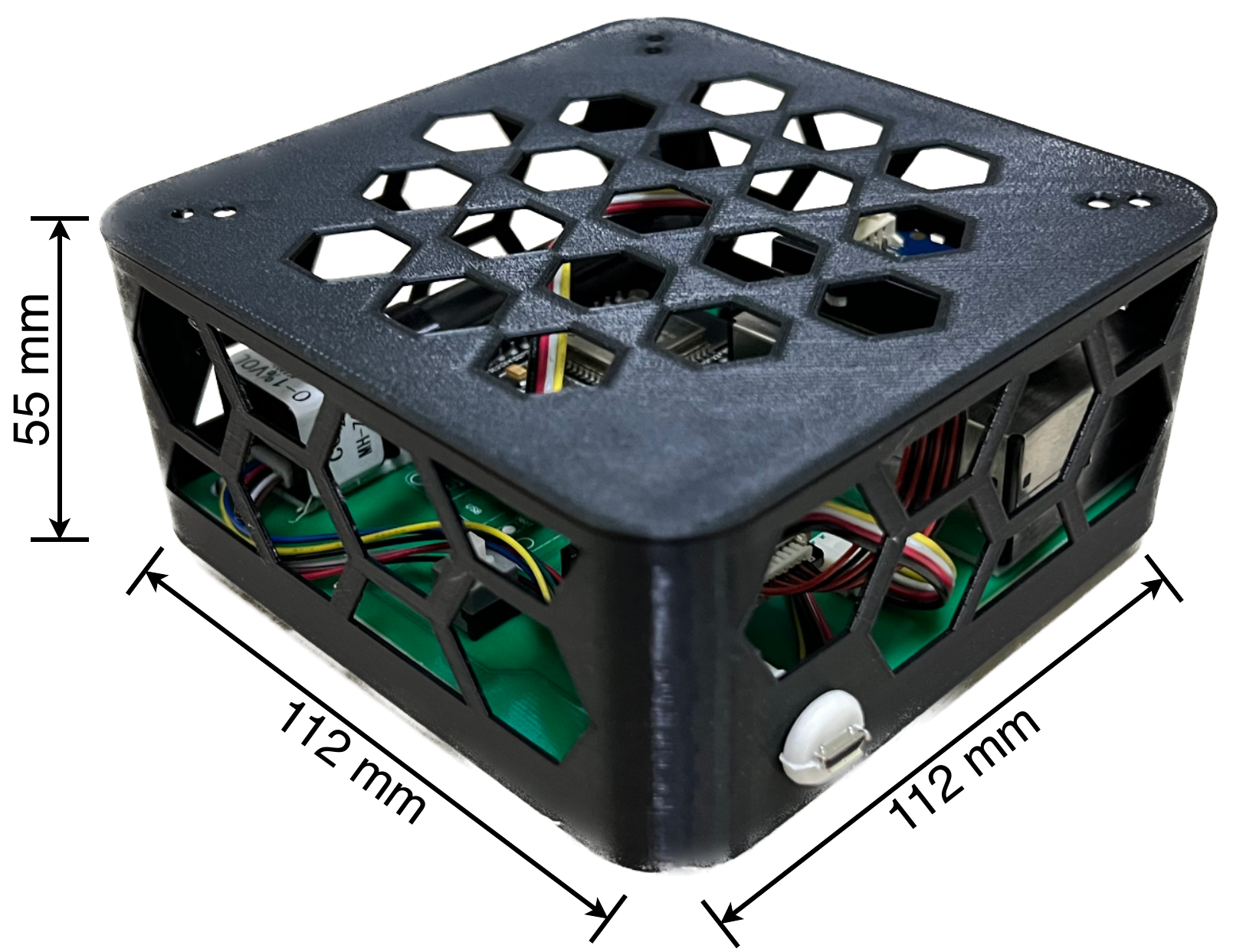}
		}
	\end{center}
	\caption{\ourmethod{} platform in Household scenarios. We deployed multiple instances of the platform throughout different household rooms to capture the origin, spread, and spatiotemporal diversity of the indoor pollutants, providing better actionable insights.}
	\label{fig:scenario}
\end{figure}

\textbf{\ourmethod{} platform Design:} To perform a comprehensive large-scale indoor air pollution study in developing nations, we designed an end-to-end framework named \ourmethod{} (\textbf{D}istributed \textbf{A}ir Qua\textbf{L}i\textbf{T}y M\textbf{ON}itor, \figurename~\ref{fig:scenario}) that can operate in a decentralized manner and provides a better picture of indoor pollution dynamics while employing an app-based interaction mechanism with the user. We developed an in-house module that is equipped with sensors to measure particulate matter (PM\textsubscript{x}), carbon oxides (CO\textsubscript{x}), volatile organic compounds (VOC), ethanol (C\textsubscript{2}H\textsubscript{5}OH) along with temperature (T), and relative humidity (RH). Further, we performed a large-scale study for \datamonths{} with \numsites{} low to middle-income households scattered across \numcities{} different cities, engaging \numoccupants{} participants, with custom-made air monitoring devices deployed across all the rooms. We observed that there are several dynamic pollution sources, and the pollutants spread and linger within the rooms depending on several factors, like the airflow across the rooms, the indoor-outdoor ventilation, the number of occupants and their activities, etc. Notably, different pollutants like CO, CO\textsubscript{2}, VOC (Volatile Organic Compounds), etc., show different spread, contamination, and lingering patterns. Indeed, there are also seasonal impacts; for instance, the occupants are more sensitive to humidity and temperature; therefore, they often switch on the exhaust only when they feel uncomfortable. However, there are instances when the temperature is low, but VOC gets trapped or lingers within the room, which the occupants fail to realize, thus impacting their health conditions significantly. Our extensive experimentation with the \ourmethod{} platform suggests that the system can identify such harmful pollution events and alert the user to take counter-measures and prevent long-term exposure.

\textbf{Contributions:} The primary contributions of this paper are listed as follows:
\begin{enumerate}

    \item We develop a low-cost air quality monitoring platform named \ourmethod{}, specifically designed to operate in scale, incorporating the indoor events and activity labels from the end user and improving upon observed system-level challenges from our extensive real-world deployments.

    \item To ensure minimal labeling fatigue, we developed a change-point-based sensor grouping mechanism to associate air pollution context with indoor events and activities, annotated by the end user via a user-friendly speech-to-text Android application.
    
    \item We performed a large-scale study for \datamonths{} with \numsites{} low to middle-income households in \numcities{} cities in India with prototypes of \ourmethod{} deployed across all the rooms. We observed that there are several dynamic pollution sources, and the pollutants spread and linger within the rooms depending on several factors, like the airflow across the rooms, ventilation, floor plan, the behavior of the occupants and their activities, etc.
	
    \item An extensive survey on the usability (PSSUQ-score $2.04$) of the \ourmethod{} platform with \numoccupants{} participants of the study ironed out the primary strengths and shortcomings of the current design. It estimated the resiliency of the platform in real-world scenarios such as network outages, power cuts, fall damage, etc., along with portability, effectiveness, and user-friendly design of the platform in providing indoor air quality-related actionable insights.
\end{enumerate}

\section{Literature Survey}
\label{sec:relatedwork}
In this section, we review existing literature on indoor air quality in developed and developing countries. Our analysis reveals multiple detailed exploratory studies in developed nations, which can be categorized into air monitoring platforms, pollution exposure alerts, health effects of indoor activities, and impact of indoor pollutants. In contrast, developing countries have limited research due to governmental inaction and low awareness. We underscore the necessity for comprehensive studies in developing countries, considering cultural, socio-economic, and architectural differences compared to developed nations.

\subsection{Studies on Developed Countries}
\subsubsection{Air Monitoring Platforms}
Over the last decade, several studies related to indoor air pollution have been conducted in developed countries like the USA, China, Korea, United Kingdom etc., that proposed real-time visualization tools like pollution overlays in educational~\cite{brazauskas2021real,brazauskas2021data}, office building~\cite{mujan2021development,ji2019indoor}, and rural~\cite{li2022field} setup and provided a medium to understand the pollution dynamics over time. Works like~\cite{kim2013inair,cheng2014aircloud,qin2023system,zhong2021complexity,snow2019performance,yang2021distributed,sun2022c} have deployed sensors across buildings to visualize, analyze, and forecast pollution to plan preventive measures. Moreover, authors in~\cite{zhu2022dynamic,yang2021distributed,ding2019octopus} proposed a method to adjust the ventilation rate to the household as a countermeasure triggered by the monitoring platform. 
The work~\cite{10.1145/2030112.2030150} incorporates indoor tracking with WiFi footprints to compute personalized pollution exposure indoors from static air monitors. Whereas in work~\cite{maag2018w}, authors have developed a wearable sensing module that can be placed on the user's body to track personalized pollution exposure. 

\subsubsection{Notifications \& Alerts}
Indoor comfort also correlates with indoor air quality. For instance, the work~\cite{mujan2021development,SHEIKHKHAN2021111363} found that temperature and indoor air quality significantly correlate with reported indoor environment quality. In~\cite{jiang2011maqs,zhong2020hilo}, the authors have developed a mobile sensing module that can be placed at any location of the indoor space to measure pollutants, where~\cite{zhong2020hilo} utilized the smartwatch to connect to the monitoring platform and trigger pollution alerts in terms of vibrations, therefore, sustaining the comfort levels of the indoor space. The work~\cite{9278913,8003185} developed a WiFi-enabled custom monitoring device and performed initial experiments to trigger exposure alerts and control the air purifier on a scale as a proof of concept. These works showed that timely alerts about pollution exposure can induce awareness and proactive measures from the user end to improve the air quality of the indoor space. 

\subsubsection{Indoor Activities}
\label{sec:indoor_act}
Several studies~\cite{moore2018managing,zhong2020hilo,kim2020awareness,ingelrest2010sensorscope} considered user interactions and activity annotation to associate with pollution dynamics. For example,~\cite{moore2018managing} engaged six families to annotate their daily activities with small text messages while measuring air pollutants from a multi-monitor deployment. Moreover, they also interviewed the participants to understand their reasoning behind increased pollution levels in their daily activities. For example, one participant in this study mentioned that using olive oil during cooking produces more pollutants than avocado oil. Studies have also shown that occupants' activities and significant events, such as lunch breaks, meetings, etc., influence indoor air quality significantly. ~\cite{fang2016airsense} proposed a machine learning-based approach to detect occupant activities like cooking, smoking, and spraying in small apartments based on sensing the air pollutants. In~\cite{verma2021racer}, the authors have inferred limited indoor activities such as cooking, window opening or closing, corridor walking, etc., from air quality.

\subsubsection{Health Impacts}
The work~\cite{10.1145/3494322.3494704} has developed a sensing device to estimate the health impact of indoor air quality on asthmatic patients. The work~\cite{9703317} has forecasted the effects of real-time indoor PM\textsubscript{2.5} on Peak Expiratory Flow Rates (PEFR) of Asthmatic Children in Korea between 8–12 years of age. Moreover, the work~\cite{10.1145/2786451.2786481} observed that several environmental factors, like smoke, industrial emissions, car emissions, etc, impact our lungs and cause health problems. In work~\cite{WOLKOFF2018376}, authors observe that indoor air further causes sensory irritation symptoms in eyes and airways, fatigue, and headache, reducing work performance (i.e., Indoor Productivity Index) in office environments, hampering sleep quality. Low humidity and cold temperatures can also cause increased virus survival rates; thus, viruses like Influenza can live longer and cause infection in occupants' respiratory tracks~\cite{THAM2016637}. The above building-related illnesses are referred to as sick building syndrome, which is attributable to various causes like low ventilation, VOCs, and moisture.

\subsection{Studies on Developing Countries}
Several works from the literature have studied outdoor pollution dynamics in different developing countries~\cite{patel2022samachar,saini2020comprehensive,rai2017end,mahajan2022design}. For instance, \cite{pramanik2023aquamoho} utilized the government-deployed air monitoring stations in the city to estimate air quality at several locations with local thermal and humidity signatures, and~\cite{rakholia2022ai} has hourly forecasted particulate matter levels in the city. Whereas,~\cite{abidi2022complexity} analyzed the effectiveness of government-enforced traffic control policy with the particulate matter measurements at the Delhi-NCR region in India. \changed{Several studies~\cite{jha2022expert,dutta2017towards,barot2020qos,ma2020fine,wu2020sharing, iyer2022modeling} have utilized static and mobile low-cost air monitors to measure fine-grained outdoor air quality and build services like pollution heat-maps, alerts in case of dangerous pollution levels, etc.} Further,~\cite{gulia2020ambient} has shown that unreliable low-cost air monitoring devices~\cite{concas2021low,das2022low} coupled with air monitoring stations and satellite-based remote sensing potentially estimate regional scale air quality. Other works~\cite{liu2018third,zhang2016outdoor,zhang2016estimating} have shown that camera images, weather, and course-grained data from air quality monitoring sites can estimate air quality at personal scales.

\subsubsection{Limited Studies on Indoors}
\label{sec:indoor_studies}
In contrast to outdoors, indoors shows significantly different pollution patterns and mainly depend on indoor activities, room structure, and ventilation, as conveyed by extensive studies in developed countries. Therefore, the pollution dynamics vary from house to house and at different times of the day, requiring multiple monitors to measure the changes in pollutants across the household. Due to the lack of involvement of governmental bodies and less awareness among the general public~\cite{afolabi2016awareness} about the severe health impacts~\cite{raju2020indoor,kurmi2012indoor,ali2021health} of bad indoor air, such studies are very limited in developing countries. However, over the last decade, there have been a few case studies to analyze specific scenarios like the danger of arsenic exposure through inhalation from the burning of cow dung cakes~\cite{pal2007additional}, particulate matter variation in single-side and cross-ventilated rooms~\cite{bedi2023exploratory}, and indoor air quality measurement for commercial buildings~\cite{kumar2016indoor}. Very few studies deployed particulate matter~\cite{patel2017spatio} and Carbon-Oxide~\cite{indu2023sensor} sensors in low to middle-income households to analyze pollutant spread during cooking. However, these works obtained limited observations due to the small scale of experiments, less household diversity, and measurement of only a few pollutants like particulate matter, Carbon oxides, etc. The outcomes of studies in developed countries are not directly transferable in a developing setting due to different infrastructural, societal, and cultural reasons. The primary dissimilarities are described in the following.

\subsubsection{Dissimilarities from Developed Countries}
Due to the way houses are built without considering ventilation~\cite{holden2023impact}, pollutants accumulate frequently and remain in an indoor space for an extended period. The worst-case scenario is that it gets trapped within a room and remains there until it is ventilated. Due to rapid and unplanned urbanization, developing countries like India, China, Chad, etc., have densely packed neighbourhoods~\cite{dave2010high,shah2021causes,peng2023identification} in residential areas. Thus, pollution can also spread from one house to another~\cite{weaver2019air}. Moreover, air quality mainly depends on the underlying activity being performed in the indoor environment. Unlike Developed countries, developing countries exhibit divergent practices and activities, such as lighting candles and incense sticks due to religious practices that increase particulate matter and VOC contamination~\cite{derudi2014emission,shrestha2020incense}. Moreover, daily cooking is more common in developing countries whereas, in developed countries, people mostly rely on pre-cooked food or restaurants~\cite{wolfson2016public,mognard2023eating}. Further, in developing countries, people use raw food ingredients to prepare a meal, whereas people in developed countries mostly use processed or packaged food items~\cite{lavelle2016barriers} to save preparation time and reduce cleaning efforts. Therefore, the disposable food residue in developing countries behaves as an additional pollution source~\cite{wu2010emission,zhang2020malodorous} for VOC, Ethanol, and methane if left open in the household. Estimated 70\% of households in developing countries use fuels such as wood, dung, and crop residues for cooking~\cite{international2010household}. Studies have shown that emissions like particulate matter, carbon oxides, VOC, ethanol, etc., from such energy sources hugely impact our health~\cite{ezzati2002health,mocumbi2019cardiovascular,guercio2021exposure}. More importantly, infants, little children, elderly women~\cite{ali2021health}, and older adults are most affected by indoor pollution in developing countries as they spend most of their time indoors with several active pollution sources around them.

\subsection{Key Takeaways}
Considering the above dissimilarities among developing countries and the limitations of the current indoor air quality studies, here are the key takeaways to further explore unique indoor pollution patterns in developing countries.
\begin{itemize}
    \item \textbf{Deployment Scale:} As discussed in Section~\ref{sec:indoor_studies}, the measurement studies are small-scale and do not convey a holistic understanding of indoor pollution dynamics in developing countries. For instance, spread, accumulation, and lingering patterns in low to middle-income households are yet to be explored.

    \item \textbf{Overlooked Pollutants:} Most of the studies done in developing/developed countries have analyzed only the presence of particulate matter and carbon oxides in household air. However, unlike in developed countries, VOCs and ethanol are more prominent and frequently occurring pollutants in the households of developing countries. Thus, further studies are required to understand the dynamics of such explicit pollutants.

    \item \textbf{Spread of Pollutants:} Unlike well-explored outdoors, indoor pollution dynamics are not studied at a personal scale, especially in developing countries with very complex pollution patterns and heterogeneous pollution sources. Due to the unplanned housing construction and lack of ventilation, pollutants easily spread across the household. Multiple sensors must be deployed in a household to measure such spreading patterns. However, the literature lacks such studies in developing countries.

    \item \textbf{Impact of Activities:} As discussed in Section~\ref{sec:indoor_act}, indoor activities hugely influence the indoor air quality. Developing countries have very different types of daily practices from developed countries. To understand and correlate such activities with changes in air quality and associate different pollution events with root cause activities, occupants must participate in the study, which is yet unexplored.
\end{itemize}

\section{Development of \ourmethod{}}
\label{sec:develop}

Contemplating the socio-economic factors of the developing countries and the lack of comprehensive field experimentation, we have developed a custom hardware module that is very portable and easy to use, underscoring seamless integration with any household. Multiple such devices are deployed in different rooms to monitor the pollution signature across a household. This sensing platform is named \textit{\textbf{D}istributed \textbf{A}ir qua\textbf{L}i\textbf{T}y m\textbf{ON}itor} (\ourmethod{}). Moreover, the following section describes the primary design choices followed during the platform's development to sustain a large-scale deployment.

\subsection{Scalable Design Requirements}
To comprehend the overall pollution dynamics of a household in developing countries, we need to measure the pollutants in various locations in the indoor space. The primary reason is the co-existence of multiple pollution sources. For example, the simultaneous burning of fuel in the kitchen and ritual practices in the living space impact the air quality differently in the adjacent rooms. Moreover, due to highly congested neighborhoods in developing nations, pollutants from the side by the house can also degrade the air quality. Thus, single-point measurement is not a practical approach to estimating the complex nature of pollutants and their spread. Thus, a \textit{multi-device deployment} is necessary in this scenario. However, such a multi-device system manifests several system-level challenges that must be addressed to maintain sustainability in a real-world deployment. According to our observations, the primary requirements that need to be satisfied for a viable multi-device platform are as follows:

\begin{itemize}
    \item \textbf{Cost Effective:} The sensing device must be low-cost and affordable to be widely adopted by the masses in developing countries. However, the device should measure the most frequently occurring pollutants in a general household scenario. Therefore, the set of sensors should be selected considering a holistic observation of the most frequent pollutants in developing countries and the development cost.

    \item \textbf{Portable Hardware Design:} Sensors are mounted over portable enclosures that can be deployed anywhere in the house. The device assembly should be stable enough to prevent any fall damage; however, it must not impose any bias in measurement due to obscured sensors. Thus, the packaging of the selected sensors is crucial to quickly deploy the devices in a large-scale study and measure the unbiased pollutants in an indoor space.

    \item \textbf{Remote Maintenance:} Often, incremental firmware updates must be applied to the devices to patch existing bugs or enable newly developed features. However, configuring each device manually and individually is not feasible in a large-scale deployment. Moreover, debugging a certain malfunction in a set of devices is not doable if the devices need to be physically accessed. Thus, it is necessary to have a remote management mechanism that facilitates debugging and over-the-air updates to the devices and sustains error-free deployment.

    \item \textbf{Fault Management:} Apart from faulty sensors, a device can malfunction due to electrical surges or uninitialized sensors after a power outage, resulting in wrong measurements of air pollutants. The device should have a fault detection and recovery mechanism for a resilient sensing platform.

    \item \textbf{Human-in-the-Loop Annotation:} A multi-device pollution sensing setup enables granular measurement of pollutants in a household. However, it is very challenging to identify the sole reason behind perturbed pollutants without knowing indoor activities. Thus, a human-in-the-loop design in which the users can provide feedback to the sensing platform is necessary to obtain this activity context of the indoor environment.

    \item \textbf{User-friendly Interaction:} The system should smoothly integrate into one's daily lifestyle without incurring a significant cognitive load due to the human-in-the-loop design. Thus, the interaction interface must be easy to use and user-friendly for wide adaptability and better user participation.
\end{itemize}

\noindent To fulfill the above design requirements, we have accordingly designed the sensing device and the backbone IoT infrastructure of the sensing platform, incorporating remote debugging, updating firmware, and several data processing and fault recovery mechanisms. The hardware module and the IoT backbone design are explained as follows.

\begin{wrapfigure}[11]{r}{0mm}
	\centering
	\includegraphics[width=0.5\columnwidth,keepaspectratio]{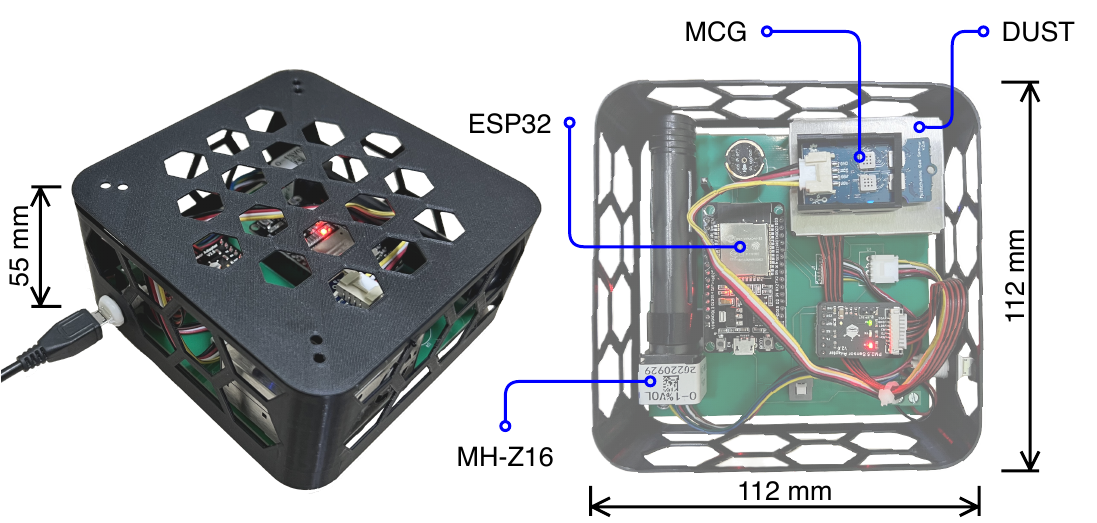}
	\caption{Prototype of \ourmethod{} sensing device.}
	\label{fig:box3}
\end{wrapfigure}

\subsection{Low-cost Portable Hardware Design}
The hardware prototype of \ourmethod{} sensing device is shown in \figurename~\ref{fig:box3}.
It is a portable lunchbox size (112 mm $\times$ 112 mm $\times$ 55 mm) module, equipped with multiple research-grade sensors that together measure the concentration of most occurring pollutants, such as \textit{Particulate matter} (PM\textsubscript{x}), \textit{Nitrogen dioxide} (NO\textsubscript{2}), \textit{Ethanol} (C\textsubscript{2}H\textsubscript{5}OH), \textit{Volatile organic compounds} (VOCs), \textit{Carbon monoxide} (CO), \textit{Carbon dioxide} (CO\textsubscript{2}), etc. in a household of developing countries, along with \textit{Temperature} (T) and \textit{Relative humidity} (RH). We utilize the ESP-WROOM-32 chip as the on-device processing unit that packs a dual-core Xtensa 32-bit LX6 MCU with WiFi $2.4$GHz HT40 capabilities. \tablename~\ref{tab:ovl_spec} details the sensing device's overall specifications. The connectivity board is a two-layer printed circuit board (FR4 material). The outer shell of the module is a 3D printed (PLA+ material) hollow structure with honeycomb holes so that the air within the module is the same as outside, resulting in unbiased measurement of pollutants (at a sampling frequency of $1$Hz). The overall cost of assembling the module is around \$250. Further, multiple replicas of such sensing devices are built to conduct a large-scale measurement study in \numsites{} households across \numcities{} cities in India, involving \numoccupants{} participants over  \datamonths{}.

\begin{table}[]
	\centering
	\scriptsize
	\caption{Overall specifications of \ourmethod{} sensing device}
	\label{tab:ovl_spec}
	\begin{tabular}{|l|l|l|l|l|l|c|c|c|c|c|} 
		\cline{1-3}\cline{5-11}
		\multicolumn{3}{|c|}{\textbf{System Specification}}                                                              & \multicolumn{1}{c|}{\textbf{}} & \multicolumn{2}{c|}{\multirow{2}{*}{\textbf{Sensor}}} & \multicolumn{5}{c|}{\textbf{Operational Details}}                                                                                                                                                                                                                                                                                                                                                         \\ 
		\cline{1-3}\cline{7-11}
		\multicolumn{2}{|l|}{Microprocessor}   & \begin{tabular}[c]{@{}l@{}}Xtensa®32-bit LX6\\Clock 80\textasciitilde{}240 MHz\end{tabular} & \multicolumn{1}{c|}{\textbf{}} & \multicolumn{2}{c|}{}                                 & \textbf{Range}                               & \textbf{Resolution}  & \textbf{Error Margin}                                                                                                                   & \begin{tabular}[c]{@{}c@{}}\textbf{Response}\\\textbf{Time}\end{tabular} & \begin{tabular}[c]{@{}c@{}}\textbf{Operational }\\\textbf{Temp \& RH}\end{tabular}                               \\ 
		\cline{1-3}\cline{5-11}
		\multirow{2}{*}{Memory} & ROM          & 448 KB                                                                  &                                & \multirow{3}{*}{\changed{DUST~\cite{dust}}}   & PM\textsubscript{x}                         & 0\textasciitilde{}500 $\mu g/m^3$                    & 1                    & \multicolumn{1}{l|}{\begin{tabular}[c]{@{}l@{}}$\pm$ 10 $\mu g/m^3$ @0\textasciitilde{}100 $\mu g/m^3$\\$\pm$ 10\% @100\textasciitilde{}500 $\mu g/m^3$\end{tabular}} & \multirow{3}{*}{$\leq$10 s}                                                   & \multirow{3}{*}{\begin{tabular}[c]{@{}c@{}}-10\textasciitilde{}60 \textdegree C\\0\textasciitilde{}99\%\end{tabular}}  \\ 
		\cline{2-3}\cline{6-9}
		& SRAM         & 520 KB                                                                  &                                &                         & RH                          & 0\textasciitilde{}99 \%                      & \multirow{3}{*}{0.1} & \multicolumn{1}{l|}{$\pm$ 2\%}                                                                                                             &                                                                          &                                                                                                                \\ 
		\cline{1-3}\cline{6-7}\cline{9-9}
		\multicolumn{2}{|l|}{Connectivity}     & Wi-Fi 2.4GHz                                                            &                                &                         & T                           & -20\textasciitilde{}99 \textdegree C                 &                      & \multicolumn{1}{l|}{$\pm$ 0.5 \textdegree C}                                                                                                       &                                                                          &                                                                                                                \\ 
		\cline{1-3}\cline{5-7}\cline{9-11}
		\multicolumn{2}{|l|}{Scan Rate (Hz)}    & 1                                                                       &                                & \multirow{4}{*}{\changed{MCGS~\cite{mcgs}}}   & NO\textsubscript{2}                         & 0.1\textasciitilde{}10 ppm                   &                      & \multirow{4}{*}{--}                                                                                                                      & \multirow{3}{*}{$\leq$30 s}                                                   & \multirow{6}{*}{\begin{tabular}[c]{@{}c@{}}-10\textasciitilde{}50 \textdegree C\\0\textasciitilde{}95\%\end{tabular}}  \\ 
		\cline{1-3}\cline{6-8}
		\multicolumn{2}{|l|}{Max Power (W)}        & 3.55                                                                       &                                &                         & C\textsubscript{2}H\textsubscript{5}OH                      & \multirow{2}{*}{1\textasciitilde{}500 ppm}   & \multirow{2}{*}{1}   &                                                                                                                                         &                                                                          &                                                                                                                \\ 
		\cline{1-3}\cline{6-6}
		\multicolumn{2}{|l|}{Max Current (mA)}     & 760                                                                     &                                &                         & VOC                         &                                              &                      &                                                                                                                                         &                                                                          &                                                                                                                \\ 
		\cline{1-3}\cline{6-8}\cline{10-10}
		\multicolumn{2}{|l|}{Dimensions(mm)}   & 112 $\times$ 112 $\times$ 55                                                          &                                &                         & CO                          & 5-5000 ppm                                   & 0.5                  &                                                                                                                                         & $\leq$10 s                                                                    &                                                                                                                \\ 
		\cline{1-3}\cline{5-10}
		\multicolumn{2}{|l|}{Weight (g)}       & 160                                                                     &                                & \multirow{2}{*}{\changed{MH-Z16~\cite{mhz16}}} & \multirow{2}{*}{CO\textsubscript{2}}        & \multirow{2}{*}{0\textasciitilde{}10000 ppm} & \multirow{2}{*}{1}   & \multirow{2}{*}{$\pm$ 100ppm +6\%value}                                                                                                    & \multirow{2}{*}{$\leq$30 s}                                                   &                                                                                                                \\ 
		\cline{1-3}
		\multicolumn{2}{|l|}{Power Adapter}    & DC (5V, 15W)                                                            &                                &                         &                             &                                              &                      &                                                                                                                                         &                                                                          &                                                                                                                \\
		\cline{1-3}\cline{5-11}
	\end{tabular}
\end{table}

\subsection{Remote Maintenance of \ourmethod{}}
In this section, we describe the primary features of the IoT backbone and IP-independent design of the \ourmethod{} platform to enable remote management and firmware upgradation on the fly to sustain real-world deployment.

\subsubsection{IP-agnostic Design}
To manage a real-world, large-scale IoT sensing infrastructure that spans multiple local area networks, a public IP address for each module is not always attainable due to the constraints enforced by the Internet Service Provider (ISP) and the limited domain expertise of the end user. Thus, we choose a publisher-subscriber-based IP-agnostic approach where any sensing module can be uniquely identified by its ID. Such a design simplifies the initial setup procedure for the end user. Moreover, the sensing modules can set up an asynchronous communication channel among themselves by their respective device ID.

\subsubsection{Over-the-Air Admin Control}
\begin{wrapfigure}[15]{R}{0mm}
	\centering
	\includegraphics[width=0.5\columnwidth,keepaspectratio]{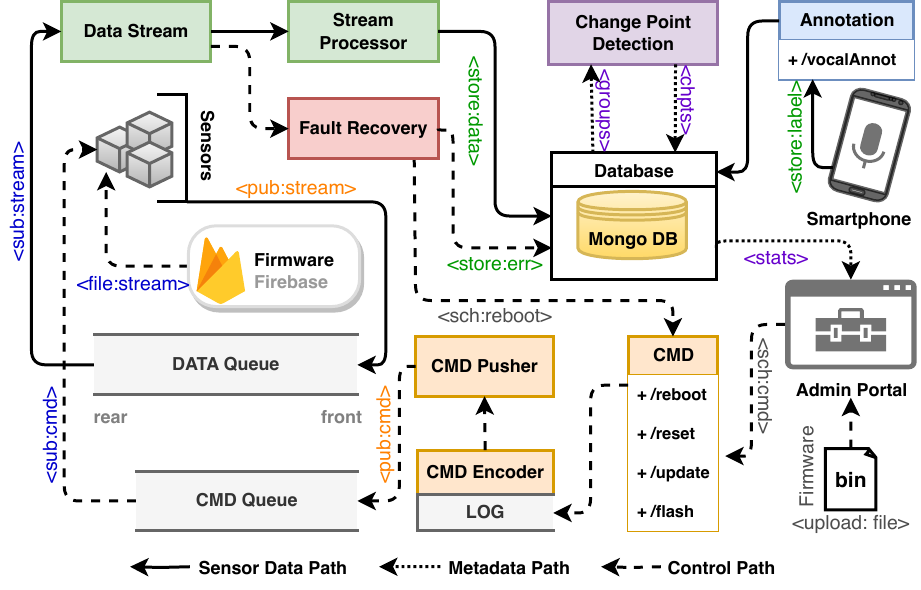}
	\caption{IoT backbone microservices of \ourmethod{} platform. The solid arrow represents streaming data. The dotted arrow represents metadata such as data statistics, change points, etc. The dashed arrow represents error handling and control signals.}
	\label{fig:backend3}
\end{wrapfigure}

\changed{To enable Over-the-Air control, the IoT backbone exposes endpoints for remotely executing commands (i.e., reboot, reset, update, flash, etc.) in the sensing modules. Upon querying a specific endpoint along with the device ID, the \textit{CMD Encoder} keeps a log and encodes the command in a format that the sensing modules understand. Subsequently, the \textit{CMD Pusher} publishes the encoded command to the \textit{CMD queue} as shown in \figurename~\ref{fig:backend3}, reliably broadcasting to all sensing modules via the asynchronous channel. At the device end, only the desired recipients of the command decode and execute further, while others drop it.}

\subsubsection{Over-the-Air Firmware Upgrade}
Moreover, we can upgrade any device's firmware from a web interface by uploading the latest bin file and pressing the flask button. The IoT backbone uploads the firmware to a \textit{firebase storage bucket}\footnote{\url{https://firebase.google.com/docs/storage} (Accessed: \today)} with a unique file descriptor and triggers a flash command via the \textit{CMD pusher} with the associated file descriptor so that the sensing modules can initiate an HTTP stream to the \textit{firebase storage bucket} and upgrade the firmware.

\subsection{Error Handling and Fault Management}
Here, we describe the stream processing pipeline, auto fault recovery mechanism, and remote debugging capabilities of the IoT backbone to ensure reliable data transfer from all the deployed devices. 

\subsubsection{Reliable Data Storage}
The IoT backbone employs multiple microservices, each responsible for sub-tasks as follows. \textit{Queuing} service hosts a MQTT broker\footnote{\url{https://mosquitto.org/} (Accessed: \today)} that manages the data queue. Queuing of the data is necessary to ensure first-in-first-out (FIFO) and one-time delivery in the underlying asynchronous wireless channel utilized by the modules. Each module publishes the pollutant measurements to a specific topic via the broker, producing a reliable data stream. Moreover, the queuing service allows both-way indirect communication among sensing modules and the rest of the IoT backbone. \textit{Data Stream Processor} subscribes to the mixed data stream of the queuing service and decouples it into individual streams corresponding to each module, storing the data in the \textit{Data Storage} as shown in \figurename~\ref{fig:backend3}.

\subsubsection{Liveness Portal}
Using a web-based liveness portal, we list all the live sensing modules sending sensor readings into the IoT backbone. Moreover, the user can find information about the disconnected modules, such as the latest timestamp when the module was live, location of deployment, view error log and plot live data to track down and resolve any problem with the module.

\subsubsection{Auto Recovery}
Upon detecting any anomaly in the streaming sensor data, the stream processor triggers the \textit{Fault Recovery} microservice with the affected device ID; the fault recovery service determines the type of fault and suitable recovery action (i.e., reboot) for the device. Consequently, it stores the error log in \textit{errorlog} collection of the database service and queries the \textit{Command Pusher} to schedule the recovery action.

\subsection{Human-in-the-Loop Labeling}

\begin{figure}[]
        \centering
	\captionsetup[subfigure]{}
		\subfloat[Change points in bedroom's device\label{fig:changepoints}]{
			\includegraphics[width=0.45\columnwidth,keepaspectratio]{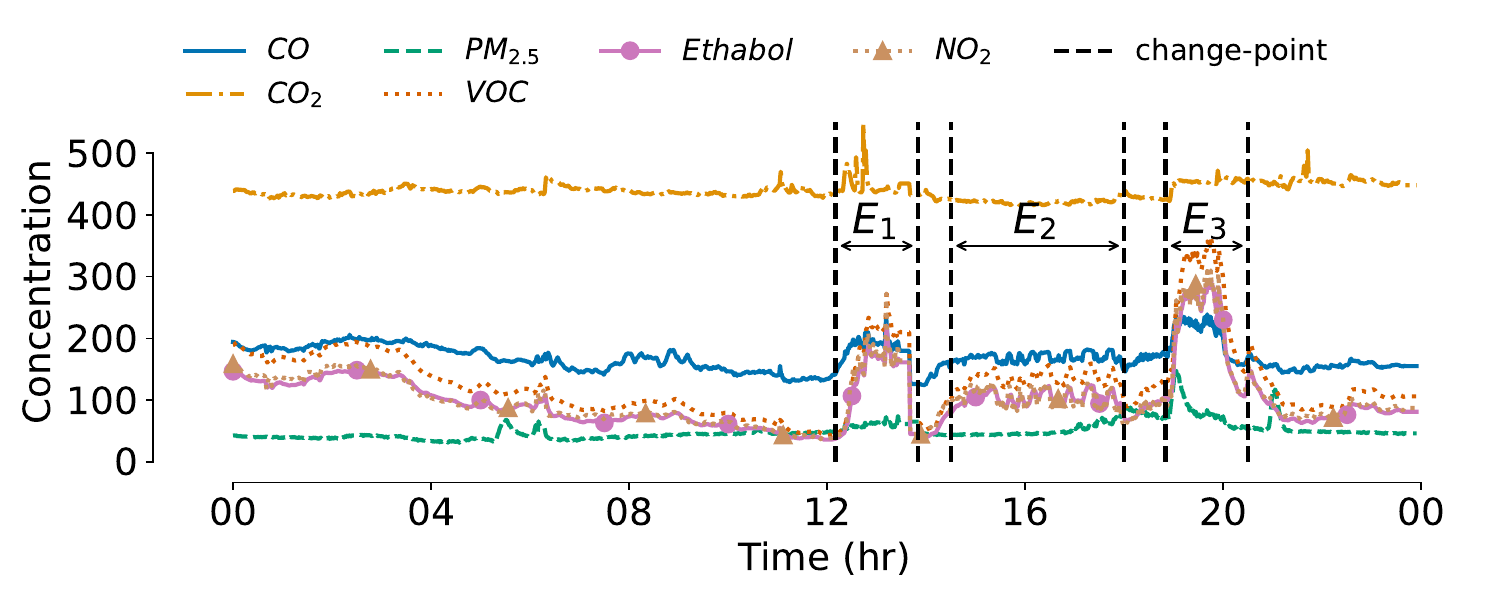}
		}\
		\subfloat[Change points in adjacent device ($m$) in kitchen\label{fig:changepoints_kit}]{
			\includegraphics[width=0.45\columnwidth,keepaspectratio]{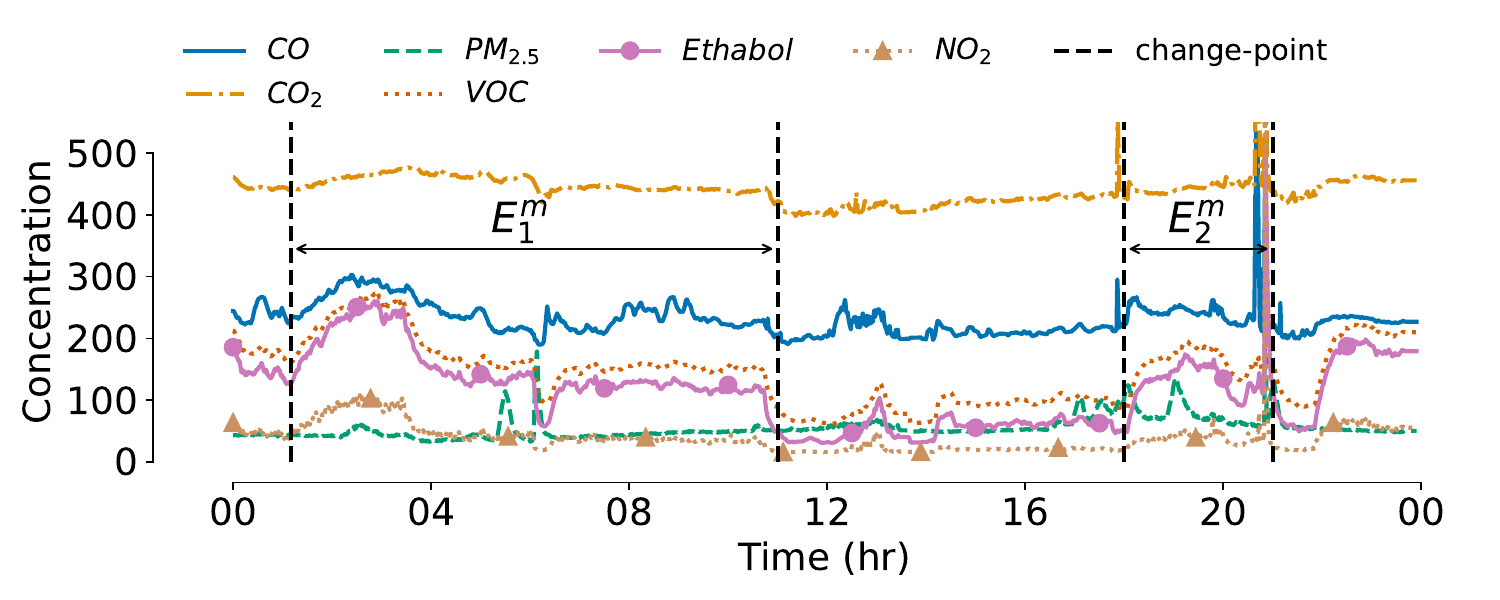}
		}\
	\caption{\changed{Detected time segments from the change points computed using the KLCPD algorithm for two adjacent devices in the bedroom and kitchen. The event $E_2^m$ in the kitchen's device and $E_3$ in the bedroom's device are associated with significant time overlap.}}
	\label{fig:chpts}
\end{figure}

We require human-in-the-loop ground-truth annotations of indoor activities within the household to associate the measured pollution data. The microservices responsible for simplifying user annotations are as follows.

\subsubsection{Change-Point Detection}
Not every activity generates pollutants; thus, asking the users to annotate all the activities will significantly waste their effort. Hence, we developed a \textit{sensing-aware} solution to collect minimally-required information by asking them to provide the activity labels corresponding to the higher concentration of pollutants observed by our developed device. A naive threshold-based solution is not a good approach here, as it may ask for frequent annotations with every peak in the measured pollution level. For example, during cooking, the sensors may observe periodic short-duration pollution peaks depending on the kitchen's ventilation and what is being cooked. Thus, we utilized a change point detection algorithm \textit{Kernel Change Point Detection} (KLCPD)~\cite{chang2019kernel} to compute change points in the pollutant concentration of the indoor. Notably, \ourmethod{} platform is oblivious of the KLCPD algorithm and can be used with other change point detection algorithms~\cite{killick2012optimal,celisse2018new,fryzlewicz2007unbalanced}. For example, \figurename~\ref{fig:changepoints} shows an instance of change points calculated for all the pollutant measures for one sensing module over the measurements of the whole day. The change points are only considered when the module senses a significant difference in air quality, reducing the number of noisy events (when the sensor data keeps varying in a small range as shown in \figurename~\ref{fig:changepoints}, event $E_2$).

\subsubsection{Sensor Association to Reduce Labeling Effort}
We deploy multiple sensor modules in a household to capture the spatiotemporal diversity of the pollutants; however, annotating data individually for all the sensing modules further increases the users' cognitive load. Therefore, depending on the time segment (pair of change points) overlaps and adjacency of the sensing modules on each floor, we associate the pollution events to a subset of modules given that the members of the subset experience similar trends of pollutants and thus, enabling us to identify different spatiotemporal groups of modules within an indoor space. For example, we associate change points in the adjacent sensing module ($m$) of the kitchen shown in \figurename~\ref{fig:changepoints_kit} with the bedroom module shown in \figurename~\ref{fig:changepoints}, for event $E_2^m$ and $E_3$ due to significant degree of time overlap. This further reduces the effective number of events detected in all the sensing modules deployed in adjacent household rooms.

\subsection{App-based User-friendly Annotation}
\label{sec:and_app}

\begin{wrapfigure}[15]{r}{0mm}
    \centering
    \includegraphics[width=0.4\columnwidth]{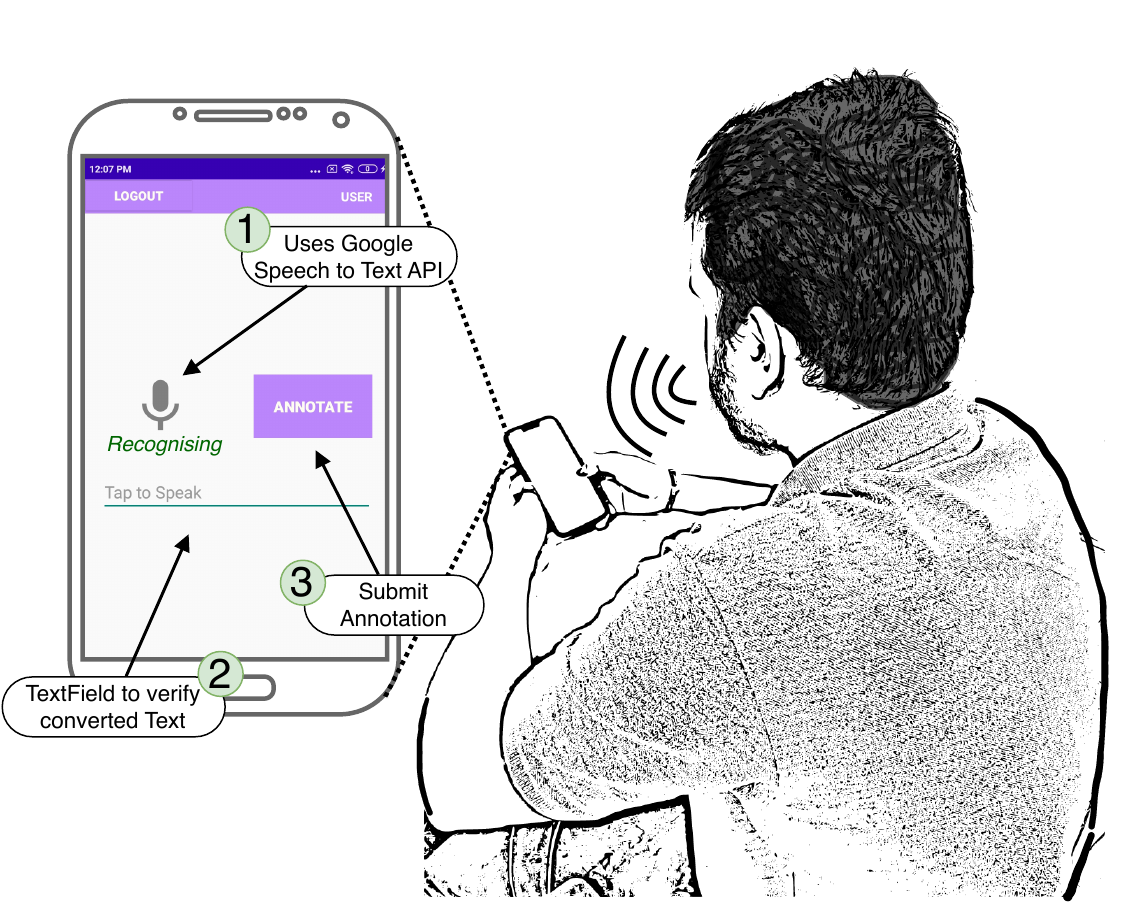}
    \caption{Developed Android application for user-friendly activity labeling. It uses Google's speech-to-text API for easy voice annotations.}
    \label{fig:app}
\end{wrapfigure}

Upon identifying a pollution event from the change point and sensor association computation, an alert is triggered to the developed Android application to annotate the causal indoor activity. The Android app is shown in \figurename~\ref{fig:app}. The app-based continuous annotation process reduces the participants' recall overhead. Moreover, the app enables vocal annotations using Google's speech-to-text API to minimize the physical and mental effort to track daily activities. After logging in with their name (only once unless intentionally logged out), the user needs to tap on the microphone icon, as shown in the figure, to activate the speech recognition and start speaking. Once the speech is correctly converted to the text, it populates the text field with the annotation text. The user can verify and edit the text if required before submitting the activity annotation with the annotate button on the app interface. The app sends a post request containing timestamped ground-truth activity labels to the \textit{Annotation} microservice of the IoT backbone, where it gets stored in the database.

\section{Key Deployment Observations}
\label{sec:eval}
In this section, we describe the field testing on the \ourmethod{} platform to evaluate its sensing capabilities in the real world. We did a long-term deployment of the platform in several types of indoor spaces (i.e., households, labs, canteens, etc.) in four cities in India, where different diffusion and spread patterns of harmful indoor pollutants such as VOCs, particulates, carbon dioxide, etc. were measured. Our data reveals several pollution dynamics, particularly for developing countries, which significantly impact the quality of a house's air but have remained unattended due to a lack of awareness and information. With its multi-point, human-in-the-loop sensing approach, \ourmethod{} shines in identifying short-term and long-term pollution events and the impact of floor plan and room structures on the spread of pollutants. Details on the field study are as follows.

\subsection{Field Study Details}
We deployed the \ourmethod{} platform in several indoor spaces, engaging the occupants in data labeling activities to collect a representative dataset on indoor air pollutants. Depending on the number of rooms and area of the space, more than one sensing devices are deployed at each measurement site. Our site selection and deployment plan is described next.

\begin{table}
\centering
\scriptsize
\caption{Deployment of the \textit{DALTON} platform and the socioeconomic background of the participants.}
\label{tab:dep_occup}
{\color{black}\begin{tabular}{|c|c|c|c|c|c|c|c|c|c|c|} 
\hline
\multicolumn{2}{|c|}{\multirow{2}{*}{\textbf{City}}} & \multicolumn{2}{c|}{\multirow{2}{*}{\textbf{Measurement Site}}} & \multicolumn{2}{c|}{\multirow{2}{*}{\textbf{Indoor Pollution}}}                                                                                                                                     & \multicolumn{5}{c|}{\textbf{Occupants}}                                                                                                                                                                                                                                                                                                 \\ 
\cline{7-11}
\multicolumn{2}{|c|}{}                               & \multicolumn{2}{c|}{}                                           & \multicolumn{2}{c|}{}                                                                                                                                                                               & \multicolumn{2}{c|}{\textbf{Gender}}                                                                                                            & \multicolumn{2}{c|}{\textbf{Education}}                                                   & \multirow{2}{*}{\begin{tabular}[c]{@{}c@{}}\textbf{Income}\\\textbf{Level}\end{tabular}}  \\ 
\cline{1-10}
\textbf{Name}          & \begin{tabular}[c]{@{}c@{}}\textbf{Locality }\\\textbf{Type}\end{tabular}               & \textbf{Site Type} & \textbf{\# Sites}                          & \begin{tabular}[c]{@{}c@{}}\textbf{Ventilation }\\\textbf{Appliances}\end{tabular}                       & \begin{tabular}[c]{@{}c@{}}\textbf{Primary}\\\textbf{Sources}\end{tabular}                    & \begin{tabular}[c]{@{}c@{}}\textbf{Female }\\\textbf{(\%)}\end{tabular} & \begin{tabular}[c]{@{}c@{}}\textbf{Male }\\\textbf{(\%)}\end{tabular} & \textbf{Degree} & \begin{tabular}[c]{@{}c@{}}\textbf{Tech }\\\textbf{Expert}\end{tabular} &                                                                                           \\ 
\hline
\bqa{}                  & Rural                       & Household          & 2                                          & \begin{tabular}[c]{@{}c@{}}Window, Vent slits, \\Exhaust Fan\end{tabular}                           & \begin{tabular}[c]{@{}c@{}}LPG, Kerosine,\\Food, Disinfectants\end{tabular}                   & 50                                                                      & 50                                                                    & Bachelor        & No                                                                      & Low                                                                                       \\ 
\hline
\dgp{}                  & Suburban                    & Household          & 2                                          & \multirow{3}{*}{\begin{tabular}[c]{@{}c@{}}Window, Vent slits, \\Split AC,Exhaust fan\end{tabular}} & \multirow{3}{*}{\begin{tabular}[c]{@{}c@{}}LPG, Microwave,\\Food,~Disinfectants\end{tabular}} & 50                                                                      & 50                                                                    & Doctorate       & Yes                                                                     & Middle                                                                                    \\ 
\cline{1-4}\cline{7-11}
\ccu{}                  & Urban                       & Household          & 4                                          &                                                                                                     &                                                                                               & 44                                                                      & 56                                                                    & Doctorate       & No                                                                      & Middle                                                                                    \\ 
\cline{1-4}\cline{7-11}
\multirow{5}{*}{\kgp{}} & \multirow{5}{*}{Suburban}   & Household          & 5                                          &                                                                                                     &                                                                                               & 60                                                                      & 40                                                                    & Doctorate       & Yes                                                                     & Middle                                                                                    \\ 
\cline{3-11}
                       &                             & Apartment          & 8                                          & Window, Vent slits,                                                                                 & Occupancy, Food                                                                               & 33                                                                      & 67                                                                    & Student         & Yes                                                                     & Low                                                                                       \\ 
\cline{3-11}
                       &                             & Food Canteen       & 2                                          & Vent slits, Exhaust fan                                                                             & LPG, Food                                                                                     & 50                                                                      & 50                                                                    & Metric          & No                                                                      & Middle                                                                                    \\ 
\cline{3-11}
                       &                             & Research Lab       & 5                                          & Split AC, Window AC                                                                                 & \multirow{2}{*}{Occupancy}                                                                    & 11                                                                      & 89                                                                    & Student         & Yes                                                                     & Low                                                                                       \\ 
\cline{3-5}\cline{7-11}
                       &                             & Classroom          & 2                                          & Split AC, Central AC                                                                                &                                                                                               & -                                                                       & -                                                                     & Student         & -                                                                       & -                                                                                         \\
\hline
\end{tabular}}
\end{table}

\subsubsection{Site Selection \& Deployment}
\begin{figure}
	\captionsetup[subfigure]{}
	\begin{center}
            \subfloat[Indoor types\label{fig:deploy_pie}]{
			\includegraphics[width=0.25\columnwidth,keepaspectratio]{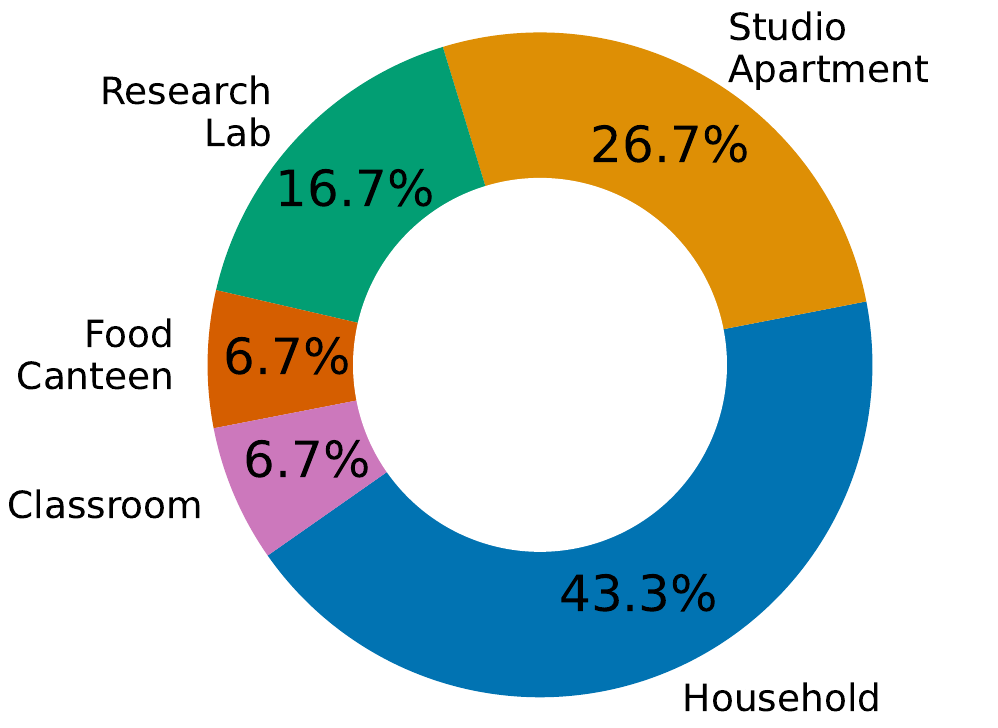}
		}
            \subfloat[Site Area\label{fig:area}]{
			\includegraphics[width=0.18\columnwidth,keepaspectratio]{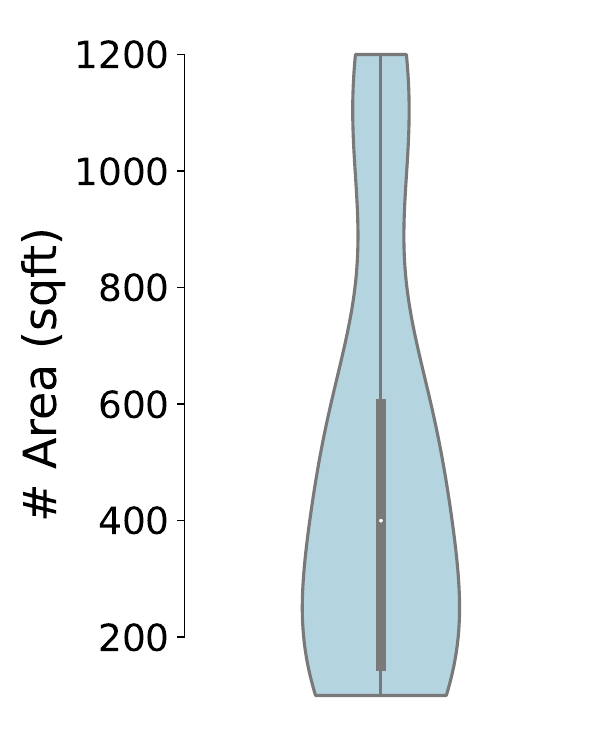}
		}
            \subfloat[Sensors per 500sqft\label{fig:sens_500}]{
			\includegraphics[width=0.18\columnwidth,keepaspectratio]{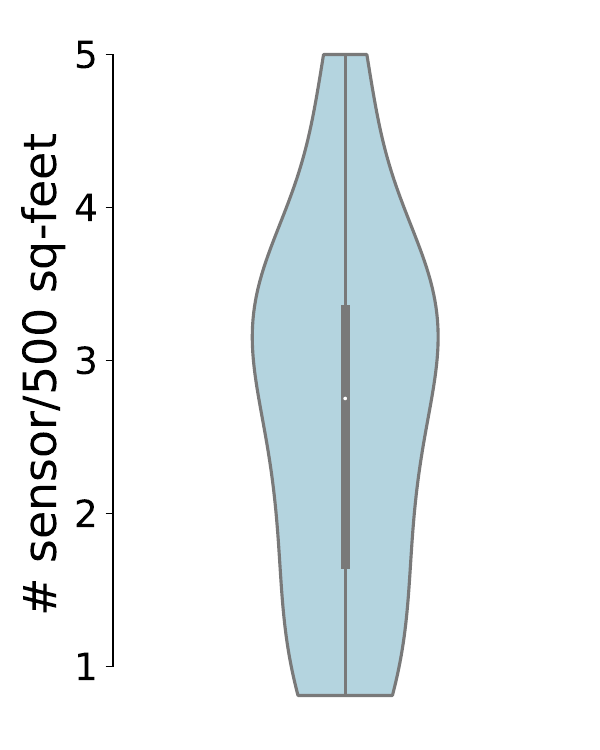}
		}
            \subfloat[Occupants per site\label{fig:occu_site}]{
			\includegraphics[width=0.18\columnwidth,keepaspectratio]{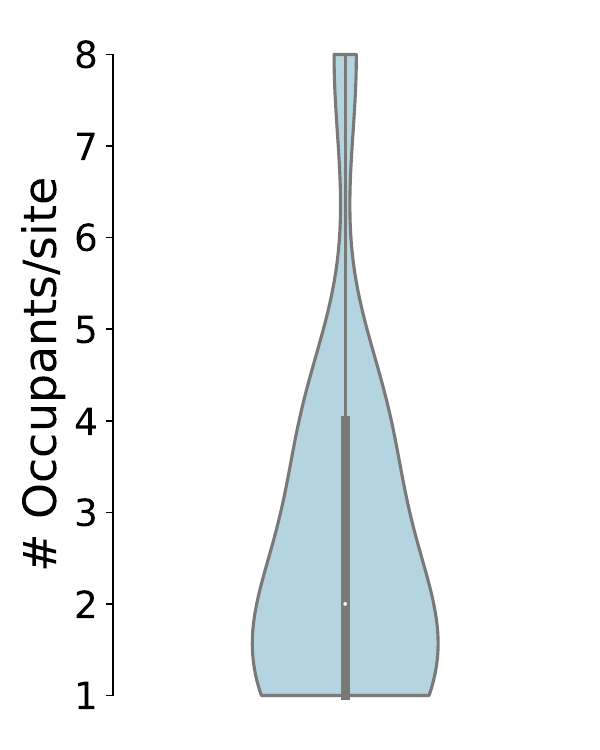}
		}
            \subfloat[Occupants' Age\label{fig:age_dist}]{
			\includegraphics[width=0.18\columnwidth,keepaspectratio]{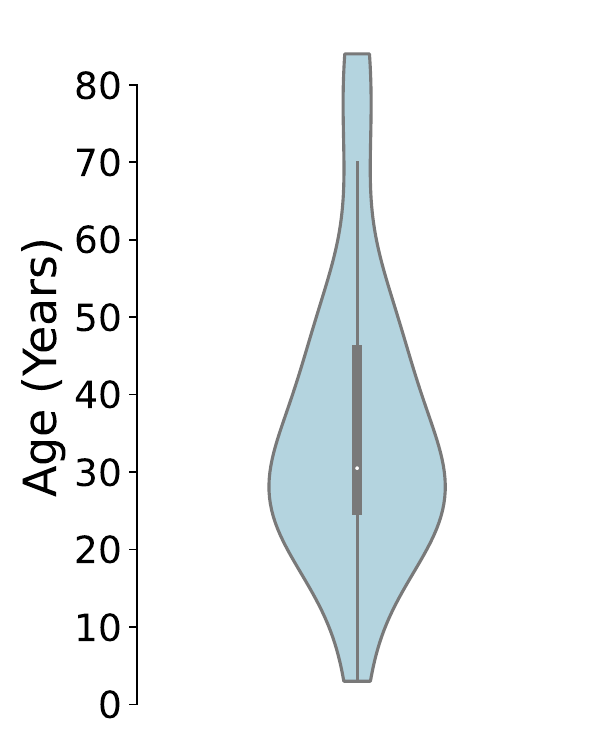}
		}
	\end{center}
	\caption{Demographics and Details of the field-study.}
	\label{fig:demog}
\end{figure}

We have collected data for \numsites{} measurement sites across four cities in India for over \datamonths{} on primarily five types of indoor environments, namely, households, studio apartments, research labs, food canteens, and classrooms. We have carefully chosen the \numcities{} cities such that they capture typical indoor pollution dynamics in the developing region. 
\changed{Notably, in \bqa{}, most building constructions are unplanned and thus have congested neighborhoods. The houses are naturally ventilated, and people are accustomed to daily cooking with locally sourced food items and using LPG stoves, firewood, incense sticks, etc. Moreover, rural areas have a large body of greenery and less outdoor pollution. In contrast, \dgp{} is a well-planned industrial city with several operational steel and sponge iron factories, resulting in significant outdoor pollution. The \ccu{} is a metropolitan city where most of the population is office goers and are habitual to air conditioners, packaged food ingredients, LPG gas stoves, induction and microwave cookers, etc. Lastly, \kgp{} is a university town consisting of student apartments, faculty housing, canteens, restaurants, etc. \tablename~\ref{tab:dep_occup} summarizes the measurement sites, ventilation appliances, and primary pollution sources for each city and indoor type.} The percentage of each indoor type in the overall deployment is shown in \figurename~\ref{fig:deploy_pie}. The number of rooms and areas of the deployment sites vary from a studio apartment to a household. The studio apartments mostly have one room that is approximately $150$ sqft in size. Meanwhile, in a typical household, there can be three to six rooms spanning about $600$ to $1100$ sqft. \figurename~\ref{fig:area} shows the area distribution of the sites.

We have placed at least one sensor in each room to effectively measure the spread of pollutants between rooms in an indoor space. Moreover, the devices are deployed approximately at chest height ($1.5$ meter from the ground) to measure the actual exposure level regarding the occupants. Thus, we have deployed $1$-$2$ sensors per studio apartment, $3$-$4$ sensors per classroom and lab, and $3$-$6$ sensors per household. The \figurename~\ref{fig:sens_500} depicts the number of deployed sensing devices per $500$ sqft area. The occupants of these measurement sites are requested to install the Android application described in Section~\ref{sec:and_app} and participate in the field study, providing feedback to the platform. The details on the user demographics are presented below.

\subsubsection{User Demographics}
\figurename~\ref{fig:occu_site} shows the distribution of the number of occupants per site, varying from one in a studio apartment to eight in a household. \changed{In total, \numoccupants{} occupants from \numsites{} measurement sites participated in the study, among which \males{} participants are male and \females{} participants are female. \tablename~\ref{tab:dep_occup} shows the socioeconomic background of the participants.} The developed Android application lets the participants label indoor activities or events via voice commands. The overall age distribution of the occupants is shown in \figurename~\ref{fig:age_dist}. We can observe that most occupants are aged between $20$ to $40$ years; thus, they are accustomed to using such Android apps in between their daily activities. However, the youngest of the occupants is a $3$ years old girl, and the oldest is a $84$ years old man. They are too young or old to participate in this study actively; thus, other members of the corresponding site (i.e., household) report their indoor activities on their behalf. 

The participants are college students, university staff, professors, homemakers, canteen owners, etc. Hence, the level of expertise in handling and debugging the sensing devices in case of any failure varies drastically from participant to participant. Here, the automatic fault recovery and remote debugging capabilities of the \ourmethod{} platform become crucial for a sustainable deployment with minimal user intervention. Moreover, the Android app-based human-in-the-loop labeling mechanism was quickly adopted by all the participants despite their different technical backgrounds.

\subsubsection{Weekly Patterns in the Dataset}
\begin{figure}
	\captionsetup[subfigure]{}
	\begin{center}
		\subfloat[Kitchen VOC\label{fig:kitchen_voc}]{
			\includegraphics[width=0.33\columnwidth,keepaspectratio]{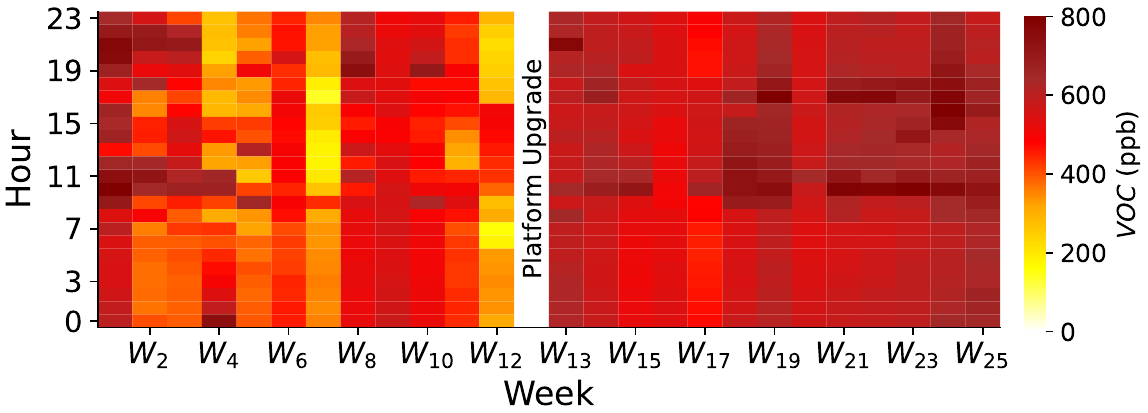}
		}
	    \subfloat[Bedroom VOC\label{fig:bedroom_voc}]{
	    	\includegraphics[width=0.33\columnwidth,keepaspectratio]{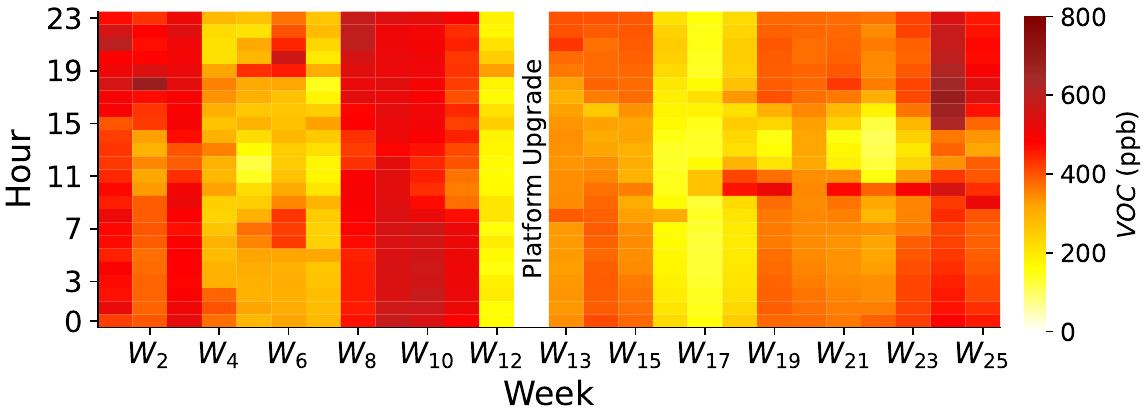}
	    }
		\subfloat[Kitchen CO\textsubscript{2}\label{fig:kitchen_co2}]{
			\includegraphics[width=0.33\columnwidth,keepaspectratio]{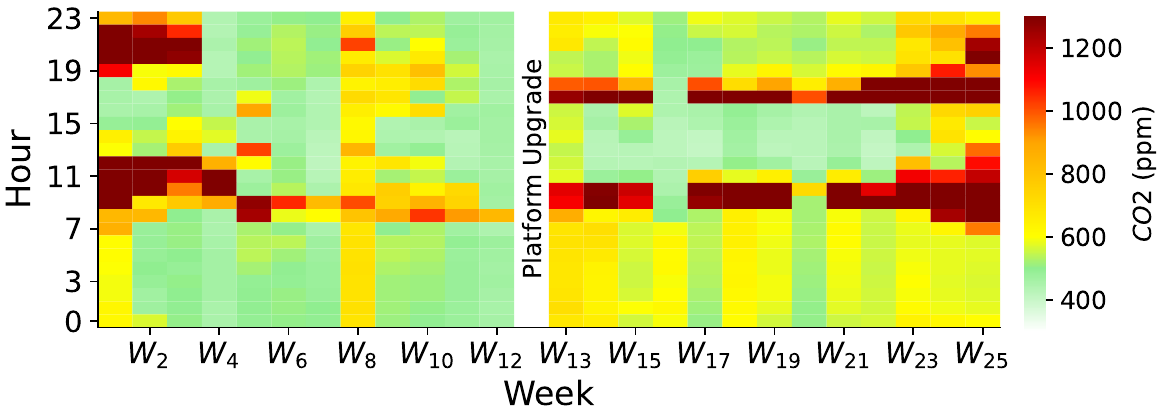}
		}\
	    \subfloat[Bedroom CO\textsubscript{2}\label{fig:bedroom_co2}]{
	    	\includegraphics[width=0.33\columnwidth,keepaspectratio]{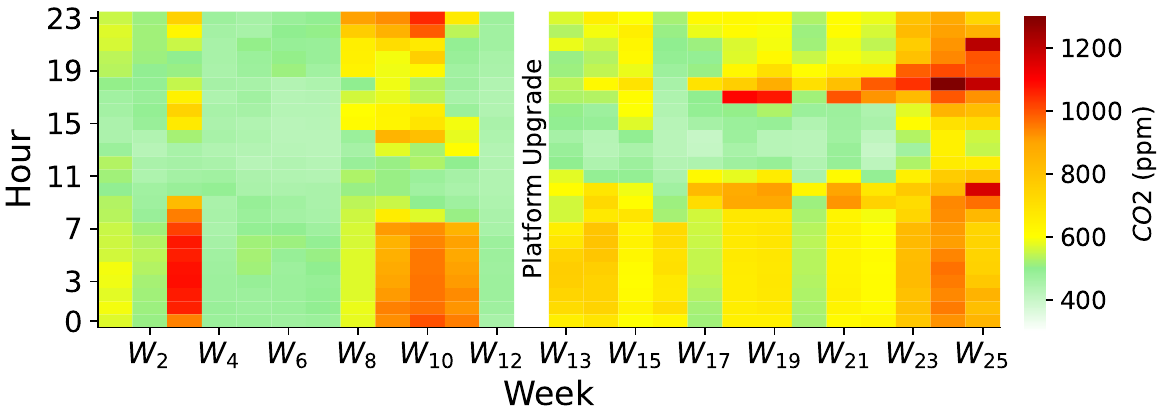}
	    }
            \subfloat[Temperature\label{fig:overall_t}]{
			\includegraphics[width=0.33\columnwidth,keepaspectratio]{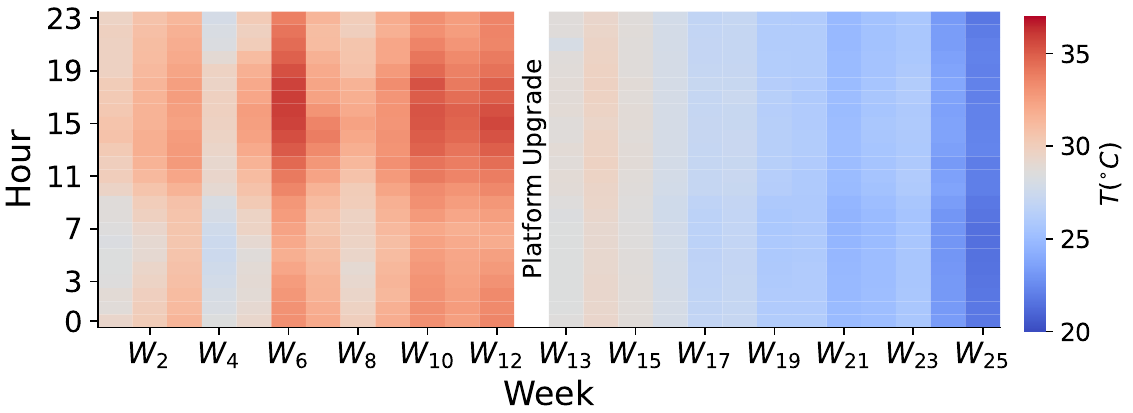}
		}
	    \subfloat[Humidity\label{fig:overall_hum}]{
	    	\includegraphics[width=0.33\columnwidth,keepaspectratio]{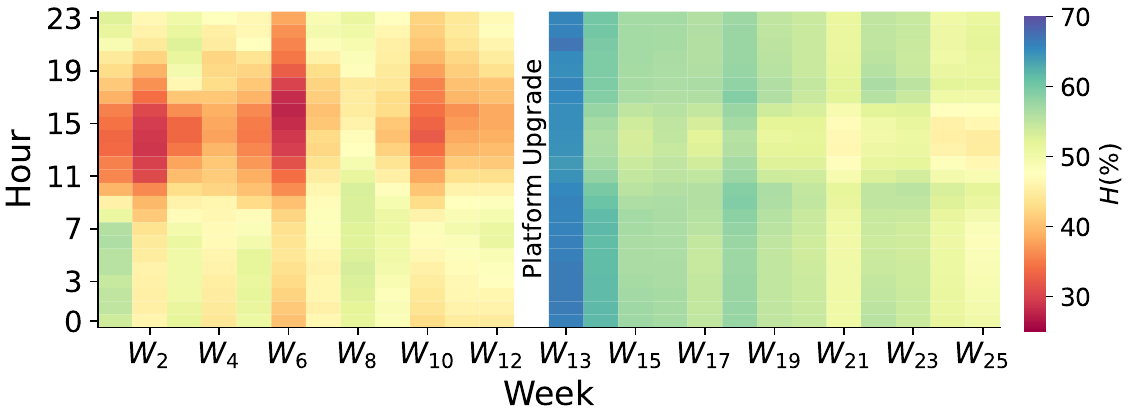}
	    }\
	\end{center}
	\caption{Daily indoor trends by week and month. We observe higher concentrations of VOC and CO\textsubscript{2} during the winter months.}
	\label{fig:pattern}
\end{figure}

Here, we show the weekly and monthly variations of pollutants in the collected dataset from the large-scale deployment of the \ourmethod{} platform throughout \datamonths{}. The \figurename~\ref{fig:pattern} shows the weekly and daily variation of VOC, CO\textsubscript{2} levels, along with temperature and humidity change throughout the dataset. The dataset is collected in both summer and winter, totaling over \datamonths{}. Notably, after the summer season, we upgraded the platform to integrate the remote management features and resume data collecting from the winter. This time gap is highlighted in all the sub-figures of \figurename~\ref{fig:pattern}. We observe a similar pattern in the maximum hourly VOC exposure for the kitchen and bedroom as per the heat-maps shown in \figurename~\ref{fig:kitchen_voc} and~\ref{fig:bedroom_voc}, which indicates that, in general, pollutants emitted from the kitchen are spread towards the bedrooms. 

During the summer (i.e., weeks W\textsubscript{1} to W\textsubscript{12}), we observe a steady rise in temperature over the weeks as per \figurename~\ref{fig:overall_t}. As shown in \figurename~\ref{fig:overall_hum}, the overall humidity also increases from W\textsubscript{7} week of the data collection. The food items and fruits degrade quickly in high temperatures and humidity, releasing excessive VOCs; thus, we observe a rise in the VOC levels in the kitchens and bedrooms from week W\textsubscript{8} onwards. Regarding CO\textsubscript{2} exposure in summer, we observe a maximum peak in the kitchen during the first month when the temperature remains relatively comfortable, as shown in \figurename~\ref{fig:kitchen_co2}. The primary reason for such observation is that we are more sensitive towards temperature change (detail explanation and analysis in Section~\ref{sec:cook_style}); thus, in comfortable temperatures, the kitchen exhaust fans are mostly turned off, resulting in poor ventilation for the emitted CO\textsubscript{2} (we observed this from the annotated labels as well). As the mean temperature increases over the months, we observe that the CO\textsubscript{2} peaks are reduced as the exhaust is turned on more frequently, providing much-needed ventilation. Interestingly, CO\textsubscript{2} in bedrooms do not significantly correlate with the kitchen, implying that CO\textsubscript{2} exposure is contained near the source, where VOC spread across the entire household.

However, during winters (i.e., weeks W\textsubscript{13} to W\textsubscript{25}), the environment becomes humid, and the temperature continues to decrease over the weeks. Therefore, occupants tend to close all windows of the indoor space to maintain above 20\textdegree C temperature. The kitchen exhaust is also unused due to low temperature, and the heat generated during cooking further improves the thermal comfort for the occupants. As a result, the exposure to pollution is drastically increased across the indoor space. For instance, the kitchen becomes the most contaminated room, and pollutants such as VOC and CO\textsubscript{2} spread toward the bedrooms. Accordingly, we observe in \figurename~\ref{fig:kitchen_voc}, and \figurename~\ref{fig:bedroom_voc}, VOC is correlated between kitchen and bedroom in winter. Unlike summer, CO\textsubscript{2} spreads further into the indoor space, and we observe a correlation between kitchen and bedroom CO2 measurements from week W\textsubscript{15} onwards. Building upon these observations, we next analyze the indoor pollution behaviors for specific scenarios such as inadequate ventilation, degree of ventilation, indoor activities, etc., along with the spatiotemporal spread of pollutants based on floor plans and room structures.

\begin{lesson}{1}{le1}
 Different pollutants show different spatiotemporal behavior in indoor environments based on types of activities. The seasonal temperature and humidity changes greatly influence occupant's activities. Winter observes a higher degree of spread between the kitchen and other rooms of a household due to compromised ventilation for maintaining a comfortable temperature.
\end{lesson}

\subsection{Inadequate Ventilation}
\label{sec:insuff_airflow}
The user-inclusive design of \ourmethod{} in labeling the indoor events and activities, along with measuring changes in pollutant levels, enables us to isolate several commonly occurring pollution instances in bedrooms, kitchens, hall rooms, etc., where the pollutants accumulate over time due to lack of ventilation. Here, we highlight its severity regarding the pollution exposure and the exposure duration. Our observations are as follows.

\begin{figure}
	\captionsetup[subfigure]{}
	\begin{center}
		\subfloat[CO\textsubscript{2} concentration \label{fig:vent_co2_time}]{
			\includegraphics[width=0.3\columnwidth,keepaspectratio]{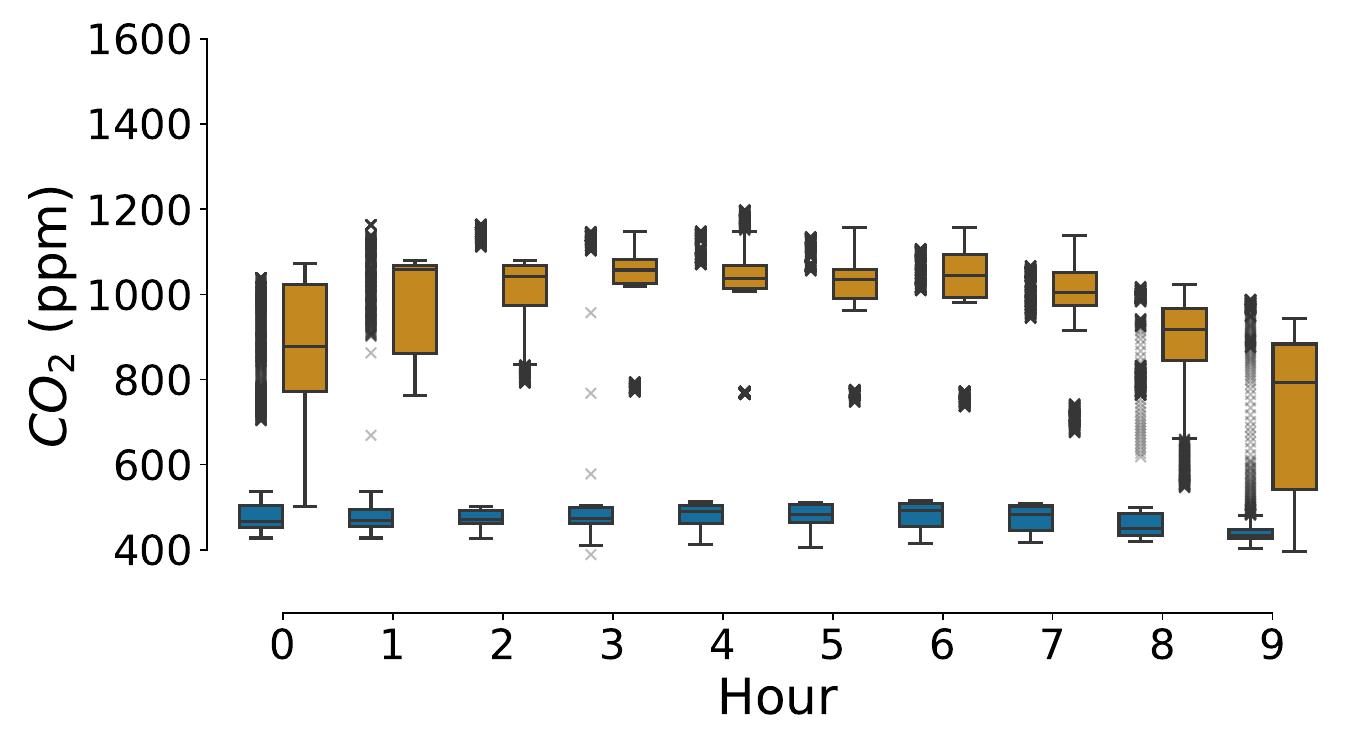}
		}
		\subfloat[VOC concentration\label{fig:vent_voc_time}]{
			\includegraphics[width=0.3\columnwidth,keepaspectratio]{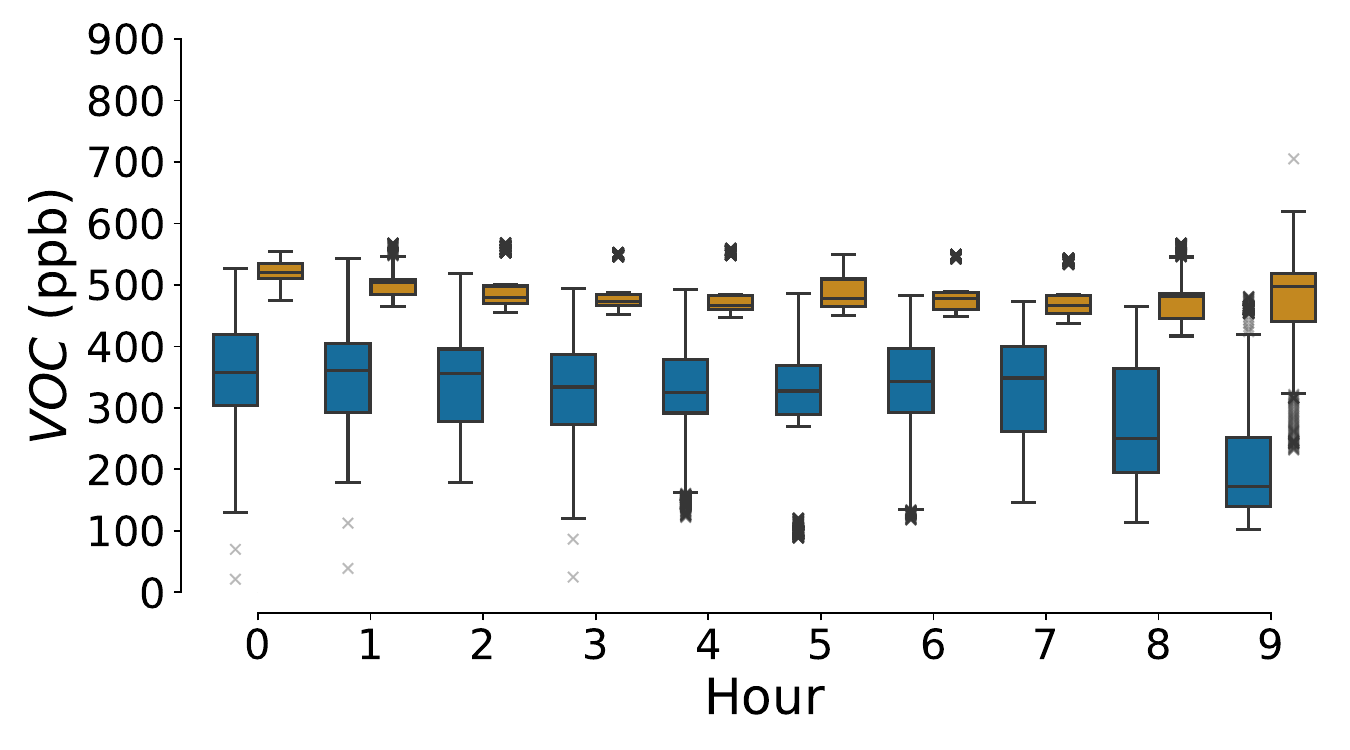}
		}
            \subfloat[Saturation of pollutants\label{fig:vent_all_time}]{
			\includegraphics[width=0.3\columnwidth,keepaspectratio]{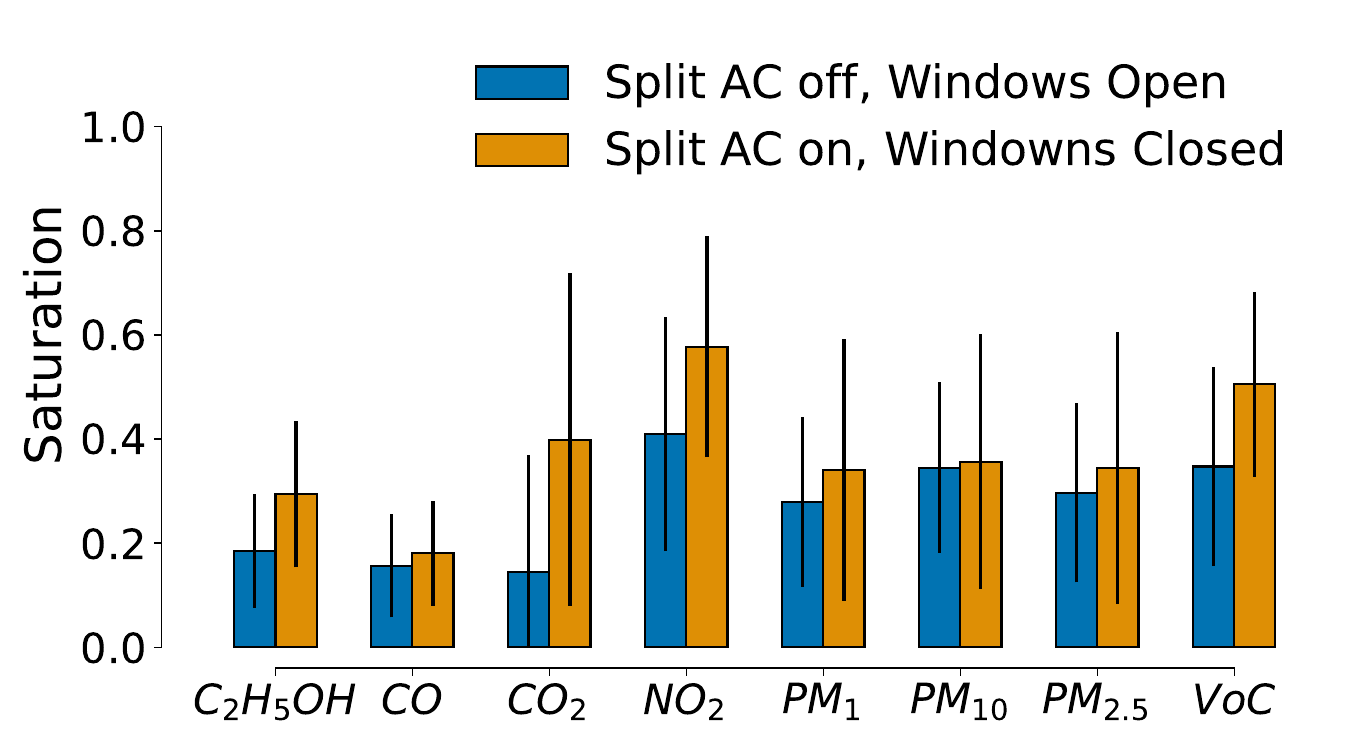}
		}
	\end{center}
	\caption{Windows are closed for effective air conditioning using split AC. Thus, CO\textsubscript{2} and VOC accumulate dangerously over the night hours in summer when the split AC is running. Sub-figure (a) shows CO\textsubscript{2} concentration almost triples compared to when the windows are open. CO\textsubscript{2} gets ventilated in the morning when the windows are opened again. However, sub-figure (b) depicts VOC persists for a little longer. The degree of accumulation for other pollutants is shown in sub-figure (c).}
	\label{fig:vent_time}
\end{figure}

\subsubsection{Bedrooms with Split AC}
To improve the power efficiency, split AC circulates the air within a room~\cite{harby2019investigation,shriram2019assessment}, rather than pulling air from outside, to ensure effective air-conditioning with minimal energy cost~\cite{ding2023integrating}. Therefore, it provides no ventilation for the airborne contaminants, leading to long-term accumulation of harmful pollutants such as VOC, CO\textsubscript{2}, Ethanol, etc. In developing countries like India, which has an extended summer season, it's very common for middle to high-income households to use split AC during the night hours (i.e., 12:00 AM to 7:00 AM) for a comfortable sleep. However, it leads to unintentional overnight exposure to pollutants. For instance, \figurename~\ref{fig:vent_all_time} depicts the degree of pollutant accumulation in the bedroom due to closed windows when the split AC is running compared to when the windows are open and the split AC is off. We observe highly elevated levels of CO\textsubscript{2}, increased VOC contamination from midnight to early morning as the occupants sleep, keeping the windows closed while using the split AC. \figurename~\ref{fig:vent_co2_time} and~\ref{fig:vent_voc_time} show the distributional changes of CO\textsubscript{2} and VOC concentration, respectively on an hourly basis for the night hours. As per the figures, the occupants experience, on average, two times CO\textsubscript{2} exposure due to poor ventilation. Similarly, VOC accumulates approximately $1.3$ times more strongly in lack of ventilation. However, In the morning, the CO\textsubscript{2} gets ventilated quickly with windows opening as indicated by the sharp dip in concentration from 7:00 AM onwards in \figurename~\ref{fig:vent_co2_time}. Meanwhile, VOC levels persist longer even with the opened window in the morning, indicating that some pollutants are more complicated to ventilate than others and, thus, more harmful in the long term.

\begin{wrapfigure}[13]{R}{0mm}
	\captionsetup[subfigure]{}
		\subfloat[CO\textsubscript{2} levels over time\label{fig:temporalco2}]{
			\includegraphics[width=0.24\columnwidth,keepaspectratio]{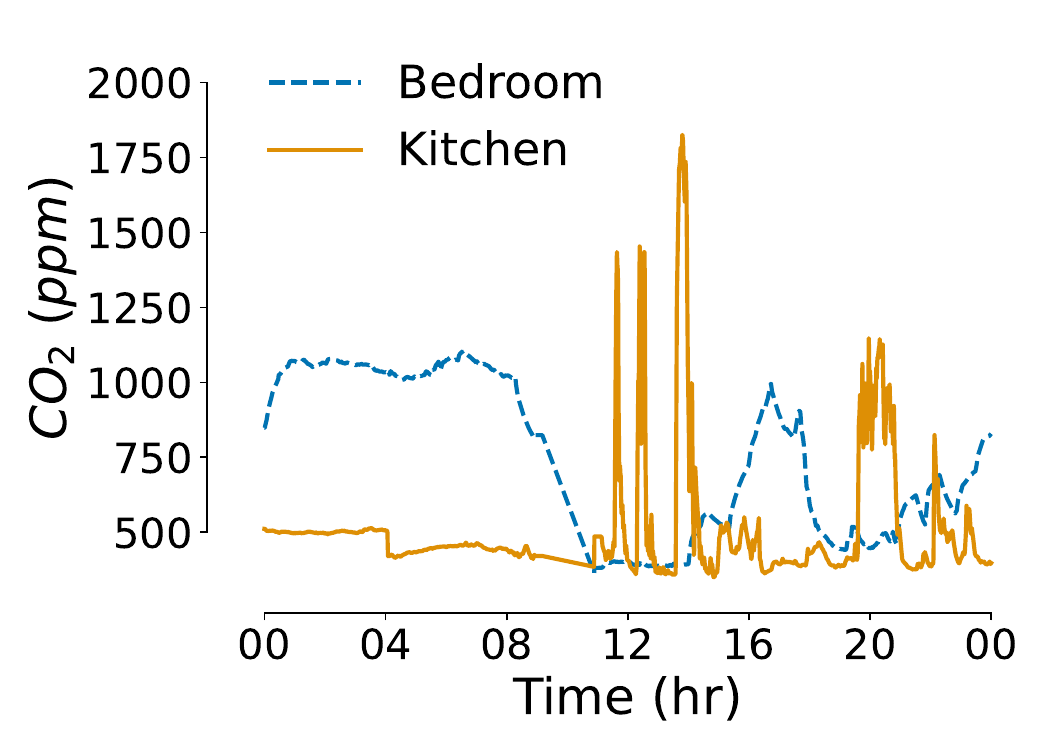}
		}\
		\subfloat[Distribution of CO\textsubscript{2}\label{fig:safety}]{
			\includegraphics[width=0.24\columnwidth,keepaspectratio]{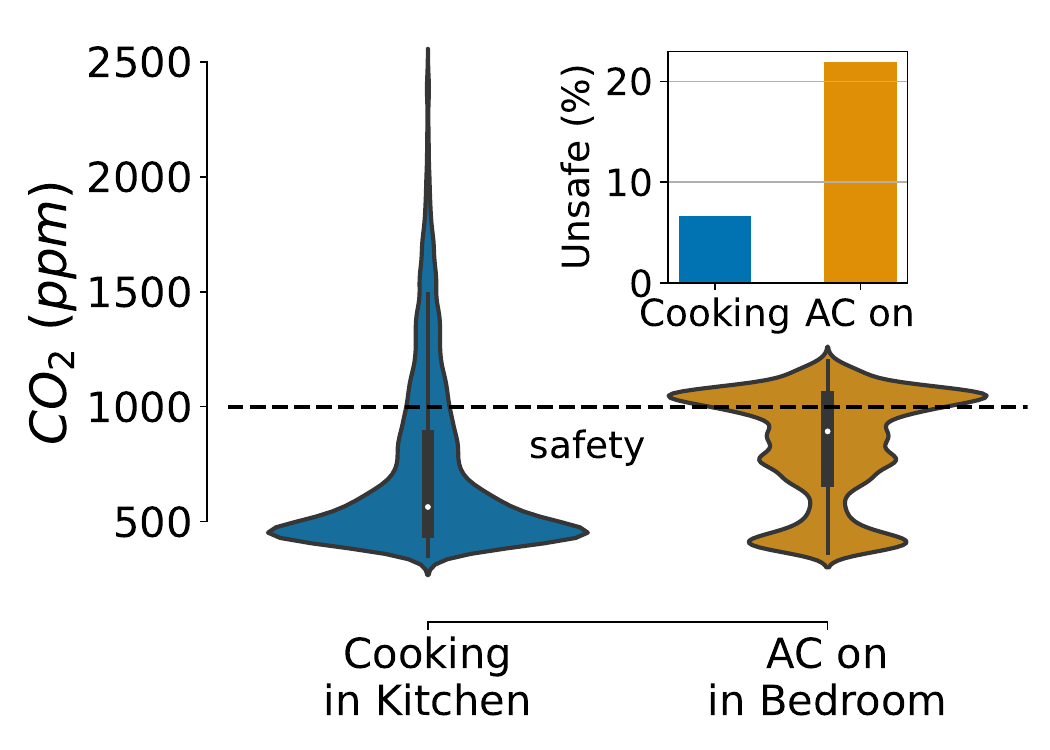}
		}
	\caption{In sub-figure (a), the kitchen shows sudden spikes of CO\textsubscript{2} while cooking. CO\textsubscript{2} accumulates in the bedroom while using split AC. On average, the bedroom is exposed to higher CO\textsubscript{2} concentration and 15.2\% more unsafe than the kitchen, as per sub-figure (b).}
	\label{fig:co2_kit_bed}
\end{wrapfigure}

\subsubsection{CO\textsubscript{2} in Kitchen Vs Bedroom}
By associating the readings from the sensors and the human-annotated event and activity data, we observe that the pollutants can either be emitted rapidly (e.g., during cooking) or accumulate at a slower rate over a long time (e.g., in a non-ventilated bedroom at night). We, as humans, are more sensitive to rapid changes in the environment and can act to reduce the exposure by turning on the ventilation system or opening up the windows. However, the primary problem arises when the emission rate is very low, but due to poor ventilation, pollutants get trapped over an extended period, for instance, while using a split AC when sleeping. For the households that use split AC in the bedroom, we compare the CO\textsubscript{2} exposure between the kitchen and the bedroom during the polluting hours. As shown in \figurename~\ref{fig:temporalco2}, the kitchen observed a sudden peak of CO\textsubscript{2}; the majority time of the day, it remained within the safety threshold ($\leq$ 1000 ppm). At the same time, the bedroom remains polluted for an extended period. Unlike developed countries where indoor spaces are ventilated with central Heating, Ventilation, and Air Conditioning (HVAC) systems, in developing nations, people tend to use low-cost alternatives like split AC for summer and room heaters for the winter season, where both require windows to be closed to work efficiently, leaving out the crucial ventilation aspect of HVAC systems. Therefore, in both seasons, indoor spaces in developing countries suffer from pollutant accumulation in bedrooms, living rooms, etc., compared to the kitchen.

Considering the overall CO\textsubscript{2} exposure in a day for this particular household, the bedroom was unsafe for $21.9\%$ of the time, whereas the kitchen was unsafe for only $6.7\%$. Notably, from the data collected over the households, we observe that the users were more sensitive towards the rapid changes in environmental temperature and humidity of the kitchen (this hypothesis is further validated in the next section) and thus turned on exhausts or opened windows to allow the contaminants to ventilate away. Therefore, we observe a tailed distribution in \figurename~\ref{fig:safety} with rapidly declining instantaneous values of CO\textsubscript{2} in \figurename~\ref{fig:temporalco2}. However, they were completely unaware of the high level of CO\textsubscript{2} getting accumulated in the bedroom when they were sleeping due to air-conditioning. Indeed, no actions were taken by the users to reduce the CO\textsubscript{2} pollution in the bedroom, leading to harmful exposure for an extended period.
\vspace{-0.1cm}
\begin{lesson}{2}{le2}
Energy saving in developing regions may come with the cost of expedited exposure to indoor pollutants. Consequently, the bedroom can be more vulnerable than the kitchen in terms of overall pollution exposure. 
\end{lesson}

\subsubsection{Cooking with the Exhaust Off}
\label{sec:cook_exhaust_off}
\begin{wrapfigure}[12]{L}{0mm}
    \centering
    \includegraphics[width=0.4\columnwidth]{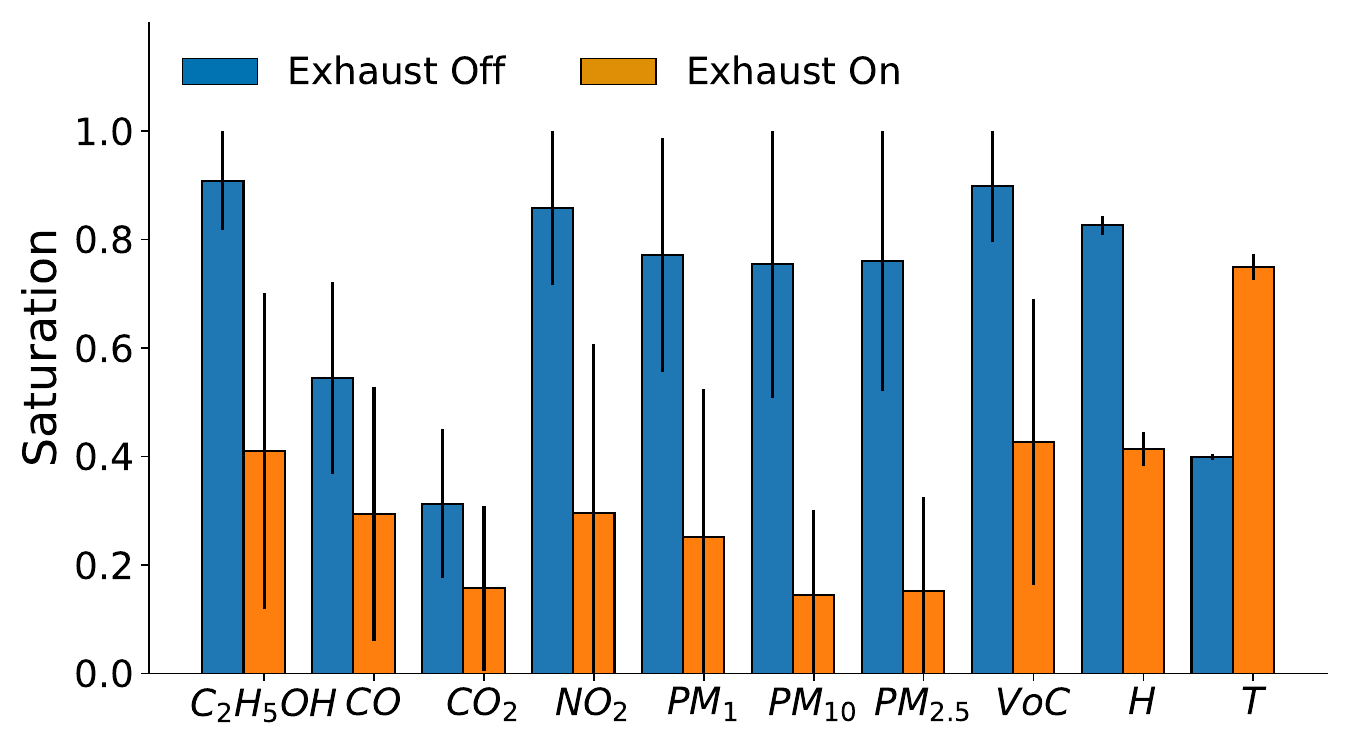}
    \caption{Saturation levels of the pollutants with exhaust off vs on. Pollutants accumulate when the exhaust fan is off during cooking. Notably, the exhaust is turned on only at high temperature.}
    \label{fig:exhaust_cook}
\end{wrapfigure}

Long-term deployment of the \ourmethod{} platform captures the general human behavior in the kitchen while choosing to turn on the exhaust fan for ventilation. Even though the concentration of pollutants in the kitchen is significantly reduced after turning on the ventilation, as shown in \figurename~\ref{fig:exhaust_cook}, from the collected data, we observe that the event of turning on the ``\textit{exhaust fan}'' is conditioned on relatively higher environmental temperature as compared to when the occupants choose not to do so. In hindsight, turning on the ventilation reduces the kitchen's humidity but does not affect the temperature significantly; thus, we hypothesize that \textit{occupants are more comfortable in a less humid environment when the temperature is high}. Conversely, the \textit{occupants ignore ventilation when the temperature is less, even though the humidity is high}. 

Geographically, most developing countries are located near the equatorial region~\cite{leastdeveloped}. Thus, the typical climate in such countries is hot and humid most of the year. Therefore, the people in these regions are more exposed to kitchen pollutants during the winter when they forget to turn on the exhaust due to lower environmental temperature. Several studies~\cite{cooktookfarm,MASSEY2012223} uncover the health impact of increased indoor pollution levels during the winter season. This typical human behavior can be seen by comparing the humidity and temperature box-plots for both ``exhaust off'' and ``exhaust on'' in \figurename~\ref{fig:exhaust_cook}. Such human behavior also highlights our limited sensory capacity to access our surrounding air quality and motivates us to conduct further human-centered field experiments with the \ourmethod{} platform.

\begin{lesson}{3}{le3}
The pollutants emitted and the general human response will vary depending on the activity. Temperature changes are more apparent to humans, so they know to turn on ventilation, but they cannot sense pollutants accumulating around them, meaning they are unknowingly exposed to them.
\end{lesson}

\begin{wrapfigure}[11]{R}{0mm}
    \centering
    \includegraphics[width=0.3\columnwidth]{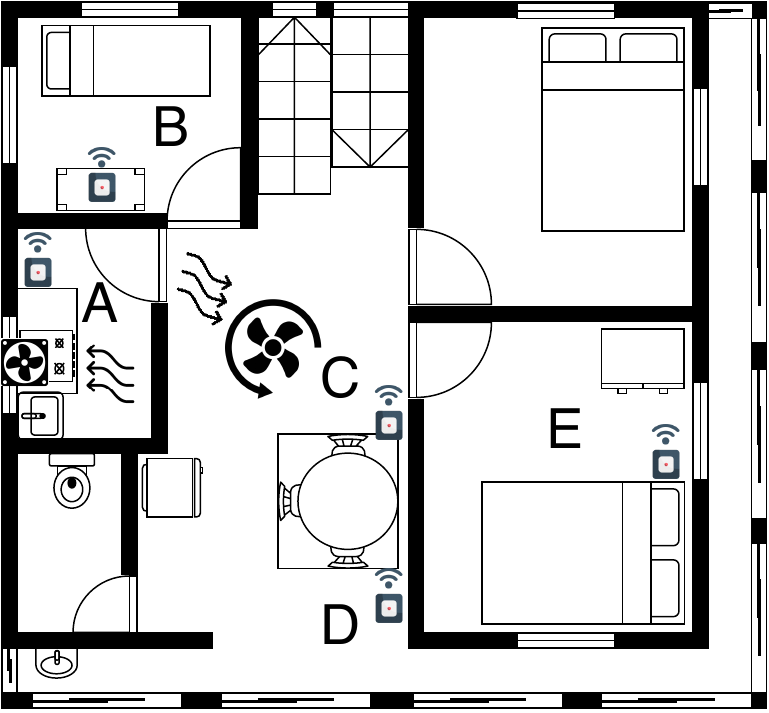}
    \caption{Household-1 (H1).}
    \label{fig:pk_house}
\end{wrapfigure}

\subsection{Spread of Pollutants across Rooms}
\label{sec:inter_spread}
Realizing the drawbacks of insufficient airflow in several scenarios of indoor spaces, we have utilized the \ourmethod{} platform to validate if increased airflow can reduce pollutant accumulation. Yet, we observed that uninformed decisions to modulate airflow can spread pollutants toward the other rooms of the indoor space. The room structure and the floor plan also act as additional factors that influence the velocity and degree of spread over the nearby rooms from the pollution source. Following are the observations.

\subsubsection{Effect of Airflow}
To describe the impact of different airflow modulations, we present an indicative middle-income household H1, where the kitchen is the acting pollution source. The floor plan of household H1 is shown in \figurename~\ref{fig:pk_house}, where the kitchen is marked as (A), and nearby bedroom and dining are marked as (B) and (C), respectively. We have observed three airflow scenarios from the long-term data of H1 when H1 has \textit{(i) Natural Airflow through Open Windows, (ii) Active Ventilation in the Kitchen, and (iii) Swirling Airflow in Dining.} A comparative analysis of the spread of pollutants for the above three scenarios is shown below.

\begin{figure}
	\captionsetup[subfigure]{}
	\begin{center}
		\subfloat[Kitchen\label{fig:kit_medium}]{
			\includegraphics[width=0.3\columnwidth,keepaspectratio]{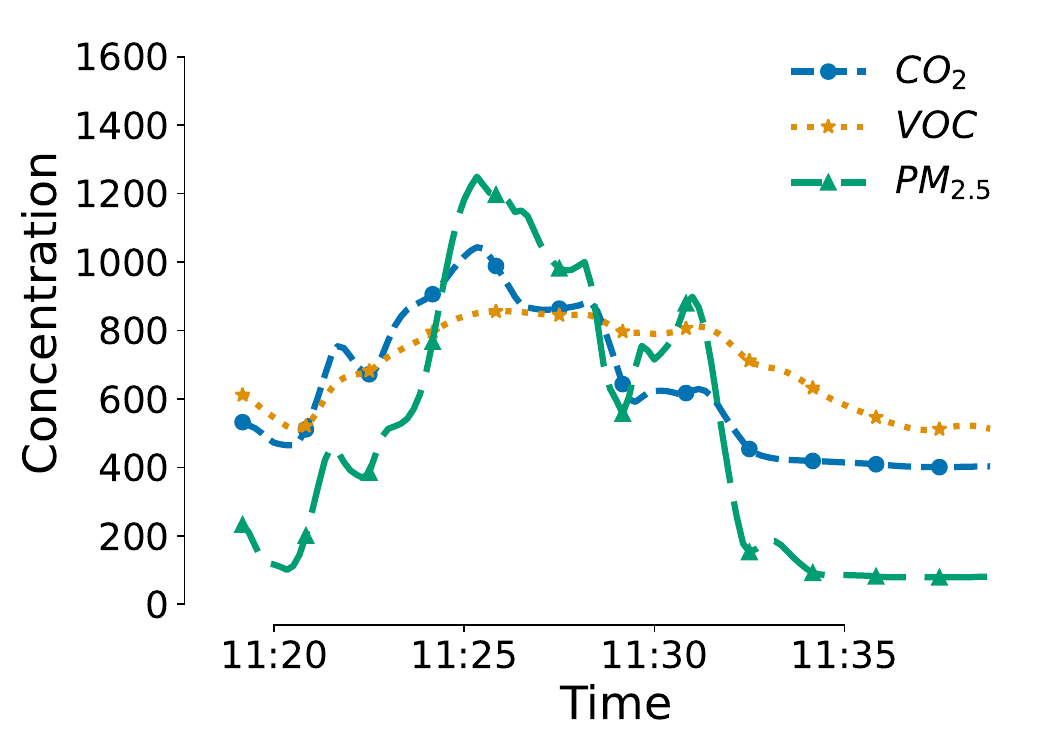}
		}
		\subfloat[Bedroom\label{fig:rm_medium}]{
			\includegraphics[width=0.3\columnwidth,keepaspectratio]{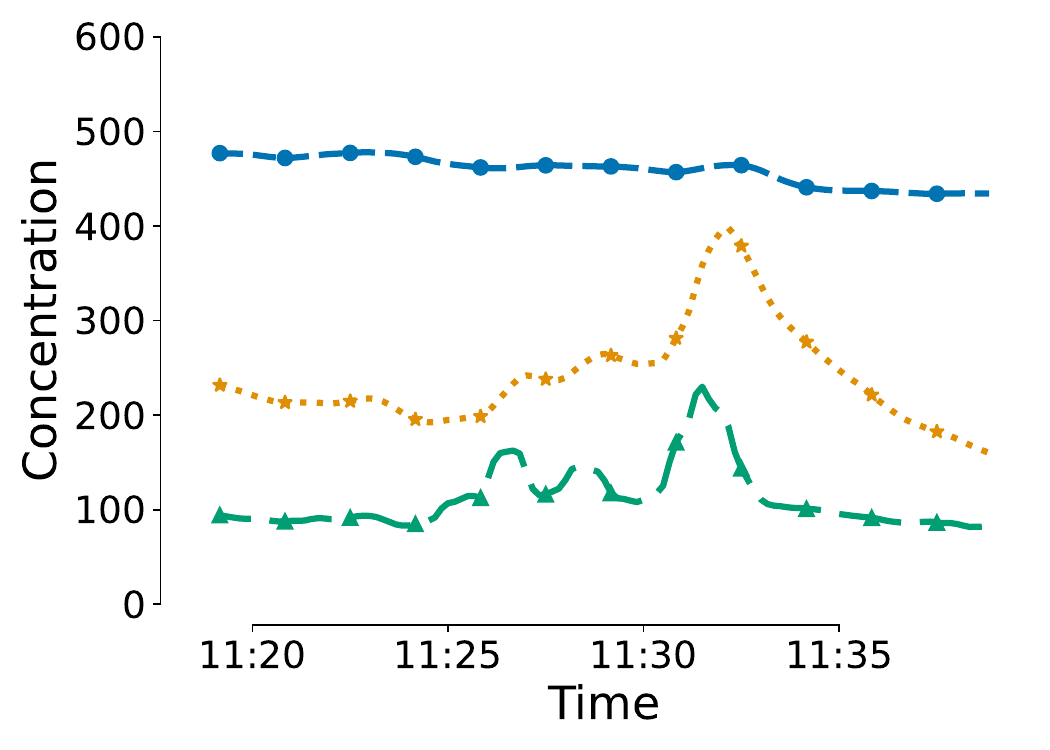}
		}
		\subfloat[Dining\label{fig:din_medium}]{
			\includegraphics[width=0.3\columnwidth,keepaspectratio]{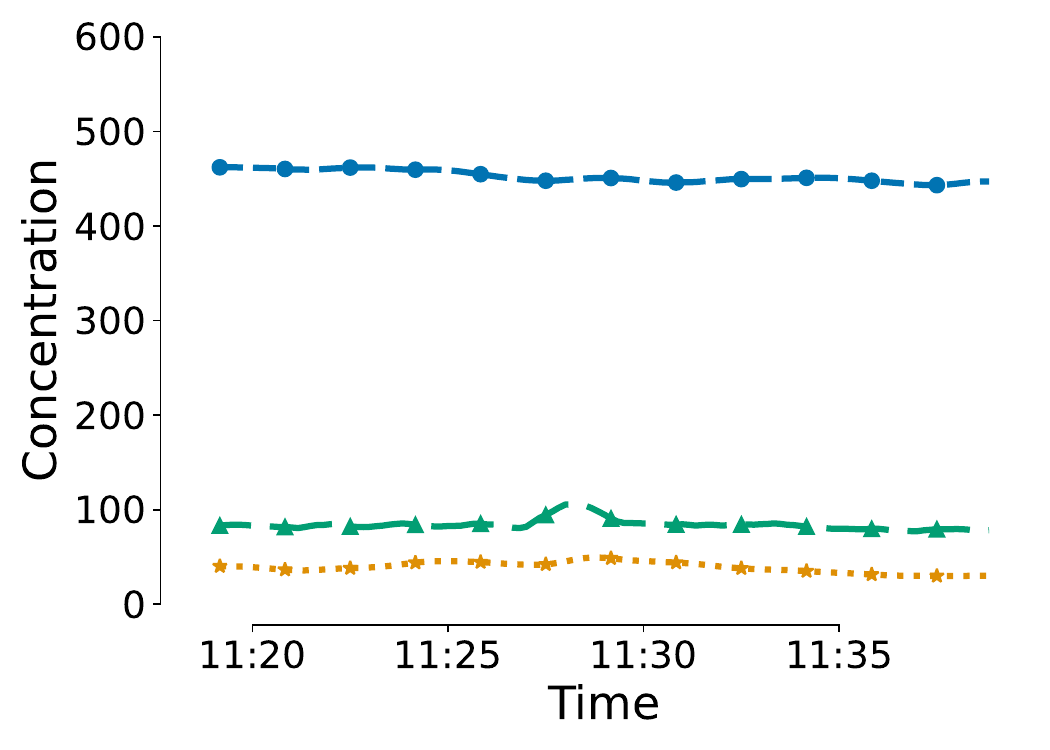}
		}
	\end{center}
	\caption{The kitchen exhaust fan is off; thus, household H1 is ventilated by natural airflow through open windows. Pollutants emitted during cooking in the kitchen (A) spread to the side by the bedroom (B). However, the dining (D) remains unaffected.}
	\label{fig:spread_med}
\end{figure}

\textit{(i) Natural Airflow through Open Windows: }
In this scenario, all the windows of H1 are open; thus, the indoor space is naturally ventilated throughout cooking in the kitchen. Due to lack of active ventilation, pollutants accumulate in the kitchen (A) and eventually spread to the nearby bedroom (B), increasing its VOC and PM\textsubscript{2.5} concentration as per \figurename~\ref{fig:rm_medium}, even after the cooking is ended. Most CO\textsubscript{2} gets ventilated through the kitchen's open window. The dining (C) is slightly impacted as shown in \figurename~\ref{fig:din_medium}. However, according to \figurename~\ref{fig:kit_medium}, the kitchen observes a moderate exposure; hence, there is scope for improvement by modulating the airflow around the indoor space.

\begin{figure}
	\captionsetup[subfigure]{}
	\begin{center}
		\subfloat[Kitchen\label{fig:kit_best}]{
			\includegraphics[width=0.3\columnwidth,keepaspectratio]{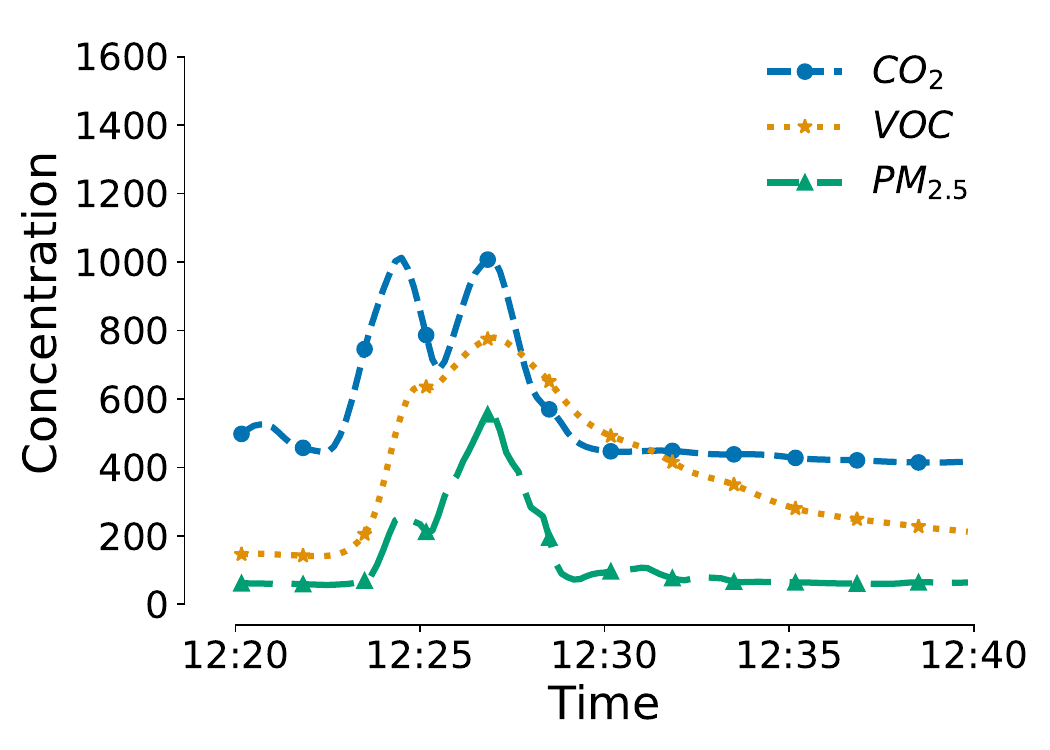}
		}
		\subfloat[Bedroom\label{fig:rm_best}]{
			\includegraphics[width=0.3\columnwidth,keepaspectratio]{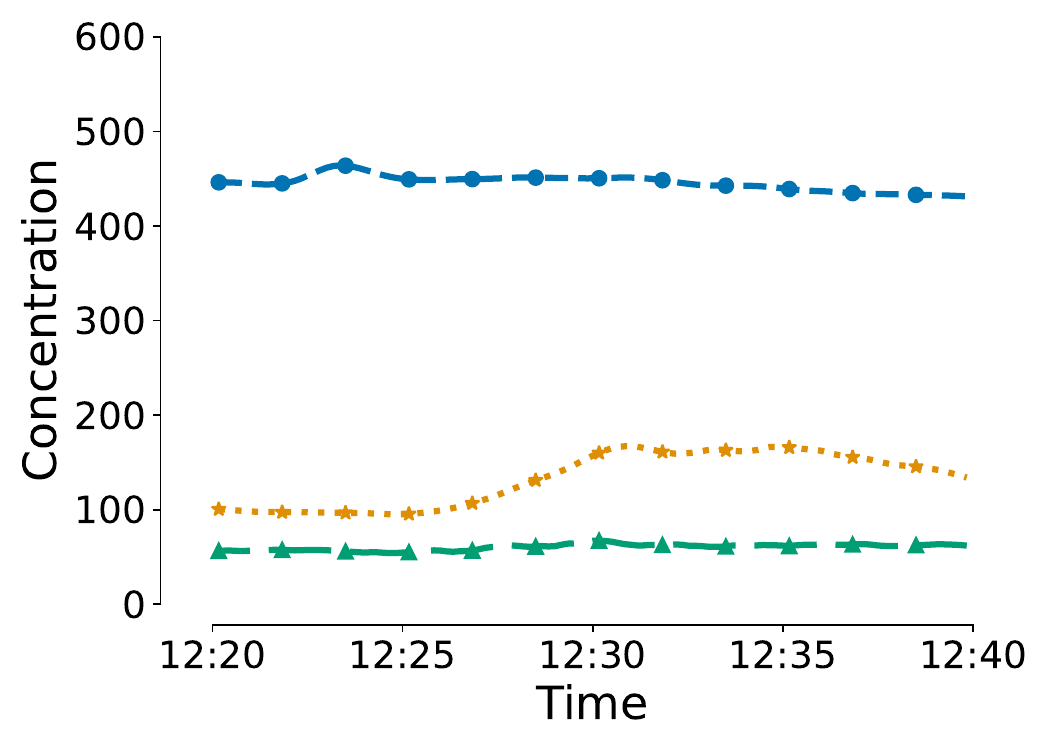}
		}
		\subfloat[Dining\label{fig:din_best}]{
			\includegraphics[width=0.3\columnwidth,keepaspectratio]{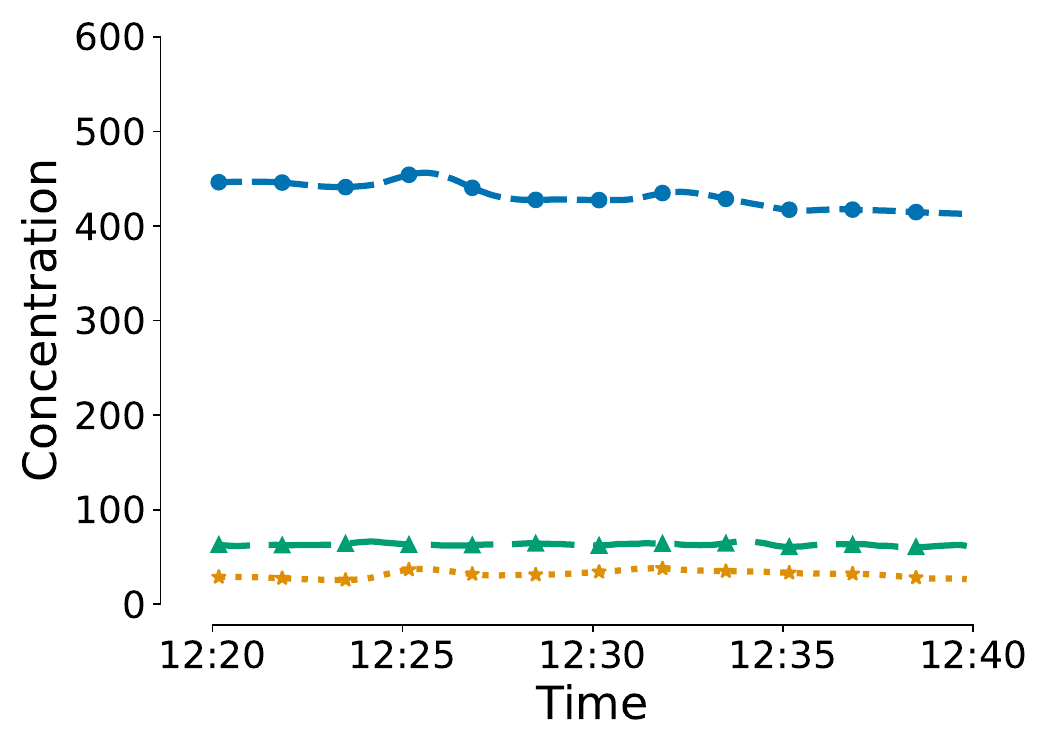}
		}
	\end{center}
	\caption{The kitchen exhaust fan is on, dispersing most pollutants from household H1 to the outdoors. Therefore, the pollutants emitted during cooking in the kitchen (A) are effectively ventilated, and the other rooms remain largely unaffected.}
	\label{fig:spread_best}
\end{figure}

\textit{(ii) Active Ventilation in the Kitchen: } 
In this scenario, the kitchen exhaust fan is turned on; thus, most of the pollutants generated during cooking are efficiently dispersed to the outdoors, keeping exposure in the kitchen (A) at its minimal level as shown in \figurename~\ref{fig:kit_best}, The nearby rooms are slightly affected as depicted in \figurename~\ref{fig:rm_best}, and \figurename~\ref{fig:din_best}. We observe a slight increase in VOC concentration in the nearby bedroom (B) as VOC is relatively complex to be entirely ventilated with airflow and eventually spreads towards other rooms from the source. However, the dining (C) remains unaffected throughout the scenario. Turning on active ventilation with an exhaust fan is the best approach to minimize pollution spread over an indoor space.

\begin{figure}[htp]
	\captionsetup[subfigure]{}
	\begin{center}
		\subfloat[Kitchen\label{fig:kit_worst}]{
			\includegraphics[width=0.3\columnwidth,keepaspectratio]{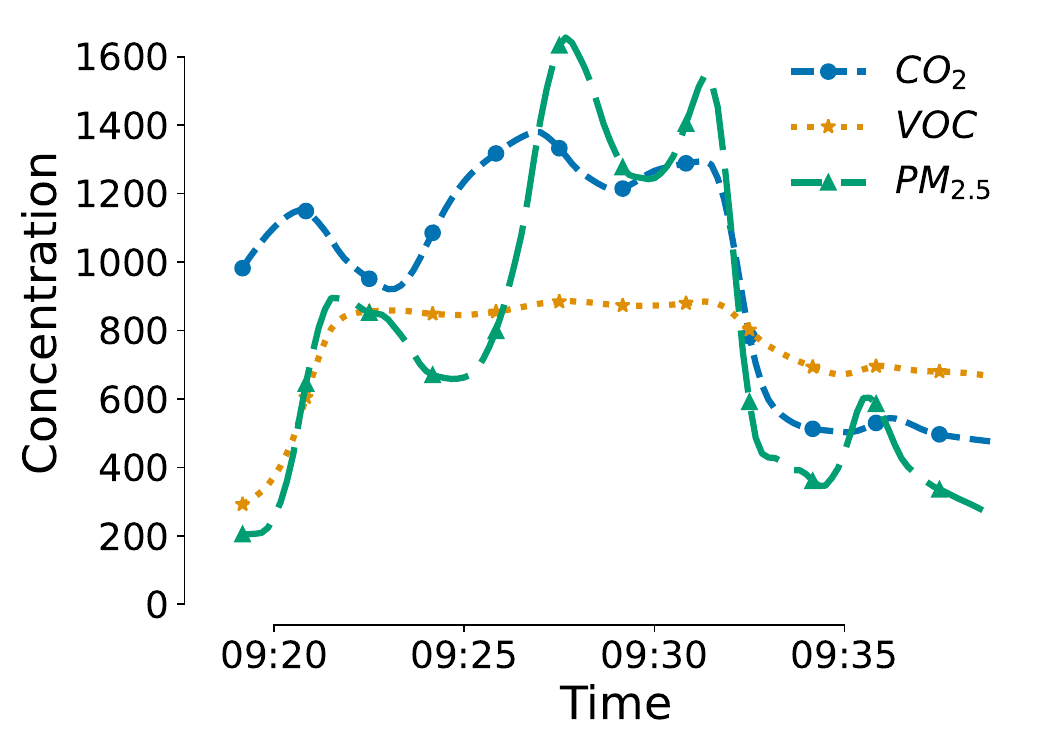}
		}
		\subfloat[Bedroom\label{fig:rm_worst}]{
			\includegraphics[width=0.3\columnwidth,keepaspectratio]{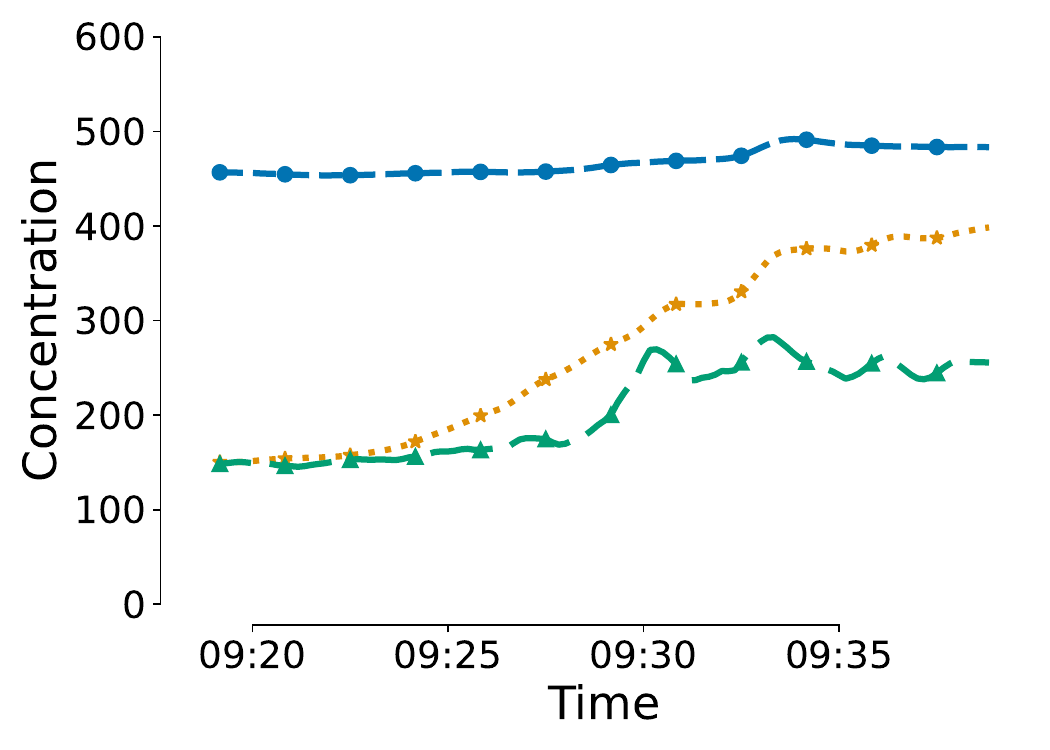}
		}
		\subfloat[Dining\label{fig:din_worst}]{
			\includegraphics[width=0.3\columnwidth,keepaspectratio]{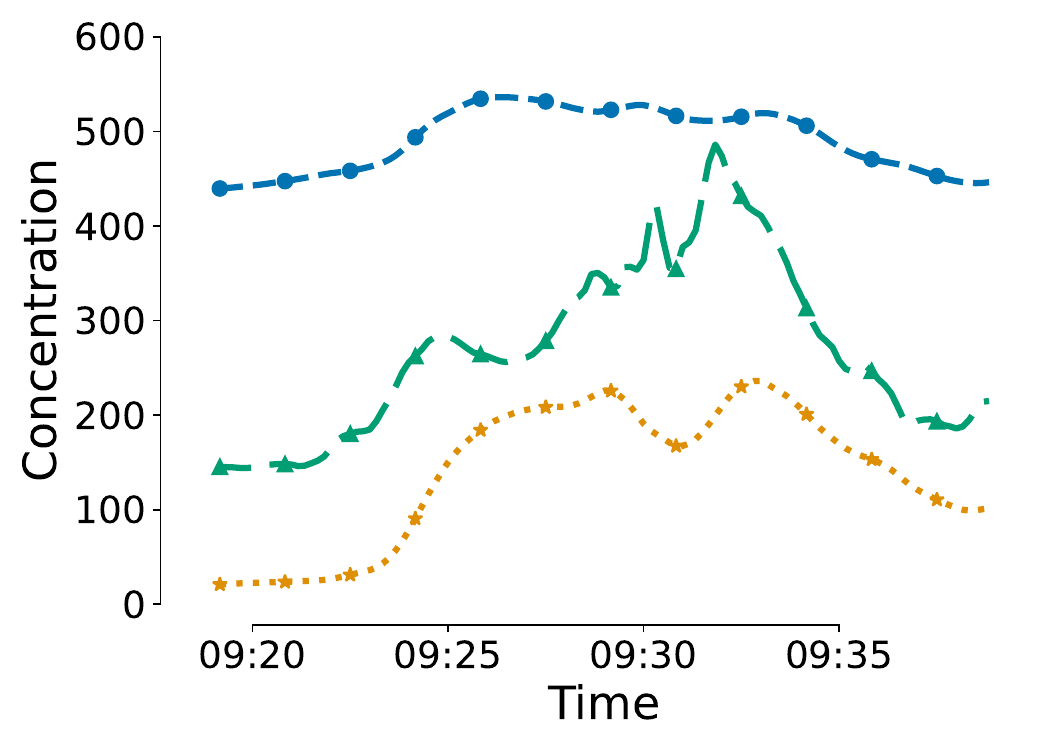}
		}\
	\end{center}
	\caption{The kitchen exhaust fan is off, while the dining ceiling fan of household H1 pulls the pollutants inward due to swirling airflow. Subsequently, it results in a worse spread where the kitchen (A), side by the bedroom (B), and dining (C) are all polluted.}
	\label{fig:spread_worst}
\end{figure}

\textit{(iii) Swirling Airflow in the Dining: }
In this scenario, the kitchen exhaust fan is off, whereas the ceiling fan in the dining area is turned on. The swirling airflow around the ceiling fan pulls pollutants toward the dining room, resulting in maximum spread across the indoor space. Pollutants from the kitchen (A) are forced not to naturally ventilate via the open windows as the ceiling fan pulls the pollutants. Therefore, the kitchen observes the worst pollution accumulation among the above three scenarios as per \figurename~\ref{fig:kit_worst}. Subsequently, dining (C) marks a sharp increase in pollutants, as shown in \figurename~\ref{fig:din_worst}. However, \figurename~\ref{fig:rm_worst} shows that pollutants gradually increase in the nearby bedroom (B) and linger for prolonged periods after cooking. Even with an opened kitchen window, CO\textsubscript{2} is pulled into the dining. Therefore, keeping the dining fan on and the kitchen exhaust off will result in the worst interior spread of pollutants and adversely affect air quality throughout the indoor space. In addition to such airflow dynamics, the room structure and the floor plan of an indoor space provide the necessary pathways for migrating pollutants, impacting the velocity of their spread over different locations of the indoor environment.

\vspace{-0.5cm}
\begin{lesson}{4}{le4}
    Although it is well known that ventilation impacts pollutants in a room, the complex air circulation patterns (due to ceiling fan, exhaust, etc.) in the households of developing countries significantly affect the spreading of the pollutants across other rooms, even when the pollution source (like, kitchen) has ventilation support. 
\end{lesson}

The following section highlights this phenomenon in detail with the help of real-world data collected from multiple measurement sites with different floor plans using the developed platform.

\subsubsection{Impact of Floor plan and Room structure}
The floor plan directly influences the velocity of the spread of pollution, and the degree of such influence varies according to specific pollutants. For instance, VOC spreads more aggressively than CO2 in indoor spaces. Moreover, we identified two crucial behaviors of indoor pollutants, namely (i) Linger and (ii) Trap, that significantly impact the overall exposure level of the occupant throughout the day. Such behaviors are generally temporally related, and lingering pollutants in a sub-optimal building structure lead to trapping the same. We define these behaviors as follows: \textit{(i) \textbf{Linger}: The pollutants keep accumulating for some time in different regions of an indoor space even after the primary pollution source is deactivated and linger for an extended period. (ii) \textbf{Trap}: Pollutants get confined into specific indoor regions due to lack of ventilation and remain trapped for a long time.} To visualize and illustrate such spread patterns of the pollutants in different indoor spaces, we have chosen three exemplar households from our dataset, each having a significantly different floor plan design. Moreover, we present a contrastive analysis of the VOC and CO\textsubscript{2} spread in these households and identify multiple architectural shortcomings.

\begin{figure}
	\captionsetup[subfigure]{}
	\begin{center}
		\subfloat[\label{fig:spread_h1_voc}]{
			\includegraphics[width=0.9\columnwidth]{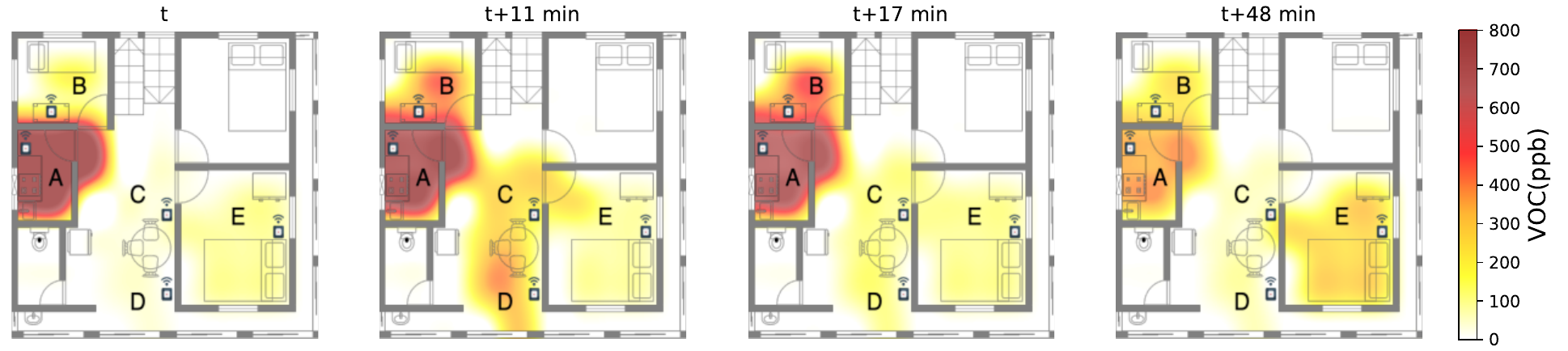}
		}\
		\subfloat[\label{fig:spread_h1_co2}]{
			\includegraphics[width=0.7\columnwidth]{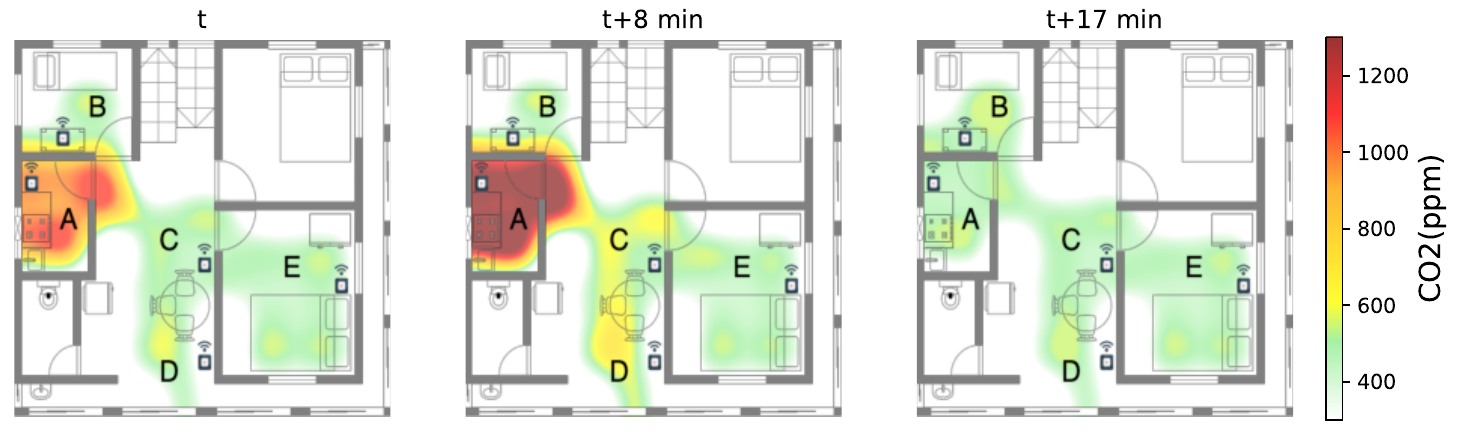}
		}
	\end{center}
	\caption{Spatiotemporal spread of -- (a) VOC, and (b) CO\textsubscript{2} from the kitchen (A) in Household H1. The pollutants spread to the side by the bedroom (B) and the dining room (C, D) with time. The cooking starts at $t$ and ends at $t$+11 minutes. The CO\textsubscript{2} normalizes within 6 minutes at $t$+17 minutes. Finally, at $t$+48 minutes, 37 minutes later, the VOC normalizes; however, trapping of VOC can be observed in the bedroom (E).}
	\label{fig:spread_h1}
\end{figure}

\textit{(i) Floor plan with well-ventilated Kitchen: }
Household-1 (H1) is an 1100 sqft indoor space with six rooms, including the restroom. The upper-right side room remains locked, and the restroom is outside the scope of this study. Therefore, we have placed five sensing devices in the kitchen (A), side by bedroom (B), dining room (C and D), and second bedroom (E). The household has large windows in front of the dining room and kitchen that naturally provide efficient ventilation for the pollutants. The \figurename~\ref{fig:spread_h1_voc} presents the spatiotemporal spread of VOC in H1 during and after cooking activity in Kitchen (A). According to the annotation from the occupant, cooking starts at $t$ time and gets over by $t$+11 mins. However, VOC continues to spread to other internal rooms of the household. At $t$+17 mins, pollutant concentrations deplete without active ventilation. Even after 48 minutes, the VOC lingered in the kitchen, and surprisingly, Bedroom (E) had a relatively higher VOC concentration than the usual scenario. Seemingly, pollutants such as VOC, Ethanol, etc., are hard to ventilate and get trapped in indoor regions with sub-optimal ventilation, leading to increased exposure in indoor environments. Notably, the dining room (C and D) is ventilated easily by the nearby windows.

In contrast, CO\textsubscript{2} is not as aggressive as VOC while spreading indoors. As per the \figurename~\ref{fig:spread_h1_co2}, CO\textsubscript{2} peaks at $t$+8 mins but mostly remains confined within the Kitchen area. The dining place is slightly impacted; however, the other two bedrooms (B and E) remain unaffected. Most importantly, we observe that CO\textsubscript{2} gets ventilated efficiently and quickly depletes to usual levels within 6 mins (see sub-figure $t$+17 min) from the end of the cooking activity; hence, does not linger for an extended time.

\begin{figure}
	\captionsetup[subfigure]{}
	\begin{center}
		\subfloat[\label{fig:spread_h2_voc}]{
			\includegraphics[width=0.7\columnwidth]{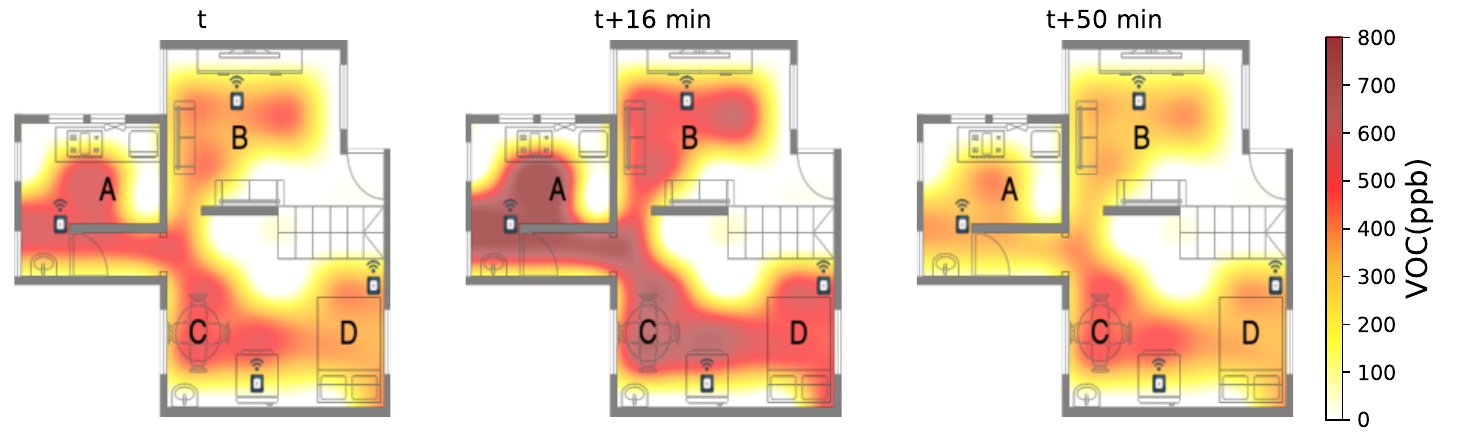}
		}\
		\subfloat[\label{fig:spread_h2_co2}]{
			\includegraphics[width=0.7\columnwidth]{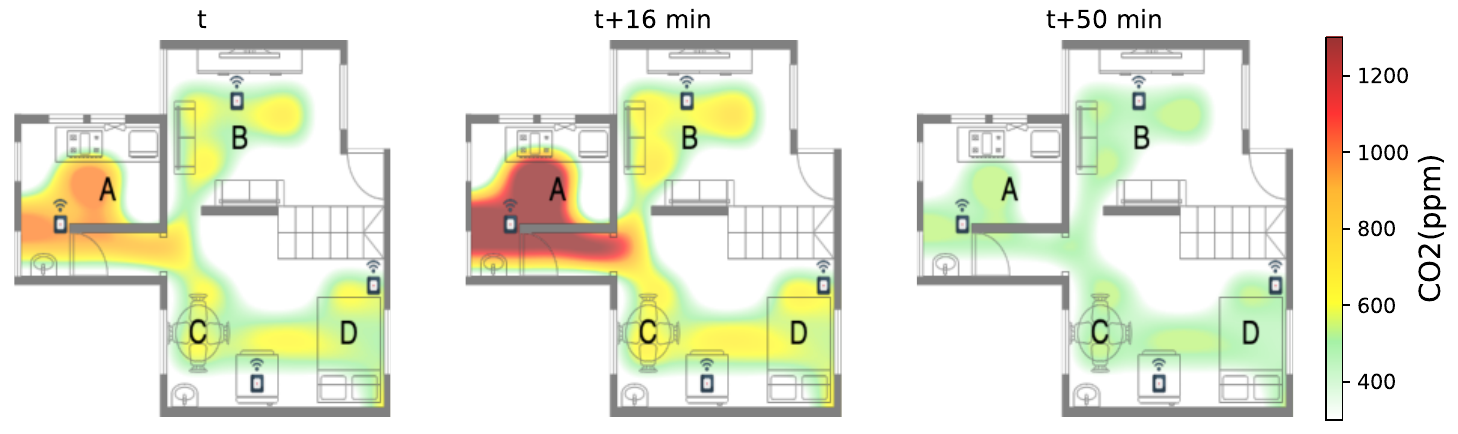}
		}
	\end{center}
	\caption{Spatiotemporal spread of -- (a) VOC, and (b) CO\textsubscript{2} from the kitchen (A) in Household H2. The cooking starts at $t$ and ends at $t$+30 minutes. Due to the connected room structure of H2, the pollutants quickly spread throughout the household. We observed a drop in pollution levels at $t$+50 minutes, 20 minutes after the cooking activity ended.}
	\label{fig:spread_h2}
\end{figure}

\textit{(ii) Floor plan with Kitchen and Hall: }
Household-2 (H2) has a kitchen (A) and a large hall room within an 1100 sqft. Interestingly, the hall is segregated into living area (B), dining (C), and bedroom (D). These hall regions share the air quality due to the absence of walls. Compared to well-partitioned floor plans, H2 has less number of windows to ventilate pollutants from the indoor space naturally. Therefore, we have selected H2 due to its open and interconnected floor plan, compromised natural airflow, and ventilation. The sensing devices are deployed in the kitchen and all segregated regions of the hall room. As depicted in \figurename~\ref{fig:spread_h2_voc}, the living area, bedroom, and dining area in the hall room get uniformly polluted throughout the cooking activity in Kitchen (A) that started at time $t$. From the beginning of the activity in Kithcen, we observe a rapid spread of the pollutants from the kitchen towards the hall room due to the open and interconnected floor plan of H2—the VOC concentration peaks at $t$+16 mins. The activity ended at $t$+30 mins; yet, the household captures significant VOC even at $t$+50 mins. In the figure, we can observe the accumulated VOC in dining even after 20 minutes from when cooking ended. Therefore, shared room designs are ineffective in ventilating VOC efficiently, primarily due to compromised ventilation.

In contrast, CO\textsubscript{2} mostly remained concentrated near the kitchen and gradually migrated towards the hall over time. As per the spatiotemporal plots in \figurename~\ref{fig:spread_h2_co2}, CO\textsubscript{2} spread at each room around $t$+16 mins due to the interconnected nature of the floorplan, giving pathways for the pollutant to migrate to other regions of the household. The CO\textsubscript{2} quickly depletes as the pollution-generating activity, cooking, ends; finally, the CO\textsubscript{2} gets normalized at $t$+50 mins.

\begin{figure}
	\captionsetup[subfigure]{}
	\begin{center}
		\subfloat[\label{fig:spread_h3_voc}]{
			\includegraphics[width=0.95\columnwidth]{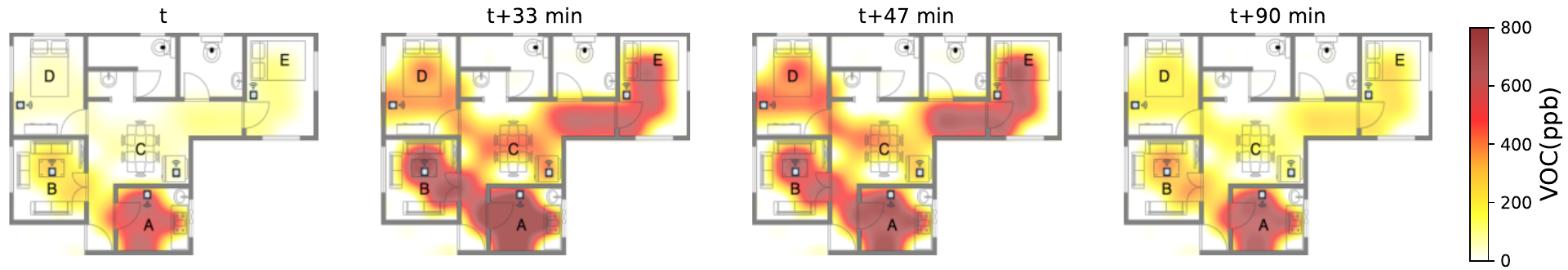}
		}\
		\subfloat[\label{fig:spread_h3_co2}]{
			\includegraphics[width=0.95\columnwidth]{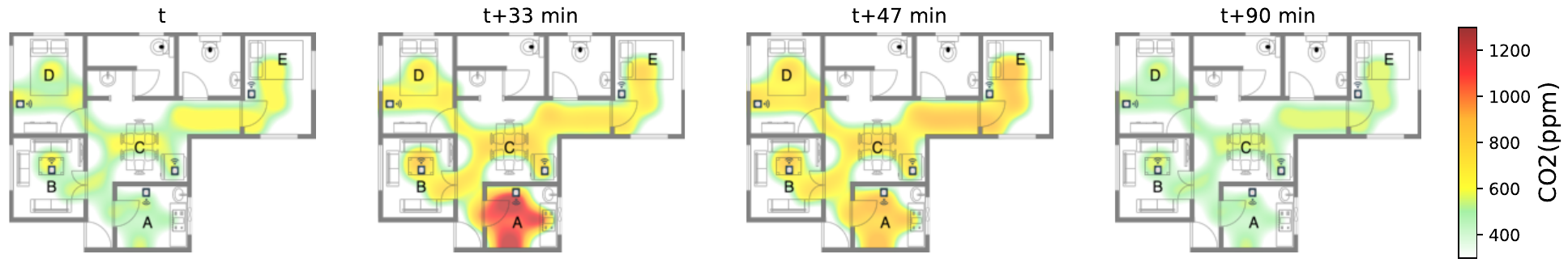}
		}
	\end{center}
	\caption{Spatiotemporal spread of -- (a) VOC, and (b) CO\textsubscript{2} from the kitchen (A) in Household H3. The cooking starts at $t$ and ends at $t$+33 minutes. Both the pollutants get trapped in H3 due to the isolated room structure. CO\textsubscript{2} normalizes 57 minutes later at $t$+90 minutes when VOC persists in the kitchen (A) and the living room (B).}
	\label{fig:spread_h3}
\end{figure}

\textit{(iii) Floor plan with isolated Kitchen: }
Household-3 (H3) is a 1,200 sq ft indoor space with seven rooms, including two restrooms. We deployed sensors in the kitchen (A), living room (B), dining (C), bedroom (D), and bedroom (E). We have chosen household H3 as the primary pollution source (i.e., the kitchen) is situated at one corner of the household; therefore, the kitchen is isolated from the other rooms. The \figurename~\ref{fig:spread_h3_voc} shows the spread of VOC over time and space. Even though the cooking activity ended at $t$+33 mins, the VOC continued lingering mainly towards the living room (B) and dining (C). From the dining room, VOC is further migrated to both bedrooms (D and E) at a slower rate than Household H2. However, with a slower spreading rate, the pollutants are also ventilated slowly. Thus, a significant amount of contaminants was trapped in a less ventilated bedroom (E); see the degree of accumulation in $t$+47 mins sub-figure even after 14 minutes of no activity in the kitchen. Meanwhile, in bedroom (D), the VOC is efficiently ventilated due to open windows. At $t$+90 mins, even 57 mins after the cooking activity has ended, we can see that VOC still lingers across different regions (i.e., trapped in the living room and kitchen) of household H3. We can observe that the VOC in the kitchen is not getting ventilated due to its cornered placement in the floor plan design.

We observe that CO\textsubscript{2} gets uniformly distributed from the kitchen to the entire household; however, it takes comparably more time to deplete in H3. The $t$+47 mins sub-figure of \figurename~\ref{fig:spread_h3_co2} shows that after 14 minutes of no activity in the kitchen, the CO\textsubscript{2} levels do not decrease in any of the rooms of H3 except in the kitchen. Further, the pollutants accumulate in the bedroom (E). However, CO\textsubscript{2} efficiently depletes to normal levels over time without active ventilation, as shown in the $t$+90 mins sub-figure. Meanwhile, VOC is much more challenging to ventilate and prone to getting trapped within less ventilated indoor regions.

\begin{lesson}{5}{le5}
Pollution levels in crucial hot-spot areas such as the kitchen and the living room are greatly influenced by the activities, floor plan, and dynamic indoor air-circulation patterns, which affect the trapping and lingering of the pollutants in different rooms.
\end{lesson}

In summary, isolated rooms are less exposed to pollutants; however, pollutants accumulate in such indoor regions due to compromised ventilation. In the worst case, an isolated kitchen can lead to trapped and long-term lingering pollutants in a household. Regarding more open and interconnected floor plan designs, they are prone to spread and lack adequate ventilation for harmful pollutants like VOC. Room structures that accommodate large windows can be very effective in recovering from a major pollution event. Lastly, a few pollutants (i.e., VOC, Ethanol, etc.) are more aggressive in spreading and challenging to ventilate, irrespective of the floor plan and room structures. However, sub-optimal room structures (i.e., proximity to the kitchen, interconnected rooms, and fewer windows) further complicate indoor pollution dynamics and lead to the trapping of pollutants for an extended time. Indoor pollution primarily depends on the events and activities the occupants perform. The following section explores the emission levels of several contaminants with daily practices and activities, highlighting the short and long-term impacts on air quality.

\begin{figure}[]
	\captionsetup[subfigure]{}
		\begin{center}
            \subfloat[\label{fig:sc_co2_time}]{
			\includegraphics[width=0.45\columnwidth,keepaspectratio]{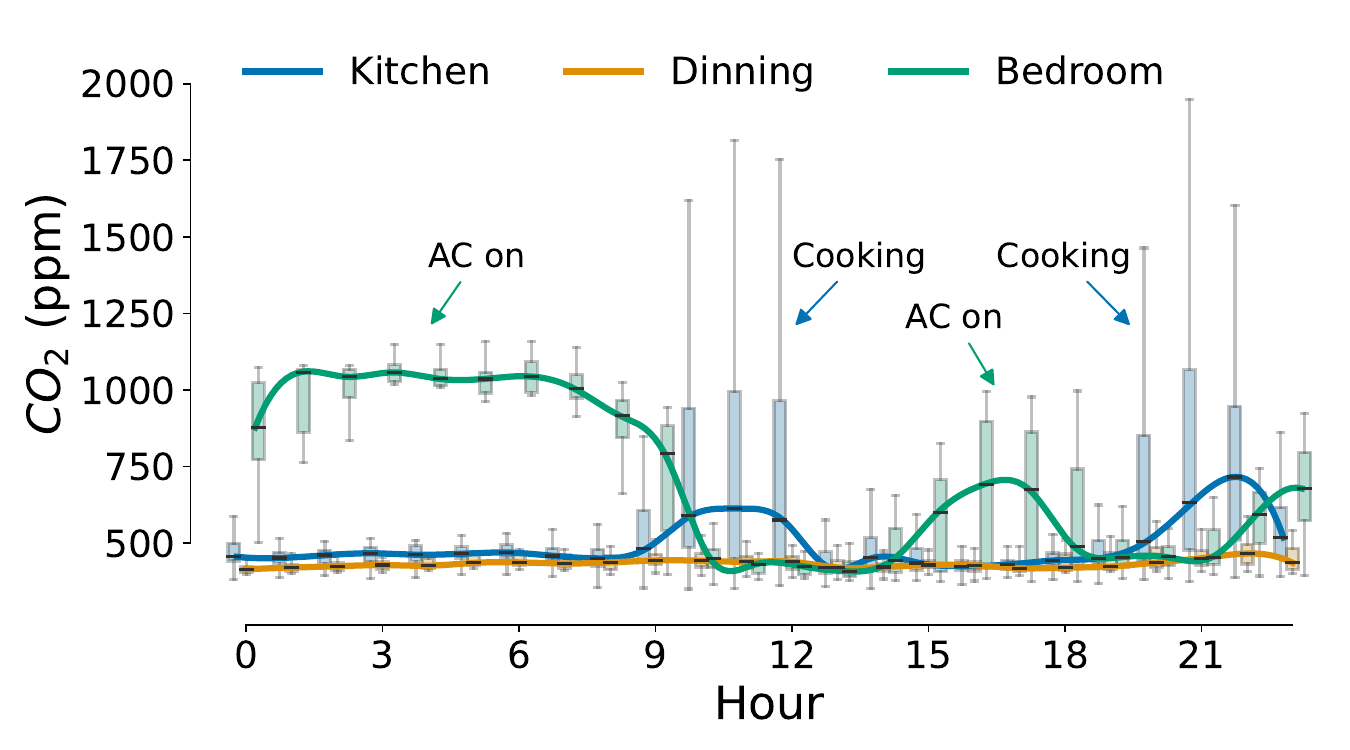}
		}
            \subfloat[\label{fig:sc_voc_time}]{
			\includegraphics[width=0.45\columnwidth,keepaspectratio]{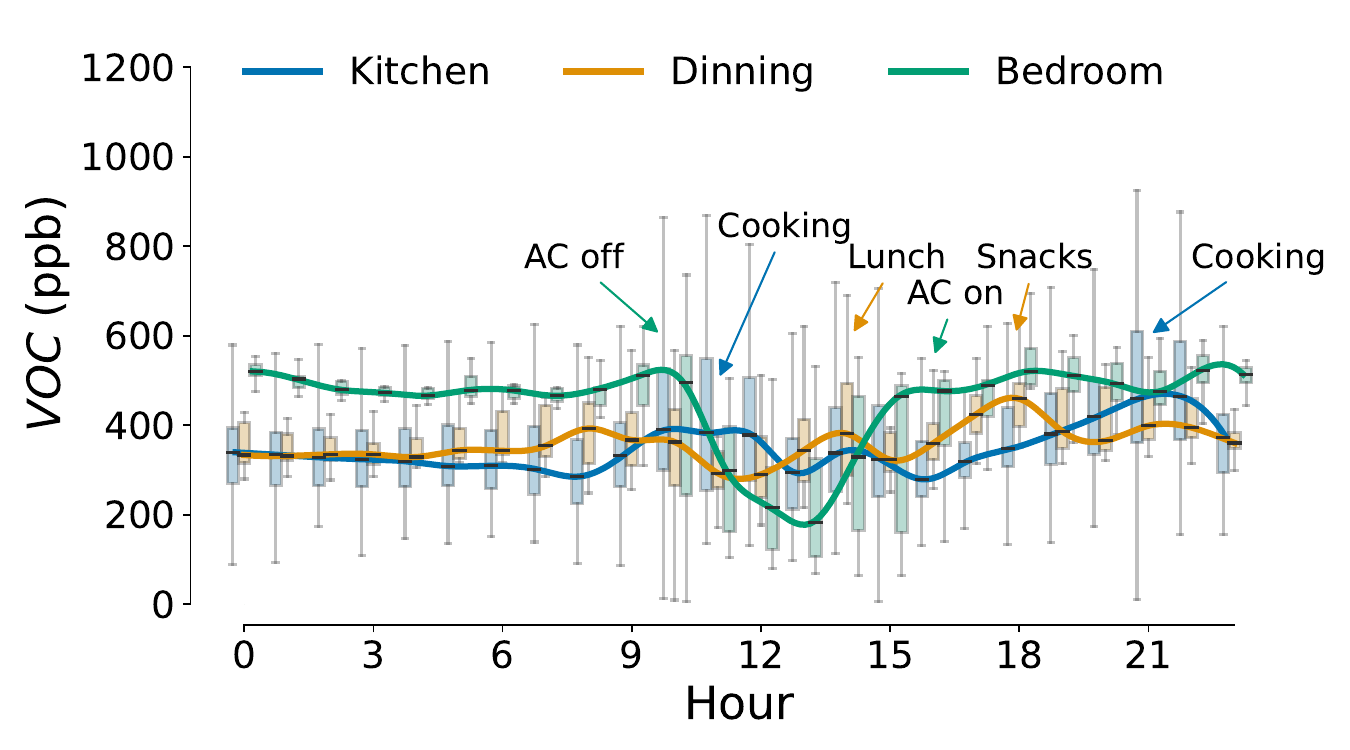}
		}
	\caption{Variations in concentrations of (a) CO\textsubscript{2}, and (b) VOC over hours of the day. Daily activities significantly influence the pollutants in the kitchen, dining room, and bedroom. For example, the bedroom gets polluted when the AC is on and the windows are closed. Similarly, the kitchen and dining area get polluted when food is prepared.}
	\label{fig:poll_time_h2}
        \end{center}
\end{figure}

\subsection{Daily Activities and Pollution}
Indoor pollutants exhibit distinct accumulation and spreading patterns based on the activity and how the activity is being performed. Therefore, indoor pollutants follow a periodic pattern in our daily household activities. Specifically, different parts of the indoors behave as acting pollution sources at different times of the day. Therefore, in \figurename~\ref{fig:poll_time_h2}, we observe that the median CO\textsubscript{2} and VOC concentrations are significantly different across the kitchen, dining, and bedroom for different hours of the day. For instance, \figurename~\ref{fig:sc_co2_time} shows that in the kitchen, CO\textsubscript{2} is emitted during cooking, and with good ventilation (exhaust fans, open windows, etc.), it quickly descends to normal levels. However, for the bedroom, the median CO\textsubscript{2} levels are high (more than even the kitchen's peak CO\textsubscript{2}) throughout the night hours, mainly due to lack of ventilation as shown in Section~\ref{sec:insuff_airflow}. Further, VOC also shows similar accumulation patterns as CO\textsubscript{2} in the bedroom and kitchen. However, unlike CO\textsubscript{2}, VOC does not deplete rapidly even with good ventilation and lingers for an extended period, resulting in long-term exposure as shown in \figurename~\ref{fig:sc_voc_time} (see bedroom from 18:00 to 21:00). Moreover, VOC is naturally emitted from fruits, vegetables, food residuals; thus in the figure, we see a steady increase of VOC in the kitchen from the evening hours until the kitchen is cleaned (see 22:00). Accordingly, the dining place also observed an increase in the VOC during lunch, which gets descended after the dining was cleaned. 

\begin{wrapfigure}[13]{L}{0mm}
	\captionsetup[subfigure]{}
            \subfloat[\label{fig:fruit}]{
			\includegraphics[width=0.3\columnwidth,keepaspectratio]{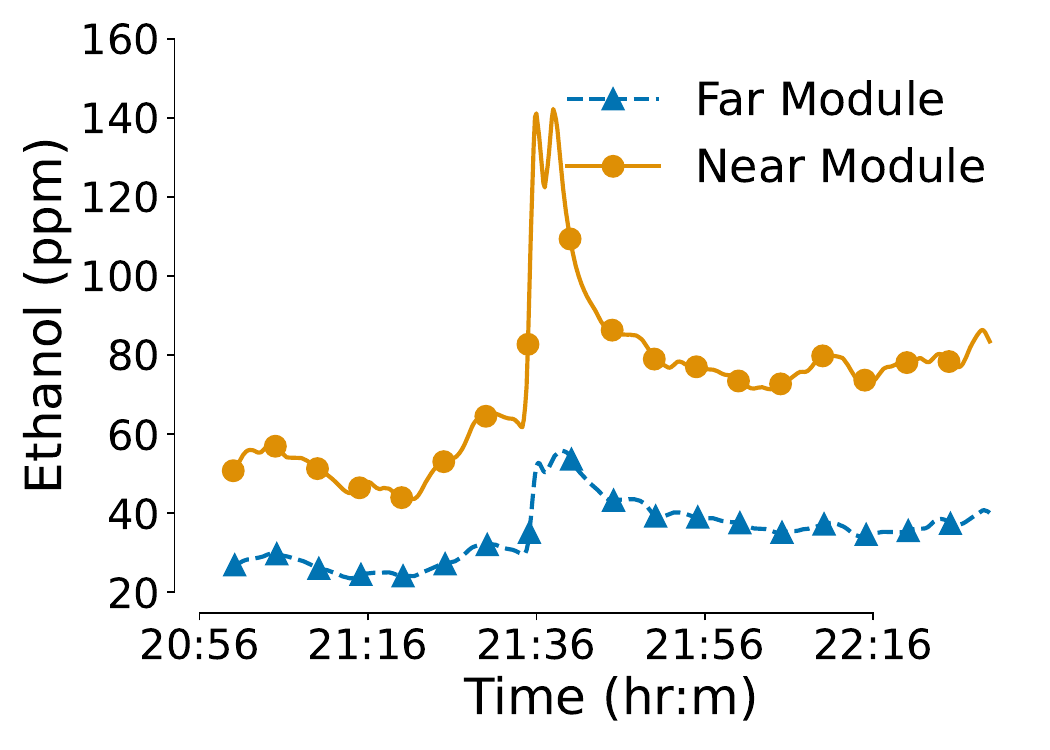}
		}
		\subfloat[\label{fig:leftover}]{
			\includegraphics[width=0.3\columnwidth,keepaspectratio]{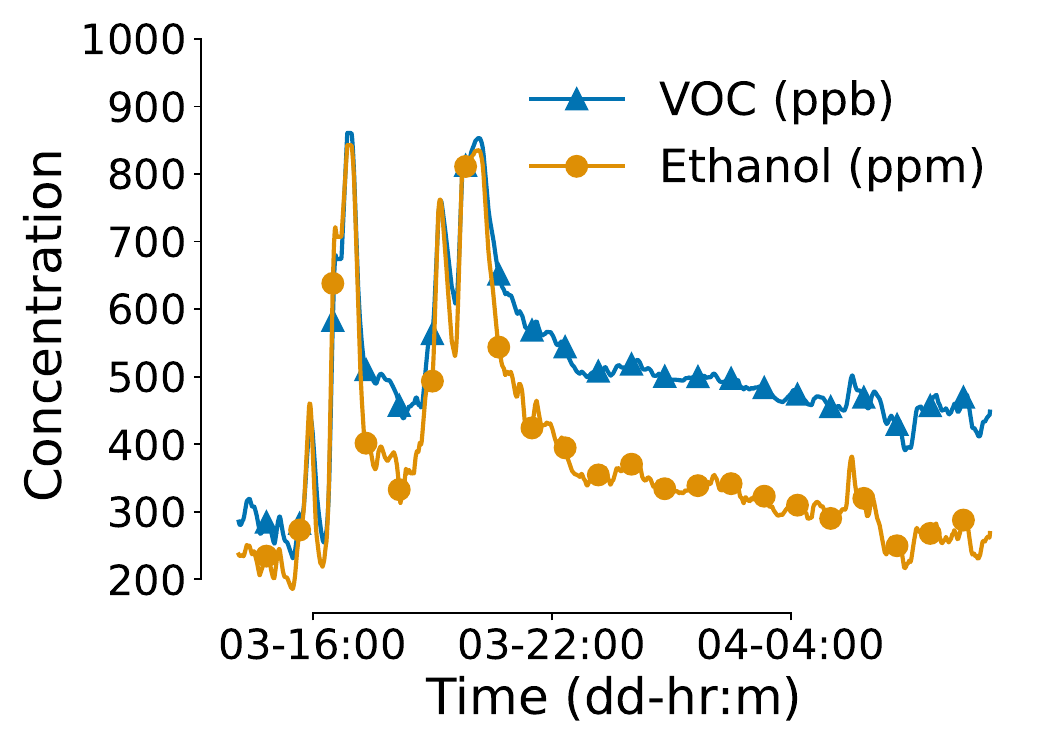}
		}\
	\caption{Emitted pollutants such as -- (a) Ethanol from fruit scraps in dining, (b) VOC and Ethanol from Food residuals and dirty dishes in the kitchen. It can be observed that the pollutants accumulate in the kitchen overnight and persist till the next day.}
	\label{fig:spread}
\end{wrapfigure}

We identified a few cases where the occupants underestimated the severity of pollution generated due to their behavior, which can lead to unintentional long-term exposure. For instance, fruit scraps and meal residuals left in an indoor location (i.e., kitchen sink) cause extended contamination, or how the food is being cooked generates different pollution intensities. Details of such cases are shown below.

\subsubsection{Fruit scraps and Food Residuals}
Due to the fine-grained activity annotation from the users, we can associate minor changes in pollutants with the root cause. For example, \figurename~\ref{fig:fruit} depicts the rise in the Ethanol concentration at the nearby sensing modules when the user cuts fruits at the dining table, and the scraps are disposed of after a while. We can observe that both the sensors capture the event at almost a similar time; however, the nearest one experiences a higher exposure. Similarly, \figurename~\ref{fig:leftover} shows the measurements from a kitchen during the night hours. The excess food residuals and dirty dishes in the kitchen sink cause elevated levels of VOC and Ethanol until the kitchen is cleaned up the next day. Notably, such association of annotated events with the pollutant readings sensed over different adjacent rooms in a household can provide the user a clear understanding of the room's healthiness, along with the spreading, trapping, and lingering nature of the pollutants across the adjacent rooms, leading to specific, actionable items for them, thus promoting healthy living.

\begin{figure}
	\centering
	\includegraphics[width=0.5\columnwidth]{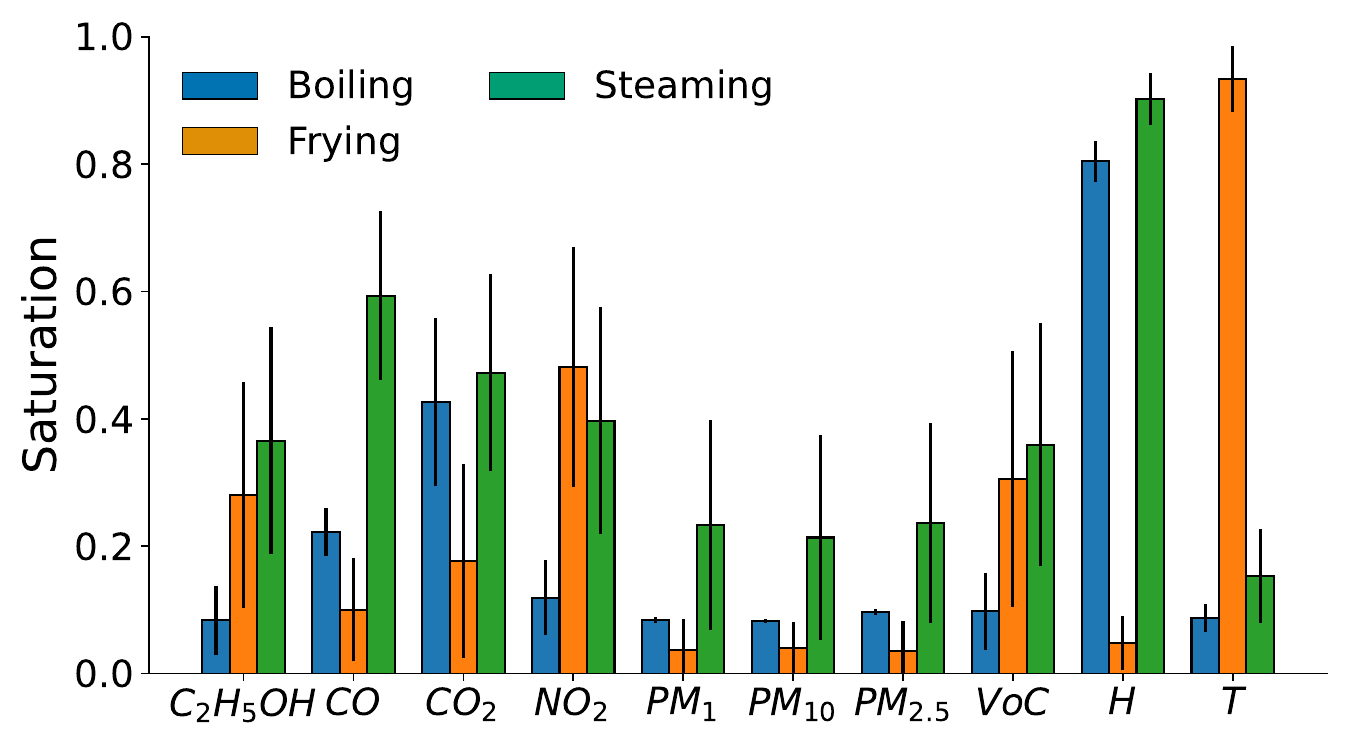}
	\caption{Saturation of pollutants for different cooking methods. Boiling emits the least pollutants except CO and CO\textsubscript{2} from the burner. Frying increases temperature, so occupants turn on the exhaust fan, improving the ventilation. Steaming emits the most pollutants among the above cooking methods.}
	\label{fig:poll_cook}
\end{figure}

\subsubsection{Cooking Method}
\label{sec:cook_style}
Pollutants can exhibit entirely different distributions based on how an activity is performed. Thus, the context of the activity like \textit{``What is being cooked''} or \textit{``Which detergent is used while cleaning the floor''} is more critical for characterizing which pollutants will majorly contaminate the indoor environment. To realize such complex pollutant dynamics, we observe three types of cooking activity, namely \textit{boiling}, \textit{frying}, and \textit{steaming}, which have significantly dissimilar pollutant signatures as shown in \figurename~\ref{fig:poll_cook}.

In the case of \textit{boiling}, we can see an increase in the humidity, while most of the pollutants are dormant except CO and CO\textsubscript{2} as the kitchen's ventilation system is usually underused, resulting in accumulation of such gases. Whereas, \textit{frying} emits a lot of C\textsubscript{2}H\textsubscript{5}OH, NO\textsubscript{2}, VOC, and increases the temperature in the kitchen; thus, the exhaust fan is generally turned on, significantly lowering the concentration of CO, CO\textsubscript{2} and particulate matters (PM\textsubscript{x}). Unlike \textit{frying}, \textit{steaming} does not increase temperature significantly, leading to under-utilization of the exhaust fan, as we observed from our dataset. However, unlike \textit{boiling}, \textit{steaming} emits lots of pollutants such as C\textsubscript{2}H\textsubscript{5}OH, NO\textsubscript{2}, PM\textsubscript{x}, VOC that accumulates throughout the activity. Moreover, due to lack of ventilation, CO and CO\textsubscript{2} also accumulate, resulting in the highest exposure among the three cooking activities. Such observations further strengthen our hypothesis in Section~\ref{sec:cook_exhaust_off} that humans are more sensitive to high environmental temperature and humidity, which leads to unintentional accumulation and spreading of harmful pollutants indoors.

\begin{lesson}{6}{le6}
Steaming may create more pollution than frying if the kitchen's ventilation is poorly controlled due to human perceptions of the environment. Consequently, correlating activities with the pollutant distributions across different rooms is vital to provide actionable insights to the users for improving the household's air quality. 
\end{lesson}

\subsection{Qualitative Analysis of \ourmethod{}}
\label{sec:survey_design}
To realize the practical utility of \ourmethod{} platform from the user's perspective, we have conducted a PSSUQ (Post Study System Usability Questionnaire) survey on the usefulness and user-friendliness of the sensing device. PSSUQ survey questionnaires primarily consist of various statements regarding the underlying system's quality, utility, and effectiveness. The study participants were asked to agree or disagree with these statements on a $7$-point Likert scale (i.e., Strongly Disagree, Disagree, Somewhat Disagree, Neutral, Somewhat Agree, Agree, Strongly Agree). The Statements that are asked about the system in the survey are as follows:

\noindent\textbf{System Usefulness (SYSUSE)}
\begin{enumerate}
    \item[UQ-1] Overall, I am satisfied with how easy it is to deploy the devices in my house.
    \item[UQ-2] It was simple to configure the devices with my home's WiFi and start using the system.
    \item[UQ-3] I could monitor my home's air quality using this system.
    \item[UQ-4] I felt comfortable using this system.
    \item[UQ-5] It was easy to reconfigure a device locally/remotely if needed.
    \item[UQ-6] I believe I could become more cautious and aware of my home's air quality using this system.
\end{enumerate}

\noindent\textbf{Information Quality (INFOQUAL)}
\begin{enumerate}
    \item[UQ-7] The device resolved the errors itself or gave me error messages that clearly told me how to fix problems.
    \item[UQ-8] Whenever I made a mistake using the system (e.g., turning off the power), I could recover easily and quickly.
    \item[UQ-9] The process to deploy and configure the devices was clearly mentioned.
    \item[UQ-10] It was easy to find the information I needed to set up the devices for the first time.
    \item[UQ-11] The configuration steps were very easy and effective for quickly setting up the devices.
    \item[UQ-12] The device's build quality is good and structurally strong.
\end{enumerate}

\noindent\textbf{Interface Quality (INTERQUAL)}
\begin{enumerate}
    \item[UQ-13] The device looked nice in different rooms of my home.
    \item[UQ-14] I liked using this system to monitor my home's air quality.
    \item[UQ-15] This system has all the functions and capabilities I expect it to have.
    \item[UQ-16] Overall, I am satisfied with this system.
\end{enumerate}

\noindent To estimate the stress and effort levels for the activity annotation using the developed Android application, we have further conducted a NASA-TLX (NASA Task Load Index) survey. The NASA-TLX survey questionnaires represent Cognitive Demand (CD), Physical Demand (PD), Temporal Demand (TD), Mental Effort (ME), Performance Effort (PE), and Frustration Level (FR) of the participant while annotating her activities throughout the day. For each survey questionnaire, the participants are asked to fill in the responses on a scale of 1 (very low task load) to 20 (very high task load). Following is the list of the survey questions.

\begin{enumerate}[TQ-1]
    \item (CD) How much speculation, decision-making, or calculation was required to perform the activity annotation?

    \item (PD) The amount and intensity of physical activity required to complete the activity annotation.

    \item (TD) The amount of time spent in completing the activity annotation.

    \item (ME) How much effort do you have to put in to perform the annotation task?

    \item (PE) How difficult was it to recall the correct events corresponding to a change point while annotating?/ How difficult was it to get the correct annotation as instructed by you to the application?

    \item (FR) How much stress were you while annotating events?
\end{enumerate}

\begin{figure}[]
	\captionsetup[subfigure]{}
	\begin{center}
		\subfloat[Looks nice in rooms\label{fig:p3_looknice}]{
			\includegraphics[width=0.3\columnwidth,keepaspectratio]{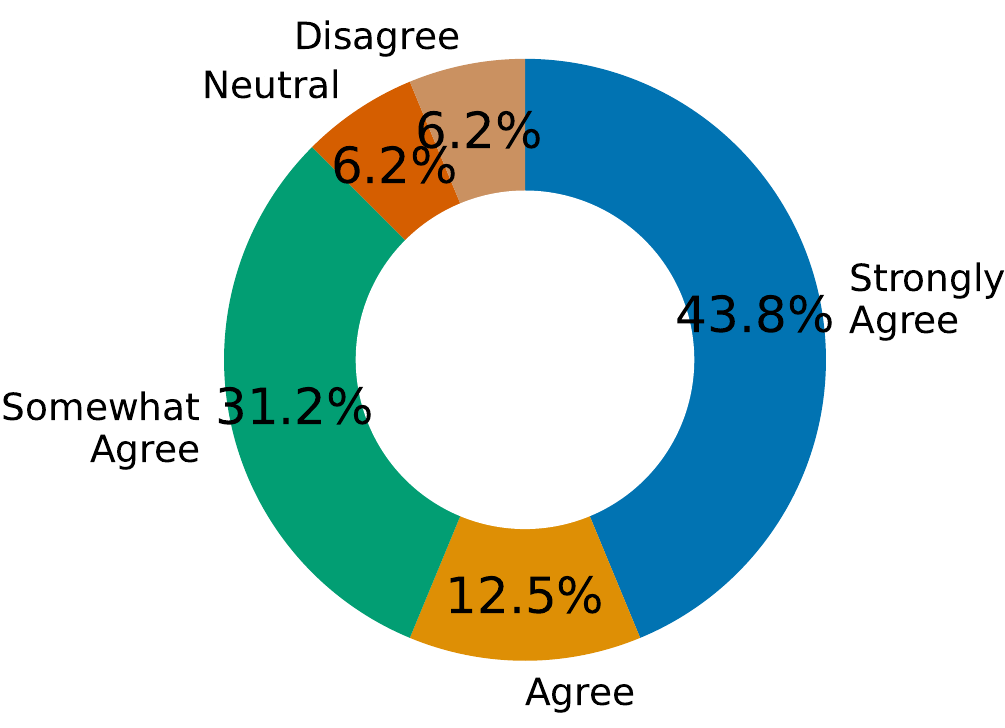}
		}
            \subfloat[Makes me cautious\label{fig:p3_caution}]{
			\includegraphics[width=0.28\columnwidth,keepaspectratio]{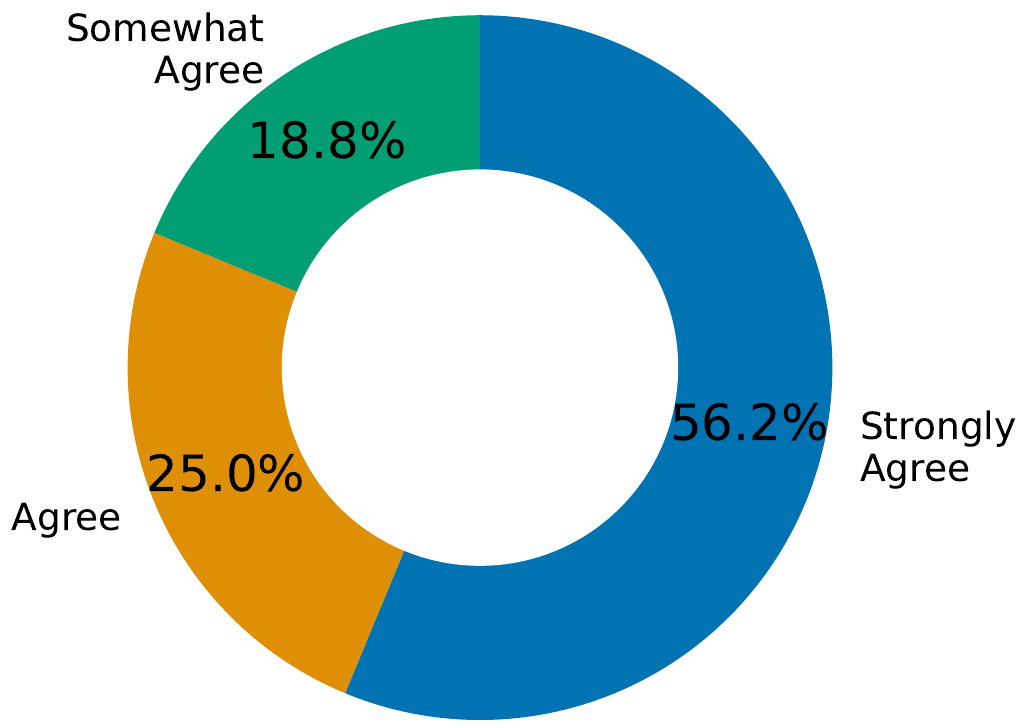}
		}
            \subfloat[Satisfied with the design\label{fig:p3_satisfy}]{
			\includegraphics[width=0.3\columnwidth,keepaspectratio]{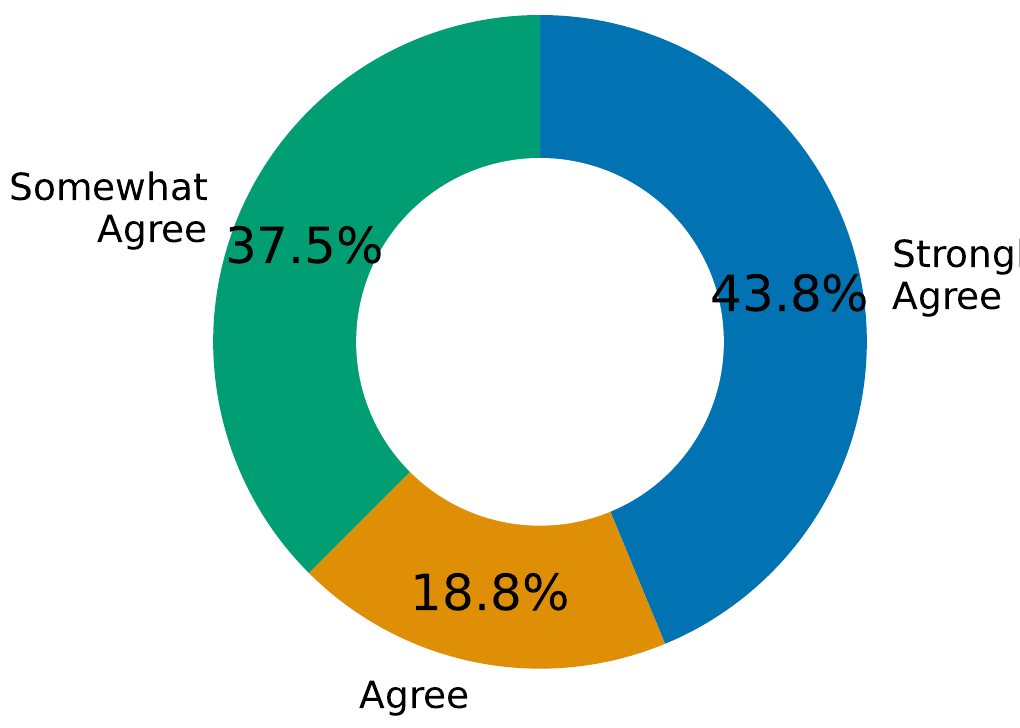}
		}\
	\end{center}
	\caption{User agreement with different survey questionnaires. In total, $87.5\%$ agree with the aesthetics, everyone agrees with an increase in awareness for indoor air quality, and are satisfied with the \ourmethod{} platform.}
	\label{fig:agree}
\end{figure}

Both the PSSUQ and NASA-TLX survey questionnaires were floated among the participants during the last week of data collection. Based on the survey responses, we present a qualitative analysis of the \ourmethod{} platform as follows:

\subsubsection{Portable Design of \ourmethod{} Platform} 
As depicted by several studies on existing air quality monitors, compactness and portability play a vital role in how quickly the monitor will be blended into the household, and the occupants also accept it as a part of their surroundings. Thus, the \ourmethod{} sensing module is designed keeping in mind the aesthetics criteria as well. As shown in \figurename~\ref{fig:p3_looknice}, from the survey responses, we found that $43.8$\% strongly agree, $12.5$\% agree, and $31.2$\% Somewhat agree, totaling $87.5$\% agreement among the participants that the sensing devices of \ourmethod{} platform look nice in the rooms of their household. Notably, only $6.2$\% of the participants disagree with the aesthetics of the platform, whereas $6.2$\% remain neutral, indicating room for further improvements.

\subsubsection{Indoor Pollution Awareness}
The effectiveness of a platform in making the end user more cautious about his surroundings is a crucial property in the case of air quality monitoring systems. As shown in \figurename~\ref{fig:p3_caution}, we observe that $56.2$\% users strongly agree, $25$\% agree, where others somewhat agree that they have become more aware of the pollution events and hot-spots, as a result, become more cautious about their house's air quality.

\subsubsection{Ease of use \& User-friendly Platform}
User satisfaction primarily depends on the user-friendly design of the system, ease of use, robustness against failures, and user interactiveness. During the PSSUQ survey, we asked whether the user was satisfied with the \ourmethod{} platform (UQ-16). \figurename~\ref{fig:p3_satisfy}, shows the user agreement responses for the satisfaction level where $43.8$\% strongly agree, $18.8$\% agree, and others somewhat agree that they are satisfied while using the platform for \datamonths{} of field study.

\subsubsection{System Usability and Quality Metrics}
\begin{figure}
	\captionsetup[subfigure]{}
	\begin{center}
		\subfloat[PSSUQ Quality Metrics\label{fig:pssuq}]{
			\includegraphics[width=0.35\columnwidth,keepaspectratio]{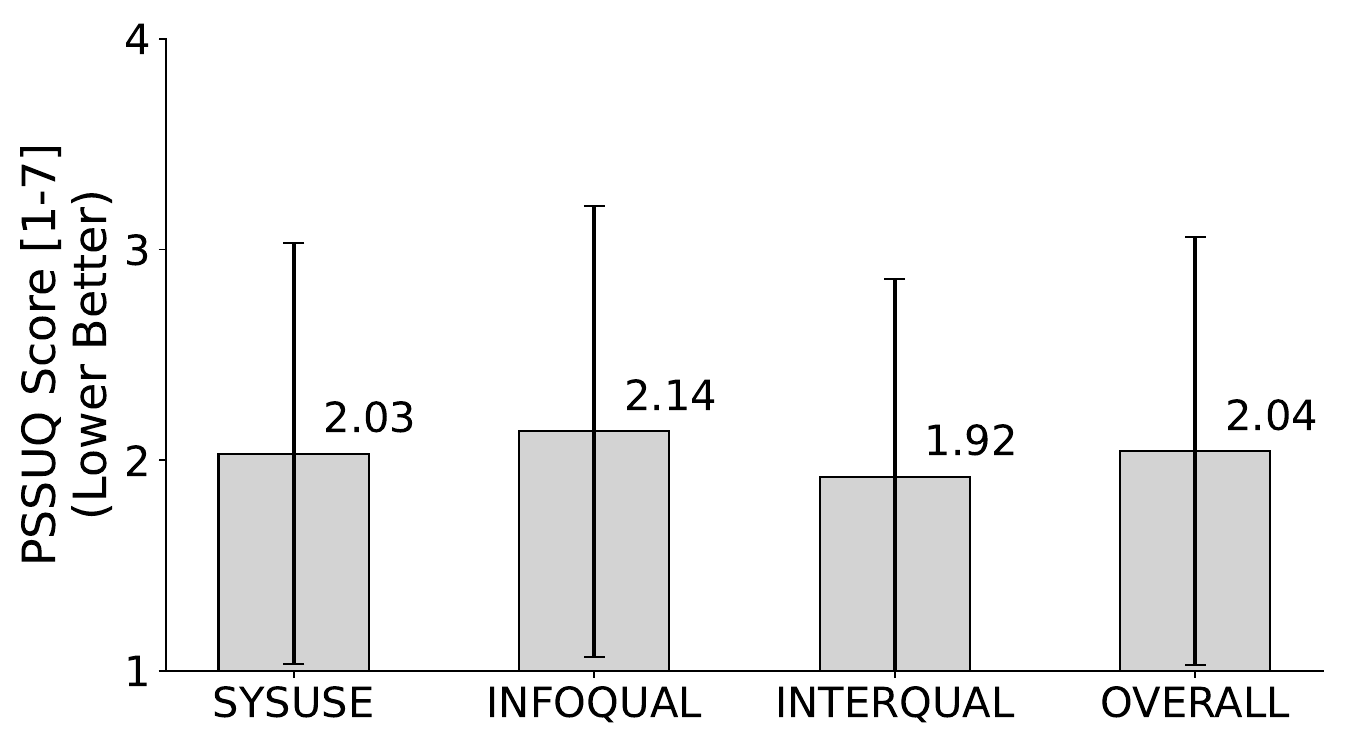}
		}\
		\subfloat[NASA-TLX Workload Demands\label{fig:tlx_score}]{
			\includegraphics[width=0.35\columnwidth,keepaspectratio]{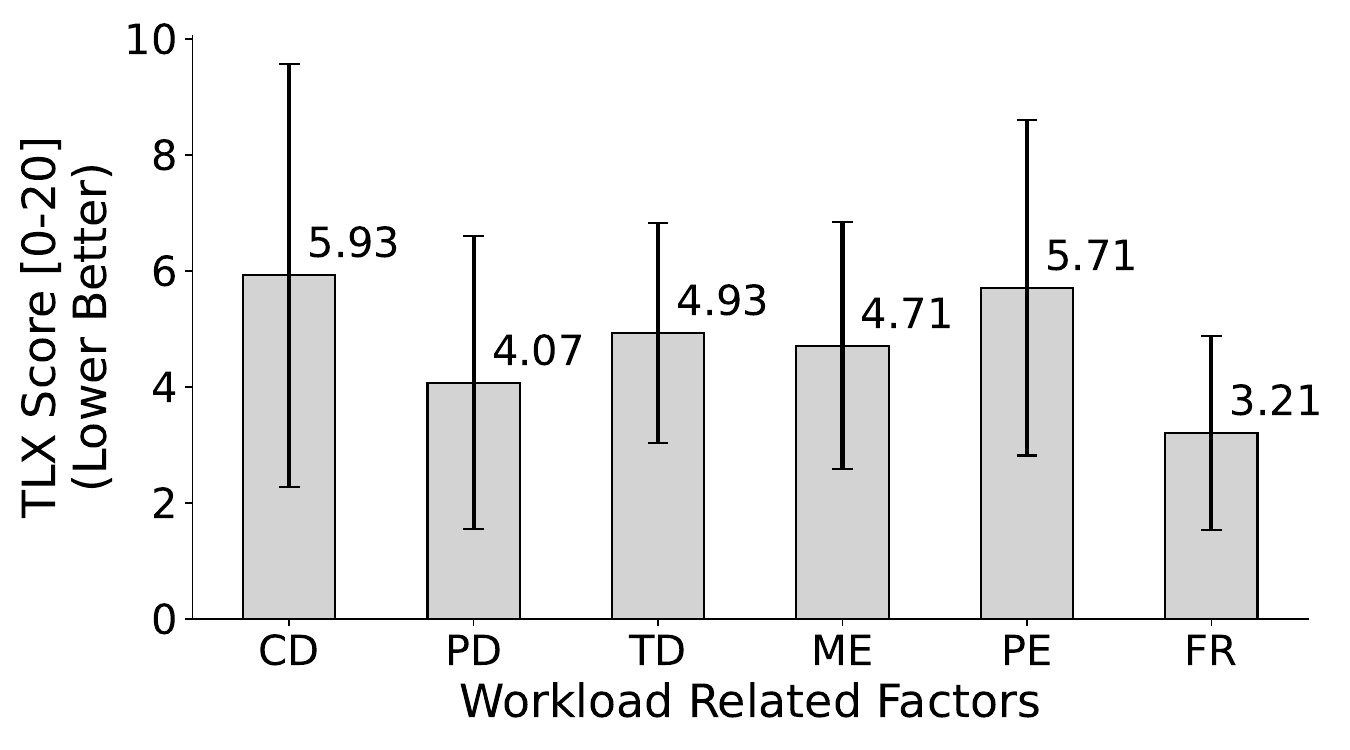}
		}\
	\end{center}
	\caption{Quality metrics of the \ourmethod{} platform and workload demand for human-in-the-loop annotation process. The overall score of the platform is $2.04$, implying a highly practical system. The reported frustration level for the annotation process is $3.21$ out of $20$.}
	\label{fig:survey_scores}
\end{figure}

The PSSUQ questionnaires are further grouped to compute metrics such as system usability (SYSUSE, UQ-1 to UQ-6), information quality (INFOQUAL, UQ-7 to UQ-12), interface quality (INTERQUAL, UQ-13 to UQ-16), along with the overall utility score (OVERALL, UQ-1 to UQ-16) as described in Section~\ref{sec:survey_design}. These metrics denote scores on a scale of $1$ (\textit{strongly agree}) to $7$ (\textit{strongly disagree}) to quantify each of the qualities mentioned above of the underlined platform. Based on the survey responses, the system usability score is $2.03$, the information quality score is $2.14$, and the interface quality score is $1.92$, resulting in an overall score of $2.04$ as depicted in \figurename~\ref{fig:pssuq}. Therefore, we can realize that the platform strikes as practical, and users overall agree on the general utility of the system to monitor the air quality of their indoor spaces.

\subsubsection{App-based Annotation Workload} 
As discussed in Section~\ref{sec:develop}, we developed an Android application to easily annotate the activities on the fly. Moreover, the ending of a particular event is detected with the help of the change-point detection module, reducing the participant's mental load. To understand the workload-related factors during annotation, we conducted the NASA-Task Load Index (TLX) survey, as mentioned earlier. \figurename~\ref{fig:tlx_score}, shows TLX-scores (between 1 to 20) for the factors Cognitive Demand (CD), Physical Demand (PD), Temporal Demand (TD), Mental Effort (ME), Performance Effort (PE), Frustration Level (FR). From the figure, we observe that the app-based annotation process incurs significantly less stress as well as physical and mental demand to perform well in the annotation task, keeping the users' frustration $5$ folds below the maximum level.

\begin{table}
\centering
\scriptsize
\caption{Features vs. price comparison between commercially available air quality monitors and \textit{DALTON} (Approximate price as of \today).}
\label{tab:price_comp}
\resizebox{\textwidth}{!}{%
\begin{tabular}{|l|c|c|c|c|c|c|c|c|} 
\hline
\textbf{Devices}         & \begin{tabular}[c]{@{}c@{}}\textbf{Remote}\\\textbf{Maintenance}\end{tabular}  & \begin{tabular}[c]{@{}c@{}}\textbf{Actionable}\\\textbf{Insights}\end{tabular} & \begin{tabular}[c]{@{}c@{}}\textbf{User}\\\textbf{Feedback}\end{tabular} & \textbf{Visualisation}     & \begin{tabular}[c]{@{}c@{}}\textbf{Pollutant}\\\textbf{Count}\end{tabular} & \begin{tabular}[c]{@{}c@{}}\textbf{Measured}\\\textbf{Pollutants}\end{tabular} & \begin{tabular}[c]{@{}c@{}}\textbf{Cloud}\\\textbf{Connected}\end{tabular} & \begin{tabular}[c]{@{}c@{}}\textbf{Price}\\\textbf{(USD)}\end{tabular}  \\ 
\hline
Pallipartners~\cite{pallipartners} & \xmark                                                                                                 & \xmark                                                                             & \xmark                                                                       & Screen                     & 4                                                                          & CO, CO\textsubscript{2}, HCHO, VOC                                                            & \xmark                                                                         & 101                                                                  \\ 
\hline
Yvelines~\cite{yvelines}             & \xmark                                                                                                 & \xmark                                                                             & \xmark                                                                       & Screen                     & 4                                                                          & PM\textsubscript{x}, CO\textsubscript{2}, HCHO, VOC                                                           & \xmark                                                                         & 113                                                                  \\ 
\hline
Smiledrive~\cite{smiledrive}          & \xmark                                                                                                 & \xmark                                                                             & \xmark                                                                       & Screen                     & 3                                                                          & PM\textsubscript{x}, HCHO, VOC                                                                & \xmark                                                                         & 119                                                                  \\ 
\hline
INKBIRDPLUS~\cite{INKBIRDPLUS}              & \xmark                                                                                                 & \xmark                                                                             & \xmark                                                                       & Screen                     & 2                                                                          & CO\textsubscript{2}, PM\textsubscript{x}                                                                       & \xmark                                                                         & 119                                                                  \\ 
\hline
ExGizmo~\cite{exgizmo}                  & \xmark                                                                                                 & \xmark                                                                             & \xmark                                                                       & Screen                     & 1                                                                          & PM\textsubscript{x}                                                                            & \xmark                                                                         & 119                                                                  \\ 
\hline
NETATMO~\cite{netatmo_indoor}                  & Low                                                                                                & \cmark                                                                            & \xmark                                                                       & Screen, App                & 1                                                                          & CO\textsubscript{2}                                                                            & \cmark                                                                        & 186                                                                  \\ 
\hline
INKBIRD IAM-T1~\cite{INKBIRD}           & Low                                                                                               & \cmark                                                                            & \xmark                                                                       & Screen, App                & 1                                                                          & CO\textsubscript{2}                                                                            & \xmark                                                                         & 239                                                                  \\ 
\hline
Luft~\cite{luft}                     & Low                                                                                                & \xmark                                                                             & \xmark                                                                       & Screen, App                & 3                                                                          & Radon, VOC, CO\textsubscript{2}                                                                & \cmark                                                      & 249                                                                  \\ 
\hline
Kaiterra Laser Egg~\cite{kaiterra}       & Low                                                                                                & \xmark                                                                             & \xmark                                                                       & Screen, App                & 2                                                                          & PM\textsubscript{x}, CO\textsubscript{2}                                                                       & \cmark                                                                        & 263                                                                  \\ 
\hline
Temtop LKC-1000E~\cite{temtop}         & \xmark                                                                                                 & \xmark                                                                             & \xmark                                                                       & Screen                     & 2                                                                          & PM\textsubscript{x}, HCHO                                                                      & \xmark                                                                         & 287                                                                  \\ 
\hline
AirKnight~\cite{airknight}                & \xmark                                                                                                & \xmark                                                                             & \xmark                                                                       & Screen                     & 4                                                                          & PM\textsubscript{x}, CO\textsubscript{2}, HCHO, VOC                                                           & \xmark                                                      & 299                                                                  \\ 
\hline
Airthings~\cite{airthings}            & Moderate                                                                                           & \cmark                                                                            & \xmark                                                                       & Screen, App                & 4                                                                          & Radon, PM\textsubscript{x}, CO\textsubscript{2}, VOC  & \cmark                                                                        & 299                                                                  \\ 
\hline
IQAir~\cite{iqair}                    & Low                                                                                                & \cmark                                                                            & \xmark                                                                       & Screen, App                & 2                                                                          & PM\textsubscript{x}, CO\textsubscript{2}                                                                       & \cmark                                                      & 357                                                                  \\ 
\hline
Aranet4 Home~\cite{aranet4}             & Low                                                                                                & \cmark                                                                            & \xmark                                                                       & Screen, App                & 1                                                                          & CO\textsubscript{2}                                                                            & \cmark                                                                        & 442                                                                  \\ 
\hline
Pranaair Sensible~\cite{pranaair_sensi}        & Moderate                                                                                           & \cmark                                                                            & \xmark                                                                       & Screen, App, Web           & 6                                                                          & PM\textsubscript{x}, CO, CO\textsubscript{2}, O\textsubscript{3}, HCHO, VOC                                                   & \cmark                                                                        & 705                                                                  \\ 
\hline
pranaair Sensible+~\cite{pranaair_sensi_plus}   & Moderate                                                                                           & \cmark                                                                            & \xmark                                                                       & Screen, App, Web           & 7                                                                          & PM\textsubscript{x} CO,CO\textsubscript{2}, NO\textsubscript{2}, SO\textsubscript{2}, HCHO, VOC                                               & \cmark                                                                        & 837                                                                  \\ 
\hline
\textit{DALTON} (\textbf{Our}) & High                  & \cmark                                                          & \cmark                                                    & App, Web & 6                                                        & PM\textsubscript{x}, CO,CO\textsubscript{2} C\textsubscript{2}H\textsubscript{5}OH, VOC, NO\textsubscript{2}                                & \cmark                                                      & 250                                                \\
\hline
\end{tabular}}
\end{table}

\section{Discussion \& Limitations}
We develop an interactive, multi-device platform to identify unique pollution patterns present in low to middle-income households. Due to the large-scale deployment of the \ourmethod{} for \datamonths{}, compared to commercially available single-point sensors that trigger frequent false alarms in short-term low-impact pollution spikes (e.g., kitchen), the platform will isolate high-impact long-term pollution exposures (i.e., bedroom, living room, etc.) explained in Section~\ref{sec:insuff_airflow}. This reduces the end user's workload and improves the system's overall utility. Furthermore, the platform accounts for the spread of pollutants from a source (e.g., food waste, cooking, etc.) toward the other rooms of the indoor environment, which is very common in low-income countries, leading to unintentional pollution exposure to infants and old-age people. We can further provide actionable insights to improve air quality by analyzing these spread patterns. For instance, as shown in Section~\ref{sec:inter_spread}, one should not turn on ceiling fans in nearby kitchen rooms when cooking occurs. Additionally, an extensive user study to analyze the qualitative aspects of the \ourmethod{} platform revealed an overall satisfactory platform highlighting the immense potential for improving daily life and promoting health and physiological comfort. \\

\noindent
\changed{$\bullet\;$\textbf{Comparison with Commercial Devices:} In \tablename~\ref{tab:price_comp}, we summarize the features and market price of several commercially available air quality monitoring devices compared to the \textit{DALTON} platform. Based on the table, we can group the devices into two categories as follows: \textit{(i) Low-cost ($<$ \$250):} Majority of the low-cost devices~\cite{pallipartners, yvelines,smiledrive,INKBIRDPLUS,exgizmo} do not stream pollutant measurements to the cloud. Instead, they only show the readings on the built-in display. Few low-cost devices~\cite{netatmo_indoor,luft} send data to the cloud and provide actionable insights (e.g., open windows) but suffer from the unavailability of crucial sensors (i.e., PM\textsubscript{x}, CO, Ethanol, etc.). \textit{(ii) High-cost ($>$ \$250):} On the other hand, high-cost devices~\cite{airknight,airthings,pranaair_sensi,pranaair_sensi_plus} integrate more number of sensors and provide actionable insights and maintenance capabilities similar to the \ourmethod{} platform at a price greater than \$700. Thus, \textit{DALTON} is a low-cost alternative that can be deployed in scale, considering the economic condition of low to middle-income households in developing nations. \textit{DALTON} offers the best of both categories at a low-cost price range (approx. \$250); it incorporates most of the crucial sensors (research-grade, reasonably accurate), providing actionable insights and extensive remote maintenance capabilities. However, unlike the existing commercial devices, \ourmethod{} considers the occupant's feedback to reason about indoor pollution events.}\\ 

\noindent
\changed{$\bullet\;$\textbf{Upfront \& Operating Costs:}
A typical low to middle-income household has three to six rooms. Considering each room has a sensing device, the initial upfront cost of deploying the \ourmethod{} platform is within \$750 to \$1500. In terms of operating cost, each device consumes 3.55 watts at maximum, as shown in \tablename~\ref{tab:ovl_spec}. Therefore, the total power consumption is between 0.255 kWH and 0.510 kWH, depending on the number of rooms in the household. Therefore, the operating cost of the platform is very marginal and viable for a sustainable deployment. Moreover, the platform requires wireless connectivity to offload data storage to the cloud, incurring no additional cost to the user.} \\

\noindent
\changed{$\bullet\;$\textbf{Limitations:} 
Therefore, \ourmethod{} provides a viable alternative as a low-cost, scalable, and interactive pollution monitoring platform for low to middle-income households in developing countries. However, some limitations emerged during field deployment:}

\begin{enumerate}
    \item The static sensing modules measure pollutants up to a certain distance. Hence, we can only monitor an environment up to a certain fidelity with the current platform. It would be impractical for low- to middle-income countries to deploy extensive sensing modules for improving fidelity. In our future work, we plan to integrate wearable devices with the \ourmethod{} platform for fine-grained monitoring. 

    \item The current platform is limited in querying the user and only triggers an alert when it identifies changes in pollutants. Therefore, rely on the user to provide the causal activity via the annotation application. In the future, we plan to use machine learning algorithms on bootstrapping data to formulate intelligent queries, understand the pollution context, and tailor the most probable causal activities.

    \item Sensor module placement is crucial from both a sensing (correctness, alerting) and a usability (power, connectivity) standpoint. Automatic placement of sensor modules given a layout to optimize for preventing health hazards due to pollutants will make the solution much more compelling. The task has been left for future work.

    \item In heavy outdoor pollution-ridden areas, door and window openings exacerbate indoor pollution because outdoor pollution contributes disproportionately. In that case, active measures like air cleaners~\cite{yin2023field,beswick2023room,romero2023strategies} or air filters in split air conditioners~\cite{suksuntornsiri2020effects,sepahvand2023recent} are critical. Although we have selected four \textit{diverse} cities regarding pollution exposure or dynamics, future studies on outdoor pollution-heavy cities or slums are needed.
\end{enumerate}
\section{Conclusion}
This paper introduces a robust and sophisticated IoT platform named \ourmethod{} with various pollution sensors tailor-made for precise monitoring of \textit {indoor health} at scale. We depict our progression from determining optimal system requirements for sustainable large-scale indoor deployments in developing countries to bringing the prototype to life, merging cutting-edge technology with user-centric designs. We deployed the platform in \numcities{} cities spanning over \datamonths{} with \numoccupants{} participants over \numsites{} deployment sites, each with multiple instances of the device; the platform exposed crucial pollution hot-spots that had been neglected due to a lack of information and awareness among residents in low to middle-income households in India. Beyond presenting mere data, the platform identifies the root cause indoor activities behind such precarious pollution hot-spots by introducing an Android app-based user interface, facilitating human-in-the-loop data labeling. Our comprehensive deployment and rigorous user study of the platform ensures the technology is adaptable and scalable for various indoor scenarios, scoring an overall system usability score of $2.04$. This work makes a substantial contribution to the existing literature on air quality monitoring by bringing attention to distinctive pollution patterns in developing countries. Additionally, it sets the stage for the development of closed-loop sensing solutions that prioritize user-inclusive designs.
\section{Ethical Clearance}
The institute's ethical review committee has reviewed and approved the field study (Order No: IIT/SRIC/DEAN/2023, dated 31 July 2023). All participants signed forms consenting to the use of collected pollutant measurements and activity annotations for non-commercial research purposes. The participants received \$50 per week as remuneration during the field study. Moreover, we have made significant efforts to preserve the privacy of the participants while providing necessary information to encourage future research on indoor air pollution.
\section{acknowledgement}
The authors would like to thank the anonymous reviewers and the associate editors for the constructive comments, which have helped to improve the overall presentation of the paper. The research of the first author is supported by the Prime Minister Research Fellowship (PMRF) in India through grant number IIT/Acad/PMRF/SPRING/2022-23, dated 24 March 2023. The work is also supported by Google's Award for Inclusion Research on Societal Computing 2023 for the project proposal ``\textit{AI-Assisted Distributed Collaborative Indoor Pollution Meters: A Case Study, Requirement Analysis, and Low-Cost Healthy Home Solution For Indian Slums.''}

\small
\bibliographystyle{ACM-Reference-Format}
\bibliography{reference}
\end{document}